\documentclass[a4paper,aps,superscriptaddress,nofootinbib]{revtex4}

\usepackage{graphicx}
\usepackage{amsmath,amssymb}
\usepackage{color}
\usepackage{subcaption}
\usepackage{hyperref}
\usepackage[capitalize]{cleveref}
\usepackage{xspace}
\usepackage{caption}
\usepackage{ragged2e}
\DeclareCaptionJustification{justified}{\justifying}
\begin{document}

\begin{flushleft}
    CERN-TH-2025-269  
\end{flushleft}

\title{Probing beyond the Standard Model with gravitational waves from phase transitions}
\author{Chiara Caprini}
\email{chiara.caprini@cern.ch}
\affiliation{Theoretical Physics Department, CERN, 1211 Geneva 23, Switzerland}
\affiliation{D\'epartement de Physique Th\'eorique and Center for Astroparticle Physics, Universit\'e de Gen\`eve, Quai E.~Ansermet 24, CH-1211 Gen\`eve 4, Switzerland}

\begin{abstract}

This review article is based on a seminar presented at the \href{https://www.pi.infn.it/hh2025/}{Higgs pairs workshop 2025}. 
Stochastic gravitational wave backgrounds can serve as probe of the diverse phenomenology encountered in beyond-Standard-Model scenarios featuring phase transitions in the early Universe. 
Focussing on gravitational wave production from first-order phase transitions, we present the main results of a recent analysis by the LISA Cosmology Working Group concerning the detectability of such signals with LISA. 
Strong degeneracies, both among the parameters controlling the phase transition and between these and the parameters of the beyond-Standard-Model scenario underlying the phase transition, complicate the reconstruction of the model from a potential signal. 
Nonetheless, once a specific scenario is assumed, LISA observations can supply constraints possibly complementary to those obtainable from present and future particle colliders.  

\end{abstract}

\maketitle

\section{Introduction}
\label{sec:intro}

Because of the weakness of the gravitational interaction, the Universe is ``transparent'' to gravitational waves (GWs) throughout its evolution. 
Indeed, particles with weaker interactions decouple from thermal equilibrium at correspondingly higher energy scales: supposing a weak interaction cross section for gravitons of the type 
$\sigma\sim G^2 T^2$ where $G$ is the Newton constant, providing an interaction rate $\Gamma(T) =\sigma \,n \,v \sim G^2 T^5$ where $n$ is the particle's number density at thermal equilibrium $n\sim T^3$, and $v=1$ (we work with units $\hbar=c=k_B=1$), the condition for thermal equilibrium in the early Universe reads $\Gamma(T)/H(T)\sim G^2T^5/(T^2/M_{\rm Pl})\sim (T/M_{\rm Pl})^3$, where $H(T)$ is the Hubble factor and $M_{\rm Pl}$ the reduced Plank mass \cite{Maggiore:1999vm}. This shows that gravitons are decoupled at temperatures smaller than the Planck scale.
Therefore, GW emission processes in the early Universe form a fossil radiation, similarly to the Cosmic Microwave Background (CMB), but whose detection has the potential to bring direct information from stages of the Universe's evolution to which we have no access through electromagnetic radiation, photons getting out of thermal equilibrium at the much lower temperature $T_{\rm dec}\simeq 0.26$ eV \cite{Durrer:2020fza}. 
For example, such a fossil GW radiation could be generated 
in the phase of thermal evolution of the Universe comprised between the end of Inflation and Big Bang Nucleosynthesis (BBN).
The detection of such a signal might constitute a direct observational probe of processes like reheating after Inflation, spontaneous symmetry breakings, baryogenesis, the production of dark matter. 

GW signals from the early Universe have therefore an amazing discovery potential in cosmology, comparable to the one of the CMB sixty years ago, but 
linked to higher energies. 
This naturally raises the question of whether they can be used as a new probe of high energy physics, complementary to particle colliders. The answer depends both on the capabilities of the available GW observatories, and on the characteristics of the potential GW sources. 
As we will show, current GW observatories are serendipitously well suited to probe most interesting energy scales in the early Universe, provided that appropriate sources exist with the right features (see e.g.~\cite{Caprini:2018mtu,Maggiore2,Athron:2023xlk}), and that the challenges related to the 
detection of the signal can be properly addressed (see e.g.~\cite{Romano:2016dpx,LISA:2024hlh,Baghi:2023qnq,Karnesis:2019mph,Santini:2025iuj,Pozzoli:2024hkt,Boileau:2022ter,Hindmarsh:2024ttn}). 

This review article is based on a seminar given at the \href{https://www.pi.infn.it/hh2025/}{Higgs pairs workshop 2025}, and focusses on the generation of GWs from first-order phase transitions (PTs), possibly occurring in the early Universe. 
PTs are a compelling GW source, with diverse consequences and relevant connections with high energy physics. 
After a description of the general properties of the GW signal in \cref{sec:general}, mainly based on \cite{Caprini:2018mtu}, in 
\cref{sec:PT} we briefly overview possible GW sources linked to PTs, as well as some of the proposed beyond-Standard-Model (BSM) scenarios realising them.   
In 
\cref{sec:FOPT} we focus on first-order PTs. 
In \cref{sec:sources} we briefly review some of the main GW generation processes operating at first-order PTs, as well as the features of the GW signal: this section is based on \cite{Caprini:2024ofd}. 
In \cref{sec:EWPTatLISA} we discuss in particular the case of the Electroweak (EW) symmetry breaking, testable by LISA, and review the results on the detectability of this signal obtained in the last work of the LISA Cosmology Working Group on this topic, Ref.~\cite{Caprini:2024hue}.
This work also assessed the capabilities of LISA to constrain two particular scenarios of first-order PTs that are particulary compelling: we review these results in \cref{sec:twopp}.
We conclude in \cref{sec:conc}. 
The background metric of the Universe is the Friedmann Lema\^itre Robertson Walker (FLRW) metric $ds^2=-dt^2+a^2(t) dx_i dx^i$ and GWs are inserted at first-order in cosmological perturbation theory.  

\section{Motivation: characterisation and detection of stochastic GW backgrounds}
\label{sec:general}

Let us consider a GW generating process operating during the radiation dominated, thermal equilibrium phase in the early Universe, at a temperature $T_*$ and over a short time interval, typically less than one Hubble time $H(T_*)^{-1}$. 
It is natural to assume that this process 
has a characteristic length scale that is bounded, by causality, by the Hubble length: $\ell_*\leq H(T_*)^{-1}$.
This length scale corresponds to an angular size on the sky today $\Theta_*=\ell_*/d_{A,*}\leq H(T_*)^{-1}/d_{A,*}$, where $d_{A,*}$ is the angular diameter distance \cite{Durrer:2020fza}. 
For example, the angular scale corresponding to the Hubble scale at the epoch of photon decoupling $T_{\rm dec}\simeq 0.26\, {\rm eV}$ is $\Theta_{\rm dec}= H(T_{\rm dec})^{-1}/d_{A,{\rm dec}}\simeq 0.9$ deg, as measured by the first peak of the CMB angular power spectrum \cite{Durrer:2020fza}. 
As motivated in the introduction, interesting GW generation processes typically operate at higher energy scales, i.e.~earlier times, than photon decoupling: therefore,  
the angular scales corresponding to their correlation scales are typically much smaller.
For example, a GW signal produced by a process operating at the EW energy scale would be correlated on the sky today on an angular scale $\Theta_*(T_*\simeq 100\,{\rm GeV})\simeq 10^{-12}$ deg; if produced at the QCD energy scale, it would be correlated on $\Theta_*(T_*\simeq 100\,{\rm MeV})\simeq 10^{-9}$ deg \cite{Caprini:2018mtu}.
These angular scales are way smaller than the typical angular resolution of GW detectors on the sky, which is
$\Delta\Theta\sim (fL)^{-1} $ with 
$f$ the sensitivity frequency and 
$L$ the detector baseline: for example, for LISA it is about 10 deg (see e.g.~\cite{LISACosmologyWorkingGroup:2022kbp}). 
Consequently, GW detectors measure the GW signal from the superposition of many uncorrelated regions on the sky, and thereby can only access its statistical properties. 
Therefore, from the point of view of GW detectors, the signal coming from sources operating in the early Universe takes the form of 
a stochastic GW background (SGWB), and the tensor metric perturbation $h_{ij}(\mathbf{x}, t)$ must be treated as a random variable. 

The latter is in general assumed to be statistically homogeneous and isotropic (because of the symmetries of the FLRW Universe), unpolarised (if the generation process does not violate parity) and Gaussian (because of the central limit theorem). 
%
The most immediate way to characterise the present-day signal is therefore through its power spectral density $S_h(f)$, 
\begin{equation}
    \langle {h}_{r}(f, \mathbf{\hat{k}}) {h}_{p}^*(g, \mathbf{\hat{q}})\rangle =\frac{1}{8\pi} \,
\delta(f - g) \, \delta^{(2)}(\mathbf{\hat{k}} - \mathbf{\hat{q}}) \, \delta_{r p} \, S_h(f)\,,
\label{eq:PSD}
\end{equation}
where ${h}_{r}(f, \mathbf{\hat{k}})$ are the coefficients in the plane wave expansion 
$h_{ij}(\mathbf{x}, t) = \int_{-\infty}^{+\infty} df \int d \mathbf{\hat{k}} \, h_r(f,\mathbf{\hat{k}})e_{ij}^{r}(\mathbf{\hat{k}}) \,
e^{2\pi if(t- \mathbf{\hat{k}} \cdot \mathbf{x})}$,
and we sum over the polarisations $r=+,\times$.
The bracket
$\langle ... \rangle$ in \cref{eq:PSD}
denotes the ensemble average over the random variable, 
which can be substituted with 
volume (or time, from the point of view of the detector) average under the ergodic hypothesis, as customary 
in cosmology. 
We consider the present-day signal as a superposition of plane waves, since the source active in the early Universe has long ceased. 
In the cosmological context, rather than $S_h(f)$, one often uses  $\Omega_{\rm GW}(f)$,
the power spectrum 
of the GW energy density per logarithmic frequency interval, normalised by the 
critical energy density of the Universe today $\rho_c$: 
\begin{equation}
    \frac{\rho_{\rm GW}}{\rho_{\rm c}} =\frac{\langle \dot h_{ij}(\mathbf{x}, t)\dot h^{ij}(\mathbf{x}, t)\rangle}{32 \pi G \,\rho_c}
= 
\int_0^{+\infty} \frac{df}{f}\,\Omega_{\rm GW}(f)\,,~~~~~{\rm with}~~\Omega_{\rm GW}(f) =  \frac{\pi}{2}\frac{f^3 S_h(f)}{G\,\rho_c}\,. 
\label{eq:rho_omega}
\end{equation}

The spectral shape of $\Omega_{\rm GW}(f)$ depends on the properties of the GW source. 
In the case under analysis of a GW source operating over a short time interval during the radiation dominated era and characterised by the length scale $\ell_*$, the GW signal has a characteristic frequency $f_*\sim 1/\ell_*$. 
The fractional GW energy density power spectrum $\Omega_{\rm GW}(f)$ typically peaks at this frequency, redshifted to today:
\begin{equation}
    f_0=f_*\frac{a(T_*)}{a(T_0)}= \frac{1.65\times 10^{-5} }{\ell_*H_*}  \left(\frac{g(T_*)}{100}\right)^{\frac{1}{6}}\frac{T_*}{100\,\mathrm{GeV}}\, \mathrm{Hz}\,,
    \label{eq:f0}
\end{equation}
where we have expressed $f_0$ in terms of the dimensionless ratio ${\ell_*H_*}\leq 1$, $a(T)$ is the scale factor and $T_0\simeq 2.7$ K is the temperature of the Universe today, and $g(T_*)$ denotes the number of relativistic degrees of freedom at temperature $T_*$. 
\cref{eq:f0} connects the present-day characteristic frequency of a GW signal $f_0$, with the epoch in the early Universe during which the sourcing process was active, parametrized by $T_*$.
It thus also provides a way to associate each GW observatory - defined by its operational frequency range - with the minimal energy scale in the early Universe to which it is sensitive (see also \cref{fig:Tbeta}). 

Let us first consider the currently operating network of Earth-based interferometers LIGO, Virgo and KAGRA (LVK) \cite{LIGOweb,Virgoweb,KAGRAweb}. 
Their frequency range of detection spans $1\,{\rm Hz}\lesssim f \lesssim 1000\,{\rm Hz}$. 
Setting ${\ell_*H_*}=\mathcal{O}(1)$, \cref{eq:f0} therefore selects the temperature range 
$10^6 \,{\rm GeV}\lesssim T_*\lesssim 10^{10}\,{\rm GeV}$. 
Interestingly, this range corresponds to the energy scale of the Peccei-Quinn PT \cite{Peccei:1977hh, Peccei:1977ur,Kim:1979if, Shifman:1979if,Dine:1981rt, Zhitnitsky:1980tq} in ``post-inflationary'' models, within which, the axion decay constant $F_a$ is bounded by
$10^{7-8}\,{\rm GeV}\lesssim F_a \lesssim 10^{10-11}\,{\rm GeV}$ \cite{Raffelt:1990yz, MillerBertolami:2014rka,Ayala:2014pea,Dolan:2022kul,Bar:2019ifz, Chang:2018rso, Carenza:2020cis,Notari:2022ffe,Gorghetto:2018myk, Gorghetto:2020qws, Buschmann:2021sdq, Saikawa:2024bta}. 
While the detection of a SGWB of cosmological origin by the LVK networks is unlikely, as it will most probably be masked by the astrophysical foreground from stellar mass black hole binaries \cite{LIGOScientific:2025bgj,LIGOScientific:2025kry}, 
the situation might be entirely different for the planned 3rd generation of Earth-based GW detectors,
Einstein Telescope (ET) \cite{ETweb} and the Cosmic Explorer (CE) \cite{CEweb}. 
These should be operative around the middle of the 2030s, with an expected factor of 20 improvement in sensitivity \cite{Maggiore:2024cwf}, 
and the level of foreground
from astrophysical sources is expected to be much smaller \cite{Branchesi_2023, Bellie:2023jlq,Zhou:2022otw,Zhou:2022nmt,ET:2025xjr}.
This opens up the exciting possibility of probing the cosmological $U(1)_{\rm PQ}$ PT, and thereby exploring a possible solution of the CP problem, with GW detectors \cite{VonHarling:2019rgb,ZambujalFerreira:2021cte,Caprini:2024ofd}. 

Moving to lower frequencies, the GW detector that is expected to be first operative is the space-based interferometer LISA, planned 
for launch around 2035 \cite{LISAweb}. 
LISA will be composed of a triangular configuration of three spacecraft on  an heliocentric orbit with 2.5 million km arms, and will be sensitive to the lower frequency range $10^{-5}\,{\rm Hz}\lesssim f \lesssim 0.1\,{\rm Hz}$ \cite{LISA:2024hlh}.
Again setting ${\ell_*H_*}=\mathcal{O}(1)$, \cref{eq:f0} selects for LISA the temperature range 
$10 \,{\rm GeV}\lesssim T_*\lesssim 10^{5}\,{\rm GeV}$. 
Therefore, LISA offers the possibility to probe the
EW energy scale and beyond \cite{Binetruy:2012ze,Caprini:2015zlo,Caprini:2019egz}: in particular, as we will show, LISA could probe BSM scenarios in which the EWPT becomes of first-order, possibly in a complementary way to particle colliders \cite{Caprini:2024hue}.
That is, if the SGWB from the early Universe is higher than the numerous foregrounds of
astrophysical origin that are expected in the LISA band, both of galactic and extra-galactic nature \cite{Farmer:2003pa,Lamberts:2019nyk,Staelens:2023xjn,Korol:2021pun,Karnesis:2021tsh,Babak:2023lro,Lehoucq:2023zlt,Piarulli:2024yhj}. 

In the range of yet lower frequencies $10^{-9}\,{\rm Hz}\lesssim f\lesssim 10^{-7}\,{\rm Hz}$ lies the first observational evidence so far of the existence of a SGWB permeating the Universe. 
Indeed, in 2024, Pulsar Timing Arrays (PTAs) have  announced the presence in their datasets of a common red noise that follows the expected response of pulsars to GWs, the Hellings-Downs correlation \cite{EPTA:2023fyk,NANOGrav:2023gor,Reardon:2023gzh,Xu:2023wog}. 
This correlation allows to ascribe, with confidence ranging from about 3$\sigma$ to about $4.5\sigma$ depending on the PTA collaboration, the observed red noise to a proper GW signal.  
The current data will be combined within the International PTAs, and this analysis is expected to turn the evidence into a discovery
\cite{Antoniadis:2022pcn}.
An extremely precise characterisation of the SGWB is then expected with the advent of the Square Kilometer Array Observatory \cite{Rawlings2011TheSK,Janssen:2014dka}.
The most immediate explanation for this signal  
consists in the superposition of the GW emission from a population of inspiralling supermassive black hole binaries~\cite{EPTA:2023xxk, NANOGrav:2023hfp}.
However, cosmological sources operating in the early Universe are for the moment not excluded: rather, some of them could explain the detection equally well, if not better \cite{NANOGrav:2023hvm, EPTA:2023xxk,Figueroa:2023zhu,Ellis:2023oxs}. 
Within this scenario, applying again \cref{eq:f0} 
with ${\ell_*H_*}=\mathcal{O}(1)$, the selected energy scale for PTAs would correspond to 
$10\,{\rm MeV}\lesssim T_* \lesssim 1\,{\rm GeV}$: therefore, PTAs are ideally suited to probe the QCD energy scale. 
The PTA signal could then be due to GW sources  operating in connection with the QCD PT, offering the possibility to probe the conditions under which it actually took place in the early Universe \cite{Caprini:2010xv,Neronov:2020qrl,RoperPol:2022iel,Cline:2025bwe,Zheng:2024tib}. 

The 
aforementioned GWs observatories happen to be ideally suited to probe key energy scales in the early Universe. 
It is imperative to exploit this serendipitous coincidence, given its transformative scientific potential. 
{Moreover, the outlook is particularly promising, as the observational landscape is expected to expand significantly. Additional frequency bands are likely to be explored in the future as new detector concepts mature.
For instance, the deci-Hz window could become accessible within the next few decades through proposals based on both laser \cite{Sedda:2019uro,Perego:2025bif} and atom interferometry \cite{AEDGE:2019nxb,Badurina:2019hst,AION:2025igp,Abend:2023jxv,Abdalla:2024sst}, or by measuring Moon's vibrations  \cite{Ajith:2024mie,Jani:2025uaz}. 
Lower frequencies than PTAs
may also be probed via extremely precise astrometric measurements \cite{Book:2010pf,Mihaylov:2018uqm,Moore:2017ity,Crosta:2024udx,Klioner:2017asb,Darling:2018hmc,Garcia-Bellido:2021zgu}. 
Most intriguing from the perspective of high-energy physics, however, are detector concepts operating in the MHz-GHz range, which would enable exploration of even more extreme energy scales, potentially reaching those associated with Grand Unification and with reheating after Inflation  \cite{Aggarwal:2020olq,Aggarwal:2025noe,Domcke:2024eti,Berlin:2023grv,Domcke:2022rgu}.}

However, being in the right frequency range alone does not guarantee detection: signals must also lie sufficiently above the observatories’ sensitivity limit. 
This means, that the GW source operating in the early Universe must be sufficiently `strong'.
A rough estimate of the scaling of the SGWB signal amplitude with the characteristics of the source can be obtained from the following simple argument \cite{Caprini:2018mtu}. 
As before, we assume that the source is operating over a short time interval, and denote its characteristic size $\ell_*$. 
GW are sourced by tensor anisotropic stresses $\Pi_{ij}$, that is, the transverse traceless component of the energy momentum tensor of the source. 
The GW wave equation reads $\ddot{h}_{r}+ 3 \, H \, \dot{h}_{r}+k^2h_{r}= 16\pi G \, \Pi_{r}$, from which one can estimate the GW amplitude scaling as $ h \sim 16\pi G \,\Pi \,\ell_*^2$. 
For this order of magnitude estimate we neglect indices and identify the time evolution of the source with is characteristic length-scale $\ell_*$, i.e.~we assume a relativistic source. This is indeed a good assumption, as we will see later on.
The GW energy density therefore becomes (see the first equality in \cref{eq:rho_omega}) 
$\rho_{\rm GW}^* \sim \dot h^2/(32\pi G) \sim 8 \pi G\, \Pi^2\,\ell_*^2$.
Normalising to the total energy density in the Universe at the time of GW production $\rho_{\rm tot}^* = 3 H_*^2 / (8 \pi G)$, one has $
\rho_{\rm GW}^*/{\rho_{\rm tot}^*} \sim \left({H_*}{\ell_*}\right)^2 \, \left({\Pi}/{\rho_{\rm tot}^*}\right)^2
$, leading to the GW parameter today (c.f.~\cref{eq:rho_omega})
\begin{equation}
    \Omega_{\rm GW}\sim \Omega_{\rm rad}\,\left({H_*}{\ell_*}\right)^2 \, \left(\frac{\Pi}{\rho_{\rm tot}^*}\right)^2\,,
    \label{eq:scaling}
\end{equation}
where $\Omega_{\rm rad}=\rho_{\rm rad}^0/\rho_c\simeq 2.47\cdot 10^{-5}$ is the radiation energy density parameter today \cite{Planck:2018vyg}, and we have used the fact that GWs scale as radiation with the evolution of the Universe. 
The amplitude of the GW signal scales like the square of the source characteristic size normalised by the Hubble scale, and with the square of the 
amount of anisotropic stresses available to source the GWs, normalised to the energy density in the Universe at the source time. 
It is important to stress that \cref{eq:scaling} was obtained under highly simplified assumptions and is not intended to represent the full SGWB signal. Most notably, it does not contain the frequency dependence of the spectrum.
Nevertheless, it provides a useful proxy for 
general considerations. 
Comparing for example with the sensitivity of LISA $\Omega_{\rm GW}\sim 10^{-11}$ \cite{LISA:2024hlh},
we see that only sources which involve a sizeable fraction of the total energy density in the Universe, and have sizes comparable to the Hubble scale, are strong enough to produce detectable signals: 
$(H_*\ell_*) (\Pi / \rho_{\rm tot})_* \gtrsim 10^{-3}$. 
In the next section we describe a class of sources for which these conditions can be met.

\section{Gravitational waves from phase transitions in the early Universe}
\label{sec:PT}

PTs are one of the phenomena occurring in the early Universe with the richest phenomenology in terms of GW production. 
Many processes connected to PTs naturally lead to tensor anisotropic stresses and thereby to GWs.  
We can distinguish two broad classes of GW sources: first-order PTs (for reviews, see e.g.~\cite{Caprini:2024ofd,Caldwell:2022qsj,Athron:2023xlk,Roshan:2024qnv})
and PT producing topological defects (for reviews, see e.g.~\cite{Caprini:2024ofd,Vilenkin:2000jqa,Hindmarsh:1994re,Vachaspati:2015cma}).

\begin{itemize}
    \item {\bf first-order PTs:}
in this case, the anisotropic stresses arise from the out of equilibrium processes linked to the first-order PT.
In particular, the PT proceeds through the nucleation of bubbles where the field undergoing the PT, typically a scalar field (e.g.~the Higss field), has settled to the true vacuum. 
Towards the end of the PT, bubbles collide in order to convert the entire Universe to the broken phase. 
The collisions of the broken phase bubbles breaks the spherical symmetry of the scalar field spatial gradients, and can produce anisotropic stress of the form $\Pi_{ij}\sim[\partial_i\phi\partial_j\phi]^{TT}$, where $TT$ stands for the transverse and traceless projection \cite{Kosowsky:1991ua,Kosowsky:1992vn,Huber:2008hg,Cutting:2018tjt,Cutting:2020nla,Caprini:2007xq,Jinno:2016vai}. 
Since the scalar field is in general coupled to the particles of the surrounding early Universe plasma, the latter is set into motion by the bubble expansion \cite{Kamionkowski:1993fg,Espinosa:2010hh,Konstandin:2017sat,Ellis:2019oqb,Lewicki:2022pdb}.
Bulk fluid motion develops in the form of sound waves \cite{Hogan:1986qda,Hindmarsh:2015qta,Hindmarsh:2017gnf,Cutting:2019zws,Hindmarsh:2019phv} or kinetic turbulence \cite{Dolgov:2002ra,Caprini:2006jb,Gogoberidze:2007an,Kahniashvili:2008pe,Caprini:2009yp,Brandenburg:2017neh,Niksa:2018ofa,RoperPol:2019wvy,Auclair:2022jod}, leading to anisotropic stresses in the form $\Pi_{ij}\sim [\gamma^2(\rho+p) v_iv_j]^{TT}$, where $\rho$ and $p$ are the energy density and pressure of the fluid, and $\gamma$ the Lorentz factor (relativistic motion is typically favourable to large GW production). 
Finally, large-scale electromagnetic fields can be induced along with the bubble dynamics and/or the bulk fluid motion, that naturally possess anisotropic stress: $\Pi_{ij}\sim[-E_iE_j-B_iB_j]^{TT}$ \cite{Caprini:2001nb,Caprini:2009pr,Kahniashvili:2008er,Kahniashvili:2009mf,Durrer:2013pga}. 
These can be due, for example, to electromagnetic currents generated by particle separation at the bubble walls \cite{Baym:1995fk,Cheng:1994yr,Sigl:1996dm}, and can get then amplified up to equipartition with the kinetic turbulence energy. 
Another important source of GWs in first-order PTs can arise if the vacuum energy gets released into particle production rather than bubble walls dynamics \cite{Giudice:2024tcp,Cataldi:2024pgt}. 
The anisotropic stresses sourcing GWs are then those associated to the out of equilibrium particle distribution function \cite{Jinno:2022fom,Inomata:2024rkt}.

\item {\bf topological defects:}
in this case, the topological defects are directly the cause of the
tensor anisotropic stresses sourcing the GWs \cite{Durrer:2001cg,Figueroa:2012kw,Auclair:2019wcv,Simakachorn:2022yjy}. 
Topological defects may form when a symmetry is spontaneously broken, and their nature depends on the topology of the vacuum submanifold \cite{Kibble:1976sj,Vilenkin:2000jqa,Hindmarsh:1994re,Vachaspati:2015cma}: for example, if it is disconnected, domain walls may form, or if it contains loops which
cannot be shrunk to a point, cosmic strings may form. 
These are the two cases most studied in the literature.
The properties of the defects are determined by the action of the theory: of particular importance are for example their stability, the energy per unit length/surface, their interactions. 
Both cosmic strings and domain walls form a network in the Universe that produces all types of metric perturbations \cite{Durrer:1999na,Durrer:2001cg}, among which also tensor perturbations, i.e.~GWs 
{\cite{Saikawa:2017hiv}}.
In the case of local strings - arising e.g.~from the spontaneous breaking of a local $U(1)$ symmetry \cite{Vilenkin:2000jqa,Hindmarsh:1994re}, the closed loops are the strongest source of GWs, as opposed to the infinitely long strings extending outside the Hubble scale \cite{Damour:2000wa,Damour:2001bk,Binetruy:2009vt}. 
Global strings, on the other hand, mainly decay via particle production \cite{PhysRevD.9.2273,DAVIS1988219,1987PhLB..195..361H,PhysRevD.32.3172,Battye:1997jk,Brandenberger:1986vj,Vachaspati:2009kq,Long:2014mxa,MacGibbon:1989kk,Steer:2010jk,Steer:2010jk}. 
Both cosmic strings and domain walls networks evolve reaching a scaling regime. 
However, while the energy density in the cosmic strings network always remains a constant fraction of the Universe energy density, the decay of domain walls is instead very 
slow \cite{Vilenkin:2000jqa,Vachaspati:2006zz}. One must therefore introduce an annihilation mechanism in the domain walls model, acting before the domain walls network comes to dominate the Universe expansion: for example, a
small explicit breaking of the discrete symmetry inherent to the domain walls \cite{Kitajima:2023cek,Ferreira:2024eru}. 
This means that domain walls are in general a short-lived source, and their GW signal follows the prescription of \cref{eq:scaling}; while cosmic strings source GW throughout the Universe evolution, and therefore their GW signal does not satisfy the scaling of \cref{eq:scaling}. 

\end{itemize}

A general condition that must always be satisfied, both in the case of GW production by first-order PTs and by topological defects, is that the
 homogeneity and isotropy of the Universe observed at large scale by the CMB must be maintained. 
In the first case, the anisotropic stresses can reach an important fraction of the Universe energy density, but in general this occurs at small scales, and over a short amount of time, with respect to the Hubble length/time. CMB limits are then safe.
In the second case, the metric perturbations induced at large scales by the string or domain wall network must satisfy CMB constraints \cite{Ade:2013xla,Ringeval:2010ca}. 
Given the weakness of the gravitational interaction, the sourced SGWB is always 
far below the upper bounds on the radiation energy density derived by the CMB or BBN \cite{Caprini:2018mtu}. 

The PTs known in the context of the Standard Model (SM) do not lead to any appreciable GW production. 
Both the EWPT and the QCD PT are predicted to be cross-overs \cite{Aoki:2006we,Stephanov:2007fk,Gurtler:1997hr,Aoki:1996cu,Kajantie:1995kf,Kajantie:1996mn,Laine:1998jb}, and no other PT leading to topological defects is expected within the SM. 
GWs are indeed produced by the SM plasma in thermal equilibrium \cite{Ghiglieri:2020mhm}, but their amplitude is too small to be of observational relevance, given the capabilities of proposed GW detectors so far \cite{Aggarwal:2025noe}. 
Consequently, observational evidence of the presence of first-order
PTs and/or topological defects in the early Universe constitutes a significant test of new physics.
Several scenarios have been explored, that could provide this observational evidence in the form of GWs, sometimes even with other, very relevant, associated  signatures, such as the presence of 
dark matter candidates, baryogenesis, and solutions to the hierarchy problem:

\begin{itemize}
    \item {\bf EW sector extensions:}
one of the most notable example of such scenarios are BSM extension of the EW sector.  
Despite the absence of observational indication of new physics
near the EW energy scale, many 
scenarios leading to a first-order EWPT remain viable. 
They mainly rely on extending the SM with light scalars, that can alter the Higgs potential producing a barrier either at tree level or perturbatively. 
One can for example directly add a gauge singlet scalar field coupling at tree level to the SM Higgs field, possibly endowed with a $\mathbb{Z}_2$ symmetry that could make it stable and
thereby contribute a Dark matter candidate \cite{McDonald:1993ey,Espinosa:1993bs,Espinosa:2007qk,Profumo:2007wc,Espinosa:2011ax,Barger:2011vm,Cline:2012hg, Alanne:2014bra, Curtin:2014jma, Vaskonen:2016yiu, Kurup:2017dzf, Beniwal:2017eik, Niemi:2021qvp, Lewicki:2021pgr, Ellis:2022lft, Dev:2019njv,Craig:2014lda,Morrissey:2012db, Huang:2018aja,Azatov:2022tii,Cline:2021iff}. 
In the latter case, the PT can proceed in two steps \cite{Espinosa:2011ax, Profumo:2007wc}. 
Another scenario leading to a first-order EWPT proposes to extend the SM Higgs sector with scalar EW multiplets, as for example in the model featuring two Higgs doublets \cite{Huet:1995mm, Cline:1996mga, Fromme:2006cm, Cline:2011mm, Dorsch:2013wja, Dorsch:2014qja, Kakizaki:2015wua,
Dorsch:2016nrg, Basler:2016obg, Dorsch:2017nza, Basler:2017uxn, Bernon:2017jgv, Huang:2017rzf}, or with a  triplet \cite{Patel:2012pi,Chala:2018opy,Blinov:2015sna, Inoue:2015pza,Blinov:2015qva}.
Going beyond minimal extensions, albeit constrained, SUSY models provide a UV-complete theory 
that predicts new light particles which effect could be to change the order of the EWPT, such as for example in singlet/multiplet extensions of the MSSM
\cite{Huber:2015znp,Bian:2017wfv, Demidov:2017lzf,Georgi:1985nv, Cort:2013foa,Garcia-Pepin:2016hvs}.

\item {\bf effective approach:}
a different approach, adapted to the case in which the new physics 
influencing the EWPT 
is at higher energy than the EW scale, 
is to represent the
heavy new physics by effective higher dimensional operators  \cite{Zhang:1992fs,Grojean:2004xa,Delaunay:2007wb,Bodeker:2004ws,Harman:2015gif}.
This allows to study the impact of heavy BSM extensions without worrying about a specific theory. 
The Higgs potential is modified by adding new terms of higher power of the Higgs field, suppressed by the scale of the new physics. 
Whether this approach can correctly capture the EWPT features in complete theories 
depends on the model, however, it is clear that by adding those operators the EWPT can become strongly first-order \cite{Damgaard:2015con,deVries:2017ncy,Chala:2018ari,Huber:2007vva,Huang:2016odd}. 

\item {\bf new symmetries:}
moving away from the EW symmetry breaking, a minimal extension of the SM via
a local $U(1)_{B-L}$ symmetry is also 
very interesting, 
since it can possibly explain the neutrino masses, 
provide dark matter candidates in the form of 
right-handed neutrinos, 
explain the baryon asymmetry of the Universe via leptogenesis
and lead to detectable GWs since the PT of the $U(1)_{B-L}$ symmetry breaking in the
early Universe tends to be strongly first-order due to the classical conformal invariance \cite{Iso:2009ss,Iso:2009nw,Escudero:2018fwn,Dasgupta:2022isg,Jinno:2016knw,Marzo:2018nov,Sagunski:2023ynd}. 

\item {\bf conformal models:}
going beyond the 
scenarios of weakly coupled theories with 
polynomial potentials, 
there are models featuring strong dynamics such as warped extra-dimensions \cite{Creminelli:2001th,Randall:2006py,Nardini:2007me,Hassanain:2007js,Konstandin:2010cd,Konstandin:2011dr,Bunk:2017fic,Dillon:2017ctw, vonHarling:2017yew, Megias:2018sxv, Fujikura:2019oyi} and composite Higgs models \cite{Panico:2015jxa,Grojean:2013qca,Csaki:2017eio,Grojean:2004xa,Delaunay:2007wb,Grinstein:2008qi}.
In both cases, the 
confinement PT associated with the spontaneous breaking of the conformal invariance is strongly first-order, and in composite Higgs scenario it can be  naturally linked to the EW one. 

\item {\bf dark sectors:}
strong first-order PTs can arise in the context of 
dark sectors, which motivate the existence of dark matter (eluding any observation other than based on its gravitational interaction)
within a more complete theory. 
Dark sectors only interact with the SM through a portal, and feature all possible kinds of new interactions and
symmetries: they can therefore very naturally host strong first-order PTs \cite{Schwaller:2015tja,Jaeckel:2016jlh,Chala:2016ykx,Addazi:2016fbj,Baldes:2017rcu,Addazi:2017gpt,Tsumura:2017knk,Aoki:2017aws,Croon:2018new,Croon:2018erz,Baldes:2018emh,Madge:2018gfl,Breitbach:2018ddu,Fairbairn:2019xog}.
Particularly interesting are models in which the first-order PT is required 
for other phenomenological 
reasons, such as producing the baryon asymmetry, stabilising the dark matter candidate or providing the right relic abundance \cite{Baldes:2017rcu,Baldes:2018emh,Madge:2018gfl,Baker:2016xzo,Baker:2017zwx,Greljo:2019xan}. 

\item {\bf QCD and heavy axions:}
proposed long ago to solve the strong CP problem, the Peccei Quinn model is a 
well motivated BSM model with concrete applications \cite{Peccei:1977hh, Peccei:1977ur,Kim:1979if, Shifman:1979if,Dine:1981rt, Zhitnitsky:1980tq}.
The axion, arising as a pseudo-Nambu-Goldstone boson from the breaking of the $U(1)_{PQ}$ symmetry, 
can be a stable dark matter candidate; furthermore, 
this model can lead to a variety of GW sources. 
Indeed, the 
spontaneous breaking of the $U(1)_{PQ}$ symmetry can be 
realised via a strong first-order PT, directly sourcing GWs \cite{VonHarling:2019rgb,DelleRose:2019pgi,Dine:1981rt, Zhitnitsky:1980tq,Baratella:2018pxi,Gorghetto:2020qws,Caprini:2024ofd}. Furthermore, this model also features the formation of topological defects, 
including global cosmic strings from the $U(1)_{PQ}$ symmetry breaking (which mostly decay via emission of axions) but also of 
domain walls, produced when 
the $U(1)_{PQ}$ is broken to a discrete symmetry \cite{ZambujalFerreira:2021cte, Ferreira:2022zzo}, which are particularly interesting for GW production in the context of heavy axion models \cite{Holdom:1982ex,Treiman:1978ge, Dimopoulos:1979pp, Tye:1981zy,Holdom:1985vx, Flynn:1987rs, Choi:1998ep, Agrawal:2017ksf, Kitano:2021fdl,Rubakov:1997vp, Gherghetta:2016fhp, Gherghetta:2020ofz,Berezhiani:2000gh, Dimopoulos:2016lvn, Hook:2019qoh}.
To avoid domain wall domination, an explicit breaking of the discrete symmetry is needed, introducing a vacuum non-degeneracy that makes the domain wall annihilate \cite{Ferreira:2022zzo,Sikivie:1982qv}. 
This is one of the few models in which topological defects can lead to a detectable GW signal other than those in which the defects are generated by the spontaneous breaking of higher symmetry patterns connected to Grand Unified Theories. 

\item {\bf QCD PT:}
it is also important to mention that, although the QCDPT is also predicted to be a cross-over, this result has been obtained from lattice simulations run at
zero baryon and charge chemical potentials, as appropriate within the SM.
However, the actual conditions in which the PT took place in the early Universe might be different, for example the  lepton asymmetry in the Universe is much less constrained than the baryon one, and could be large, which might lead to a change in the order of 
the QCDPT \cite{Schwarz:2009ii,Wygas:2018otj,Middeldorf-Wygas:2020glx,Chatterjee:2025wbz}. 

\end{itemize}

A still open question, subject of interesting ongoing research, is to which level perturbative treatments of the
finite temperature potential in first-order PTs can be considered accurate, given that in many scenarios the barrier is not present at tree level, but induced by radiative and/or finite-temperature
corrections.
A promising direction is to implement dimensional reduction, a resummation technique based on matching
to a three-dimensional effective field theory, allowing non-perturbative treatments of the PT in the associated four-dimensional theory and thereby improved precision of the finite temperature
dynamics \cite{Athron:2023xlk,Ekstedt:2022bff,Ekstedt:2024etx,Kainulainen:2019kyp,Gould:2019qek,Niemi:2018asa,Croon:2020cgk,Kierkla:2023von,Niemi:2021qvp,Schicho:2021gca,Gould:2023ovu,Lewicki:2024xan,Banerjee:2024qiu}. 

{The brief and necessarily incomplete overview of GW-producing scenarios given above 
already illustrates their remarkable diversity. 
It is important to emphasise, however, that in many cases the resulting GW signal, despite arising from highly non-trivial processes, 
exhibits too few distinctive features to unambiguously discriminate between the different underlying scenarios. 
This is because the signal typically depends only on the gross properties of the tensor anisotropic stresses that source it, such as their typical time and length scales, rather than on the details of the high-energy physics scenarios producing them.
In the following, we aim at illustrating this point with a concrete example, focusing in particular on GWs generated during first-order PTs.}

\section{Focus on first-order phase transitions}
\label{sec:FOPT}

In the previous section, we have provided a very concise summary of some of the best motivated BSM scenarios that can lead to sizeable GW signals, 
because they involve
first-order PTs and/or the formation of topological defects. 
Given the variety of models, the question naturally arises whether it is possible to distinguish, from the detection of the GW signal, one scenario from the other. 
In the following 
we present the results of \cite{Caprini:2024hue}, in which an attempt is made to answer this question specifically for the case of a first-order PT 
occurring at the EW energy scale, 
considering its observability by the LISA interferometer.

\subsection{Gravitational wave sources operating at a first-order phase transition}
\label{sec:sources}

A first-order PT, in particular if related to the EW symmetry breaking, must complete within one Hubble time, so that the entire Universe settles in the new vacuum of the theory. 
Furthermore, as explained in \cref{sec:PT}, the anisotropic stresses are connected to the collision of bubbles: the GW source therefore has a well defined associated length scale, the size of the bubbles at collision, which is smaller than the Hubble scale at the PT time $H_*^{-1}$. 
Therefore it seems plausible that a first-order PT, as a GW source, fulfils the criteria of applicability of \cref{eq:scaling}. 
Let us start working from this hypothesis. 

The characteristic size of the GW source in this case is given by $\ell_*\sim v_w/\beta$, where $v_w$ denotes the bubble wall velocity, and $\beta$ is the transition rate parameter, 
which can be evaluated directly from the action $S(t)$ \cite{Coleman:1977py,Callan:1977pt,Linde:1980tt,Linde:1981zj}. 
Within the assumption of exponential nucleation, the probability of tunnelling per unit volume and time can be written as
$\Gamma(t) = \Gamma(t_*)\exp[(\beta(t-t_*)]$, with $\beta =\left.\frac{d}{dt} S(t)\right|_{t_*}$, where $t_*$ denotes a reference time, which in this case should be chosen around  the time of bubble percolation (for more detail, see the discussion on $T_*$ later in this section) \cite{Turner:1992tz,Megevand:2016lpr,Hindmarsh:2020hop}. 
The inverse timescale $\beta$ can be used as a proxy for the inverse duration of the PT (for more detail, see e.g.~\cite{Caprini:2024ofd}). 

The bubble wall velocity, also entering in $\ell_*\sim v_w/\beta$, is notoriously difficult to estimate. 
This quantity links the microscopic scales of the problem - it depends on the effective potential, i.e.~the pressure difference in the two phases, and on the interaction of the particles with the bubble wall, i.e.~with the field undergoing the PT -  to the  
macroscopic scales of the problem, i.e.~the bubble size, and the collective plasma dynamics at this size.
Indeed, the fluid profile around the wall depends on the bubble wall velocity.  
In the case of thermal PTs, it is often assumed that the bubble quickly reaches a steady state, characterised by a constant wall velocity given by the balance between the driving force (the pressure difference in the two phases) and the friction force (due to the interaction of the wall with the particles in the surrounding plasma) \cite{Ignatius:1993qn,KurkiSuonio:1995vy,KurkiSuonio:1995pp,Espinosa:2010hh,Hindmarsh:2019phv,Giese:2020rtr}. 
One then solves the fluid equations of motion imposing conservation of energy and momentum across the bubble discontinuity, and finds the fluid velocity and enthalpy profiles surrounding
the bubble \cite{Espinosa:2010hh}
(implemented e.g.~in the tool \texttt{CosmoGW} \cite{Cosmogw}, see also \cite{Ekstedt:2024fyq}).
It is possible to adopt a phenomenological description introducing a macroscopic friction parameter, in principle covering several particle theory models. 
However, it is very challenging to properly model the bubble wall velocity given the variety of possible BSM scenarios and its high model dependence \cite{Moore:1995si,Moore:1995ua,Bodeker:2009qy,Konstandin:2014zta,Kozaczuk:2015owa,Hoeche:2020rsg,BarrosoMancha:2020fay,Laurent:2022jrs,Huber:2013kj,Friedlander:2020tnq,Laurent:2020gpg,Lewicki:2021pgr,Konstandin:2014zta,Dorsch:2018pat,Jinno:2017fby,Ellis:2020nnr,Lewicki:2022pdb}. 
Therefore, $v_w$
is mostly assumed as an input in studies
of the GW signal, despite the fact that the signal does depend quite strongly on its actual value.

As presented in \cref{sec:general}, the characteristic size of the GW source $\ell_*$ determines the characteristic 
frequency of the SGWB $f_*$, which in turns connects to the frequency of the signal today $f_0$ through the temperature  $T_*$, representing the epoch in the early Universe during which the GW source was
active, see \cref{eq:f0}. 
Note that \cref{eq:f0} assumes that the Universe was in thermal equilibrium and in the radiation dominated phase. 
This means, that the PT must already have completed, or be about to complete, at $T_*$. 
Since the GW sourcing is due to the collision of the bubbles, and therefore takes place towards the end of the phase transition (in the case bulk fluid motion, even after the PT completion, as we will see), this assumption is reasonable.
One option is then to  associate to $T_*$ the percolation temperature, 
{i.e.~the ambient temperature when a connected group of bubbles spans the entire universe}  \cite{Athron:2022mmm,Athron:2023xlk}.
{Another option, particularly adapted to the case of exponential nucleation, is to associate $T_*$ to the time $t_*$ at which the fractional volume in the metastable phase has reached $P(t_*)=1/e$, so that roughly 63\% of the Universe is converted to
the broken phase \cite{Hindmarsh:2020hop}.
The two times are close, since $P(t_{\rm perc})=0.71$ \cite{Athron:2022mmm,Athron:2023xlk}.} 
In the absence of substantial supercooling, the radiation component dominates the Universe's evolution throughout the PT, 
{which, furthermore, must occur within one Hubble time:} the percolation temperature is then close to both the critical and the nucleation temperatures, 
{and the conversion between time and temperature can be performed using Friedmann's equation for a radiation dominated Universe.}
Conversely, in the case of substantial supercooling, the Universe is dominated by the vacuum energy for a short period of time: 
{the critical, nucleation and percolation temperatures may substantially differ, and it is therefore important to distinguish  them and to identify $T_*$ as close to the latter. 
Since, during the short inflationary phase, the volume is rapidly expanding,
one further needs to ensure that the fractional volume which
is still in the false vacuum is decreasing at percolation time, so that the PT can complete \cite{Turner:1992tz,Athron:2022mmm}.
Moreover, thermal equilibrium must be re-established via a 
reheating process: the easiest option is then}
to postulate that the latter is instantaneous, such that the Hubble rate during the vacuum dominated phase (and in particular at percolation) is the same as the Hubble rate in the thermal phase following the PT, $H_*$ \cite{Ellis:2019oqb,Ellis:2018mja,Ellis:2020nnr,Caprini:2024ofd}. 
{Within this setting, $T_*$ can be unambiguously identified.}

The other important term in \cref{eq:scaling} is the anisotropic stress energy fraction $\Pi/\rho^*_{\rm tot}$. 
One usually defines the GW source energy fraction $K=\rho_s/\rho^*_\mathrm{tot}$, where 
$\rho_s$ denotes the energy density of the GW sourcing process, and $\rho^*_\mathrm{tot}$ denotes the total energy density in the Universe at the PT time, assumed to be radiation to derive \cref{eq:scaling}.
When the anisotropic stresses are due to the colliding bubble walls, $\rho_s$ is the gradient energy in the scalar field, which we denote $\rho_\varphi$. 
When the anisotropic stresses are due to 
the bulk fluid motion caused by the coupling between the bubble wall and the surrounding plasma, 
$\rho_s$ is the kinetic energy of the bulk fluid motion, which we denote $\rho_v$.
The PT is a dynamical process in which the free energy of the system, given by the effective potential at finite temperature, is transformed mainly into
thermal energy (with no associated anisotropic stress), and in small part into gradient energy of the bubble walls and kinetic energy of the fluid. 
In the following, we always make the assumption that the equation of state of the system can be described by the  the bag equation of state, and that the sound speed of the fluid is constant and fixed to $1/\sqrt{3}$. 
Then, the relative importance of the potential to the thermal energy is given by the parameter $\alpha=\Delta V_0/[\pi^2g (T_*) T_*^4/30]$, where $\Delta V_0$ denotes the potential difference at zero temperature between the symmetric and broken phases. 
$\alpha$ is interpreted as a measure of the PT strength. 
The efficiency with which the potential energy is transformed into gradient energy of the bubble walls or kinetic energy of the fluid 
is given by the efficiency parameters
$\kappa_{\varphi,v}=\rho_{\varphi,v}/\Delta V_0$, so that the GW source energy fraction can be written as $K=(\kappa_\varphi+\kappa_v)\alpha/(1+\alpha)$. 
The efficiency parameters are
in general calculated using for $\rho_{\varphi,v}$  
the gradient/kinetic energy density of a single bubble \cite{Espinosa:2010hh}. 

The parameter $\alpha$ governs 
which process, among the scalar field gradients or the bulk fluid motion, dominates the GW production. 

\begin{itemize}
    \item If the PT is strong, $\alpha \gtrsim \mathcal{O}(1)$ and the vacuum energy is equal or larger than the relativistic fluid energy: 
the PT entails some amount of supercooling, possibly a short inflationary phase. 
Bubbles are assumed to accelerate until the speed of light, and the dominant GW source is bubble collisions. 
The anisotropic stresses in the scalar field energy momentum tensor are due to the breaking of the spherical symmetry connected to the bubble collisions \cite{Kosowsky:1991ua,Kosowsky:1992vn,Huber:2008hg,Cutting:2018tjt,Cutting:2020nla,Caprini:2007xq,Jinno:2016vai,Konstandin:2017sat,Ellis:2019oqb,Lewicki:2022pdb}. 

\item
If the PT is weak, $\alpha \lesssim \mathcal{O}(10^{-2})$ the potential energy is subdominant, and 
the GW production 
is dominated by the fluid motion.
Since the PT is weak, so are the velocity/enthalpy perturbations in the fluid generated by the expanding bubble walls: the system is characterised by the development of sound waves, which surround the bubbles \cite{Hogan:1986qda,Hindmarsh:2015qta,Hindmarsh:2017gnf,Cutting:2019zws,Hindmarsh:2019phv}. 
The superposition of sound waves following bubble collision leads to the breaking of the spherical symmetry and the presence of anisotropic stresses in the fluid energy momentum tensor. 
Remarkably, sound waves remain in the fluid for long after the 
bubble walls are collided. 
They therefore continue sourcing GWs for many Hubble times, an important qualitative difference with respect to bubble collision. 

\item If the phase transition is of moderate strength, $\alpha\sim \mathcal{O}(0.1)-\mathcal{O}(1)$, the GW production is still dominated by fluid motion, however, the velocity/enthalpy perturbations in the fluid can be high, leading to non-linear compressional and vortical turbulence, possibly of the magnetohydrodynamic (MHD) type, i.e.~accompanied by magnetic fields. 
The chaotic superposition of fluid velocity typical of turbulence then leads to non-zero anisotropic stresses in the fluid energy momentum tensor \cite{Dolgov:2002ra,Caprini:2006jb,Gogoberidze:2007an,Kahniashvili:2008pe,Caprini:2009yp,Brandenburg:2017neh,Niksa:2018ofa,RoperPol:2019wvy,RoperPol:2022iel,Auclair:2022jod}. 
As well as sound waves, turbulence also continues to source GWs for many Hubble times, after the bubbles are gone.

\end{itemize}

Note that the GW source energy fraction $K$ as defined in the literature and above, differs from the anisotropic stress energy fraction $\Pi/\rho^*_{\rm tot}$. 
It is indeed very difficult to estimate 
what fraction of the gradient energy in the scalar field, or of the bulk kinetic energy, is in the form of anisotropic stresses, efficient in sourcing GWs. 
The reasons are multiple.
For example, in the case of anisotropic stresses from bubble collisions, they depend on the particular realisation of the collision process. 
Although analytical estimates exists \cite{Caprini:2007xq,Jinno:2016vai,Jinno:2017fby}, the best way to tackle this is through numerical simulations of the bubble nucleation process, possibly capturing also the microscopic scale corresponding to the bubble wall, in order to study the scalar field dynamics at and after bubble collisions \cite{Huber:2008hg,Child:2012qg,Cutting:2018tjt,Konstandin:2017sat,Cutting:2020nla,Lewicki:2022pdb,Jinno:2019bxw}. 
If the anisotropic stresses are sourced by bulk fluid motion, the latter sources GW for long after the bubbles have collided, so the memory of the actual collision realisation is erased. 
However, the difficulty still stands in the fact that the bulk fluid motion can become highly non linear, especially in the case of moderately strong PTs. 
Again, analytical estimates exists \cite{Hindmarsh:2016lnk, Hindmarsh:2019phv,Cai:2023guc,RoperPol:2023dzg,Kosowsky:2001xp,Dolgov:2002ra,Caprini:2006jb,Gogoberidze:2007an,Caprini:2009yp,Niksa:2018ofa,RoperPol:2022iel,Auclair:2022jod,Giombi:2025tkv}, but it is imperative to validate them with numerical simulations of the fluid dynamics \cite{Hindmarsh:2013xza,Hindmarsh:2015qta,Hindmarsh:2017gnf,Cutting:2019zws,Jinno:2020eqg,Jinno:2022mie,Caprini:2024gyk,Brandenburg:2017neh,RoperPol:2019wvy,RoperPol:2018sap,RoperPol:2021xnd,Brandenburg:2021bvg,Correia:2025qif,Sharma:2023mao,Dahl:2021wyk}.  

Therefore, in general, 
numerical simulations are necessary 
to properly model the GW signal, because of the intrinsic randomness of the process, of the complicated fluid shells profiles surrounding the bubbles, of the non-linear fluid dynamics. 
At this stage, we can rewrite \cref{eq:scaling} in a more realistic way:
\begin{equation}
    \Omega_{\rm GW}(f)\sim \Omega_{\rm rad} \, \tilde\Omega_{\rm GW}\, K^2\,\mathcal{F}({H_*}{\ell_*})\,S(f)\,.
    \label{eq:scaling_complete}
\end{equation}
The above equation still does not capture all the complexity 
of the SGWB signal, but conveys some more information than \cref{eq:scaling}. 
First of all, to follow conventions, we have rewritten 
$\Pi/\rho^*_{\rm tot}=\tilde\Omega_{\rm GW}\, K^2$, where $\tilde\Omega_{\rm GW}$ parametrises the efficiency with which the 
energy fraction $K$ is converted into anisotropic stresses actually sourcing the GWs. This parameter must in general be determined via numerical simulations (for more detail, see e.g.~\cite{Caprini:2024gyk} and references therein). 
Second, we have inserted the generic function $\mathcal{F}({H_*}{\ell_*})$ to express the fact that the scaling as $\left({H_*}{\ell_*}\right)^2$
can be modified in the case of bulk fluid motion, since 
this kind of process, contrary to bubble collisions, can source GWs for a time 
longer than a Hubble time, deviating from the conditions under which \cref{eq:scaling}
was derived. 
At last, we have reinserted the 
frequency dependence in $\Omega_{\rm GW}(f)$ (c.f.~\cref{eq:PSD}), the normalised SGWB power spectrum $S(f)$. 
For more detail on the scaling and properties of the SGWB signal, see e.g.~\cite{Caprini:2024ofd}.

Numerical simulations allow to predict the SGWB from the  different sources operating during a first-order PT in detail. 
In particular, they allow to 
link the PT strength to the actual energy available in the GW source, that is~to provide estimates of the efficiency parameters
$\kappa_{\varphi,v}$, and in turns of $K$. 
As previously mentioned, the efficiency parameters are in general evaluated from the single bubble solution:
numerical simulations allow to check whether this is a good assumption. This is still very much work in progress, but discrepancies with the analytical single-bubble solution are often found \cite{Hindmarsh:2017gnf,Jinno:2020eqg,Jinno:2022mie,Caprini:2024gyk}. 
Furthermore, numerical simulations provide insight on 
which sources, among bubble collision, sound waves and turbulence, dominate the GW production, and how they are connected among each other. 
They also predict the signal amplitude and its spectral shape $S(f)$. 
Each source produces power spectra with different features, such as different peaks and slopes, and it is necessary to be able to link these features to the actual GW source and to the PT parameters, for a correct interpretation of the signal. 
In this respect, complementing numerical results with analytical understanding is of great help. 
Detailed predictions of the SGWB signals are still much work in progress. 
Several numerical codes exist that tackle the problem with different approaches: for example, numerical simulations performed with the SCOTTS code \cite{Hindmarsh:2013xza,Hindmarsh:2015qta,Hindmarsh:2017gnf,Cutting:2019zws,Auclair:2022jod} include the scalar field dynamics and the relativistic fluid motion, 
but not the expansion of the Universe neither the magnetic field; numerical simulations performed with the Pencil code \cite{PencilCode:2020eyn,Brandenburg:2017neh,RoperPol:2019wvy,RoperPol:2018sap,RoperPol:2021xnd,Brandenburg:2021bvg} include the expansion of the Universe and the magnetic field, but no scalar field; Higgsless simulations \cite{Jinno:2020eqg,Jinno:2022mie,Caprini:2024gyk} are the least computationally expensive, 
but rely on steady state solutions for the bubble expansion. 
Constant progress is being made by the community 
in refining SGWB predictions, in particular in the
most relevant case of strong first-order PTs.

In the following, we do not discuss the detailed form of the SGWB spectral shape 
$S(f)$. 
The precise determination of $S(f)$ depends on the specific GW sourcing mechanism: bubble collision, sound waves, (MHD) turbulence.
Much like the scaling of the signal with the PT parameters, this 
remains an active area of research, combining input from analytical approaches and numerical simulations. 
Reviews of this topic have been attempted, for example, in the analyses of the LISA Cosmology Working Group~\cite{Caprini:2015zlo,Caprini:2019egz,Caprini:2024hue}, where expressions for $S(f)$ are proposed that capture the main qualitative features 
of the frequency dependence for the various GW sources. 
However, providing an unified description that encompasses all possible dependencies and variations of 
$S(f)$ across different sourcing  scenarios remains extremely challenging \cite{Caprini:2024ofd}.
This review is based on the results of Refs.~\cite{Caprini:2024hue},
which are obtained adopting  state-of-the-art proposals for the signal spectral shape.

\subsection{Detecting signals from a first-order EWPT at LISA}
\label{sec:EWPTatLISA}

The GW signal is determined by the dynamics of the PT and of the surrounding plasma,
and can in principle be computed in the context of specific PT models. 
In practice, however, this evaluation involves many complex steps, and it is subject to significant uncertainties.
Nevertheless, it is still worthwhile to investigate the question of signal detectability using the information currently available, acknowledging that the resulting predictions inevitably carry intrinsic uncertainties.

Within state of the art SGWB signal modelling, the following parameters enter in the GW signal:
\begin{itemize}
    \item $T_*$, the energy scale of the PT (to be identified with the temperature around percolation, see discussion in \cref{sec:FOPT}); $\alpha$ the PT strength; $\beta/H_*$ the inverse PT duration, normalised to the Hubble rate at the PT time. These parameters can be determined directly by the effective potential, and can be subject to large uncertainty when related to a specific model, e.g.~due to the accuracy of perturbative treatments, or in the case of important supercooling. 
    \item $v_w$, the bubble wall velocity. This parameter 
    depends on~the finite-temperature effective potential and the interaction of the scalar field with the particles in the theory, and determines the fluid profile around the wall. 
    Is has been calculated only in some specific models, and is in general fixed to a constant value, often arbitrary, possibly introducing inaccuracy in the GW signal prediction.  
    \item $K$, the GW source energy fraction. This parameter is determined by the bubble expansion dynamics and the bubble interaction with the surrounding fluid. 
    In most cases, it is evaluated from the efficiency factors $\kappa_{\varphi,v}$ and $\alpha$, under the assumption that the bubble wall settles to a steady state and adopting the fits of Ref.~\cite{Espinosa:2010hh}.
    This introduces some inaccuracy in estimating the GW signal,
    because the effect of collisions and possibly non-linearities developing in the fluid dynamics can significantly alter $K$ with respect to the single bubble solution.
   \item $\epsilon$, the fraction of  kinetic energy in bulk motion that is
   converted to MHD turbulence. 
   This parameter is inserted to represent the fact that, if the PT is strong, non-linearities in the bulk fluid develop, 
   {
   likely leading to the onset of turbulence and possibly to the amplification of magnetic fields. 
   Accurately simulating PTs of adequate strength to follow the formation and evolution of (MHD) turbulence, and thus to reliably predict the resulting GW signal, remains extremely challenging.
Although rapid progress is being made in the development of numerical codes (see the simulations performed in \cite{Cutting:2019zws} and in particular the approach developed in \cite{Jinno:2022mie,Caprini:2024gyk}), current simulations are still unable to robustly determine the fraction of energy transferred into (MHD) turbulence as a function of the PT properties, such as its strength. 
Consequently, the actual value of $\epsilon$ and its dependence on the other PT parameters are not yet known, and $\epsilon$ must be treated as a free parameter.} 
\end{itemize}

\begin{figure}
    \centering
    \includegraphics[width=0.5\linewidth]{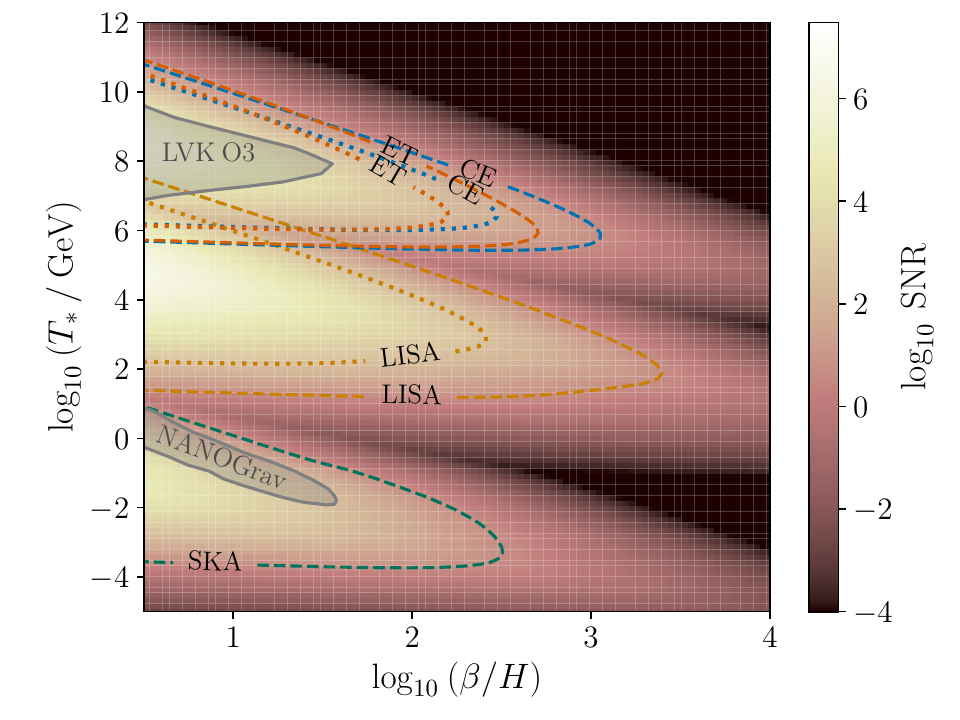}
    \caption{Taken from \cite{Caprini:2024ofd}. This figure shows the maximum signal to noise ratio (colour shading) among four GW observatories (PTAs with the SKA, LISA, Einstein Telescope and Cosmic Explorer) of the SGWB generated by a strong first-order PT in the parameter space given by the PT temperature $T_*$ and the inverse PT duration, $\beta/H_*$. 
   {The dashed lines show the SNR=5 curves for each detector, evaluated based on the detector noise only, without accounting for the presence of astrophysical foregrounds. 
    The dotted lines show instead the SNR contours of the astrophysical foregrounds relevant in each detector's frequency range. 
    More detail on the population of astrophysical sources taken into account for each detector is given in the main text. 
    Within the region delineated by the dotted contours, the SNR of the cosmological signal exceeds the one of the foregrounds, and the former should therefore be readily detectable. 
    Within the region bounded by the dashed contours, detection may still be achievable, depending on the detailed spectral shape of the cosmological signal. 
    In the case of Einstein Telescope and Cosmic Explorer, the foregrounds lie below the detectors' noise, while for LISA the Galactic foreground component is higher than the detector noise, hence the reduction in accessible parameter space due to the foreground is larger (c.f.~\cite{Caprini:2024ofd} for more detail). 
    For PTA, there is no dotted curve as the foreground corresponds to the SGWB that has been detected, and one should refer to the gray contour instead. Indeed,}
    the gray contours denote the 95\% confidence exclusion region from LVK non-detection of a SGWB \cite{Badger:2022nwo}, and the 95\% confidence region if the PTA NANOGrav signal is interpreted as originating from a strong first-order PT  \cite{NANOGrav:2023hvm}. 
    The 
     temperature intervals inferred from the intersection of 
     the SNR curves of the different observatories with the $\beta/H_*=1$ axis 
correspond to those discussed in \cref{sec:general}, representing the energy scales in the early Universe at which each observatory can probe GW sources.}
    \label{fig:Tbeta}
\end{figure}

\cref{fig:Tbeta}, taken from \cite{Caprini:2024ofd}, shows 
the reach of several GW observatories in the two-dimensional parameter space $T_*$ 
and $\beta/H_*$, 
considering the SGWB 
generated by a strong first-order PT with $\alpha\gg 1$.
In this simple case, the dominant GW source is bubble collisions, so that $\kappa_\varphi\simeq 1$, and therefore $K= \kappa_\varphi \,\alpha/(1+\alpha)\simeq 1$. 
Furthermore, the bubbles expansion can accelerate to the speed of light $v_w\simeq 1$, and there is no bulk motion, neither in the form of sound waves nor turbulence, therefore 
$\epsilon=0$. 
In \cite{Caprini:2024ofd}, detectability is assessed in terms of the signal-to-noise ratio (SNR), the colour shading in \cref{fig:Tbeta}. 
Two SNR contours are shown for each GW observatory: 
the dashed one has been produced 
accounting only for the 
instrumental noise, while the dotted one takes into account also the extra noise due to astrophysical foregrounds. 
Specifically, for the future Earth-based interferometers Einstein Telescope and Cosmic Explorer, the foregrounds are due to the mergers of compact binaries, mainly stellar mass black holes and neutron stars (note that the currently operating interferometers LIGO, Virgo, KAGRA are not yet sensitive enough to detect any astrophysical foreground).
For LISA, the main foregrounds are due to the in-spiral of compact binaries (stellar mass black holes, neutron stars and white dwarfs) both within the Milky Way and extra-galactic. 
In particular, the galactic foreground dominates the LISA noise curve in the frequency range nearby the mHz. 
For PTAs, accounting for the capability of the future radio telescope Square Kilometre Array Observatory, 
the foreground, from the point of view of cosmological signal detection, is given by the in-spiral of
super-massive black hole binaries. 
A detailed description of the foregrounds and references can be found in \cite{Caprini:2024ofd}.
From \cref{fig:Tbeta} it is clear that the presence of foregrounds can significantly affect the sensitivity of GW observatories to a possible cosmological signal.
Nevertheless, presently operating devices have already started delving into 
the parameter space: 
the gray regions 
show both the exclusion region 
from the non-observation of the SGWB from a strong first-order PT by the LVK network \cite{Badger:2022nwo}, and the parameters space region that could account for the NANOGrav 15-years data, if the SGWB is interpreted as arising from a first-order PT \cite{NANOGrav:2023hvm}. 
Incidentally, 
the temperature ranges inferred, for each GW observatory, from the intersections of the SNR curves with the $\beta/H_*=1$ 
axis, reflect the energy scales at which each detector can probe GW sources, as discussed in \cref{sec:general}. 

From \cref{fig:Tbeta}, one sees that at least part of the strong first-order PT parameter space 
lies within the reach of current and future GW observatories. However, detectability of the signal does not by itself guarantee 
that the underlying parameters can be reconstructed from the measurement. 
Within the caveats and uncertainties discussed above, most of the parameters entering the SGWB signal are known, once a  PT model is specified and complemented 
by numerical simulations of the plasma dynamics. 
Therefore, extracting these parameters from an observation would be extremely valuable, as it would enable to identify the physical origin of the GW signal from its detection.

\begin{figure}
    \centering
    \includegraphics[width=\linewidth]{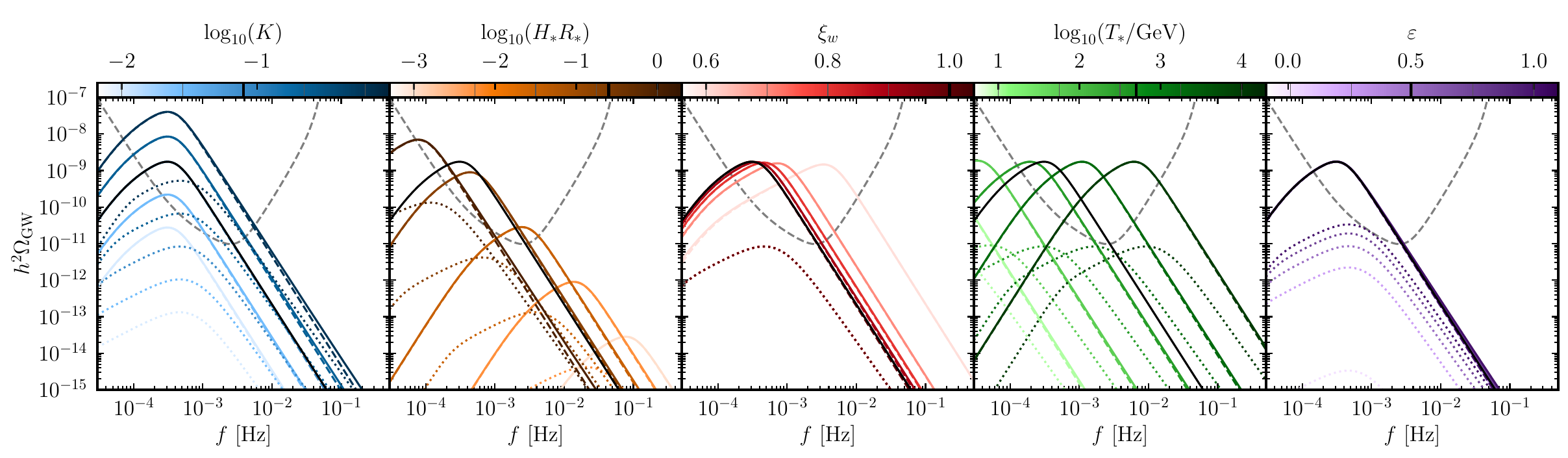}
    \caption{Taken from \cite{Caprini:2024hue}: SGWB power spectra from sound waves (dashed, coloured lines), MHD turbulence (dotted, coloured lines), and total, given by the sum of the two (solid, coloured lines), as a function of frequency. 
    The black line in each panel shows the SGWB for the benchmark values of the parameters $K = 0.08$, $H_*\ell_* = 0.25$ (the notation of \cite{Caprini:2024hue} is such that $R_*=\ell_*$), $v_w = 1$ (the notation is such that $\xi_w=v_w$), $T_* = 500$ GeV, $\epsilon = 0.5$. 
    In each panel one parameter is varied around the benchmark value, while the others are kept fixed. The value of the varying parameter can be inferred from the colour shading of the horizontal bar. The gray dashed line appearing in each panel shows the forecasted noise curve of the LISA instrument. }
    \label{fig:degeneracy}
\end{figure}

\begin{figure}
    \includegraphics[width=0.75\textwidth]{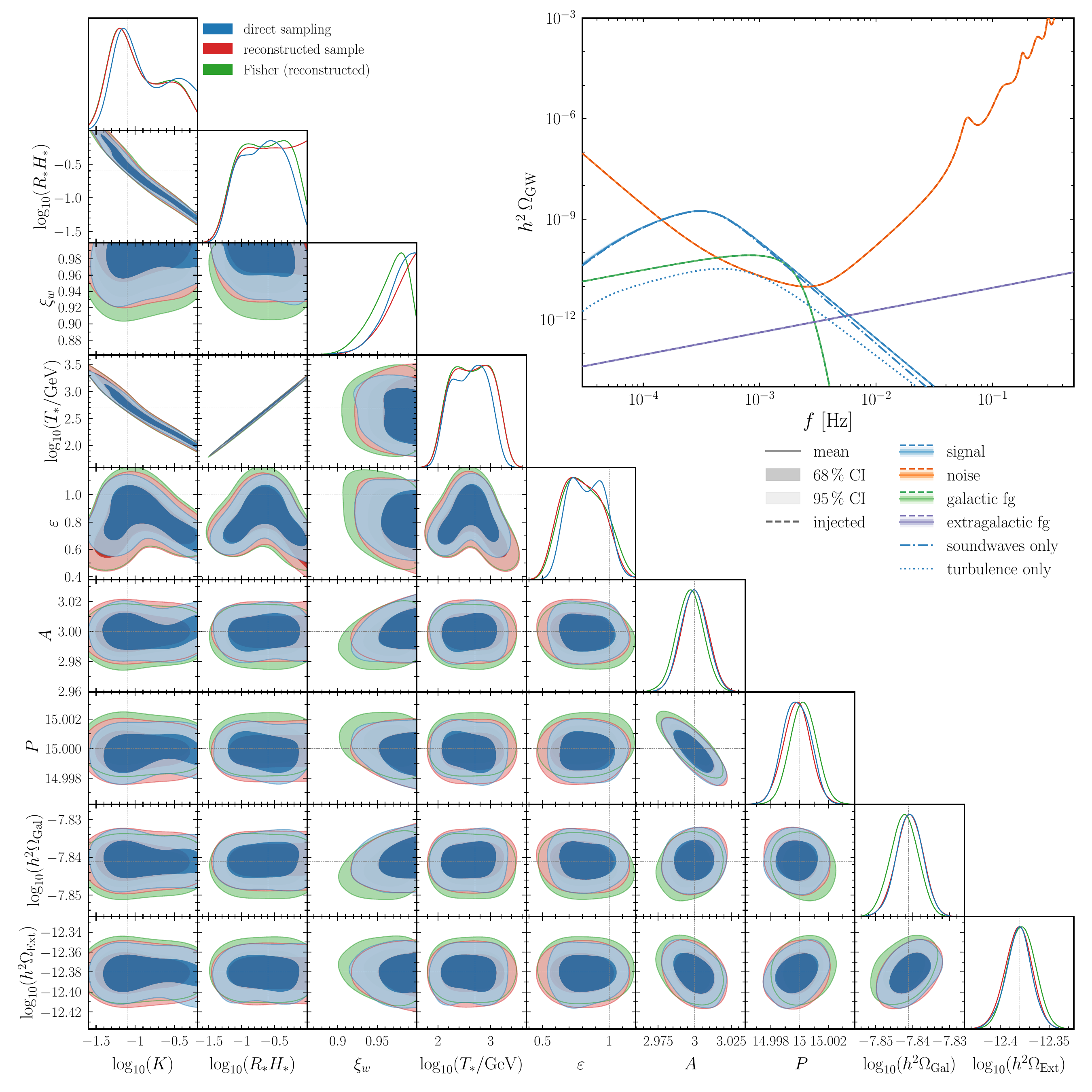}
    \caption{Taken from \cite{Caprini:2024hue}:
        Template-based
reconstruction of the thermodynamic parameters 
of a SGWB from sound waves and MHD turbulence at a moderately strong first-order PT, together with the noise and foreground parameters 
{(note that the notation of \cite{Caprini:2024hue} is such that $R_*=\ell_*$ and $\xi_w=v_w$).}
The posteriors of the direct sampling in terms of the thermodynamic parameters are shown in blue. 
The background red
contours are reconstructed from a sample in terms of the geometric parameters of the double broken power law, and translated to the thermodynamic parameters.
The green contours are obtained from a Fisher analysis in terms of the geometric parameters, also translated to the thermodynamic parameters.
The inset shows the injected noise, foregrounds and signal (according to the legend), and their the reconstruction (shaded areas around the lines). 
    }
    \label{fig:reco_sw+turb}
\end{figure}

Ref.~\cite{Caprini:2024hue} investigated the parameter-reconstruction capabilities of LISA and found that, for sufficiently strong signals, the underlying parameters can in principle be inferred. However, the resulting uncertainties are often large, owing to the strong degeneracies in the way these parameters affect the SGWB.
These degeneracies can be appreciated in the example provided in \cref{fig:degeneracy},
where the dependence of the SGWB from sound waves and MHD turbulence 
on the parameters listed above is shown. 
For example, $K$ enters in the SGWB amplitude; 
however, $(H_*\ell_*)$ also strongly affects the amplitude, making it difficult to disentangle the effect of the two parameters 
{(note that, in Figs.~\ref{fig:degeneracy} and \ref{fig:reco_sw+turb}, the parameter $\ell_*$ is denoted $R_*$, since these figures are taken from Ref.~\cite{Caprini:2024hue} and therefore follow the notation convention used in that analysis)}. 
Furthermore, $(H_*\ell_*)$
determines the SGWB peak frequency; however, 
$T_*$ does as well, introducing another 
source of 
degeneracy. 

\begin{figure}
    \includegraphics[width=0.7\textwidth]{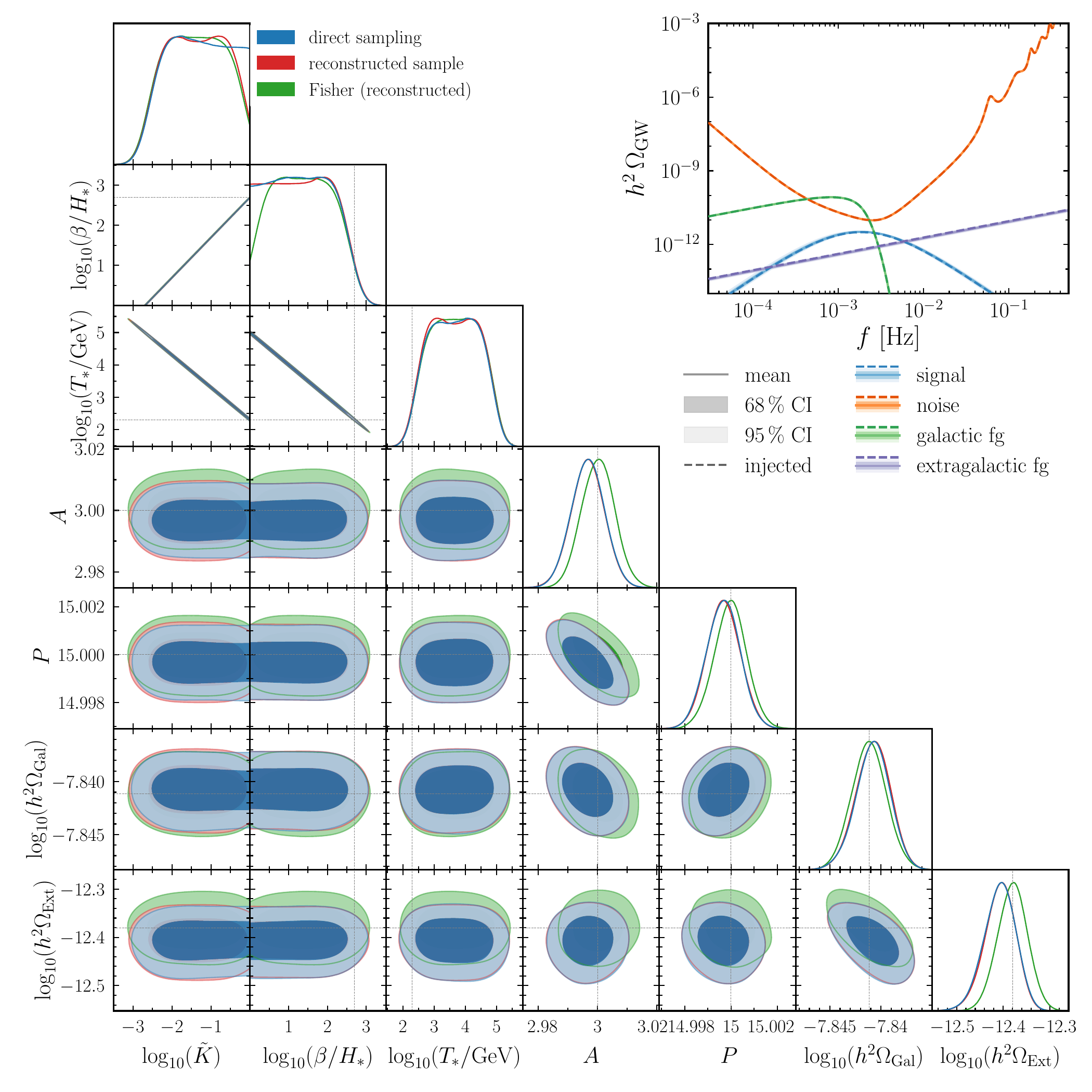}
    \caption{Taken from \cite{Caprini:2024hue}:
Template-based
reconstruction of the thermodynamic parameters 
of a SGWB from bubble collisions at a strong first-order PT, together with the noise and foregrounds parameters. 
The posteriors of the direct sampling in terms of the thermodynamic parameters are shown in blue. 
The background red
contours are reconstructed from a sample in terms of the geometric parameters of the broken power law, translated to thermodynamic parameters.
The green contours are obtained from a Fisher analysis in terms of the geometric parameters, also translated to the thermodynamic parameters.
The inset shows the injected noise, foregrounds and signal (dashed lines  according to the legend), and their the reconstruction (shaded areas around the dashed lines). }
    \label{fig:reco_bubble}
\end{figure}

To partially alleviate these degeneracies, Ref.~\cite{Caprini:2024hue} proposes an alternative parametrisation of the SGWB signal in terms of features of its spectral shape, rather than in terms of the thermodynamic PT parameters discussed above. 
In this approach, the SGWB from bubble collisions, which has the form of a broken power law, is characterised by the frequency of its spectral peak, $f_b$ and by the value of the SGWB amplitude at the peak, $\Omega_b$. 
The SGWBs from sound waves and MHD turbulence, which instead follow double broken power laws, are parametrised by the two break frequencies  $(f_1,f_2)$ (with one corresponding to the spectral maximum) and by the amplitude at the second break,  $\Omega_2$. 
These quantities are referred to as ``geometric parameters'', in contrast to the thermodynamic parameters of the PT described above.
Reconstruction in terms of geometric parameters is significantly more tractable, and the degeneracies can be largely removed when the signal is sufficiently strong. Ref.~\cite{Caprini:2024hue} forecasts the regions of geometric parameter space that LISA will be able to measure with better than 10\% precision. 
While the geometric parameters do not directly encode the physical properties of the PT, they are extremely useful for reconstructing the SGWB from LISA data; once measured, they can subsequently be translated back into constraints on the thermodynamic PT parameters, as shown in \cref{fig:reco_bubble,fig:reco_sw+turb}.
Unfortunately, the translation to the thermodynamic  parameters 
brings back the degeneracies.

The parameter reconstruction analysis of Ref.~\cite{Caprini:2024hue}
is performed on simulated data produced by the \texttt{SGWBinner} code \cite{Caprini:2019pxz,Flauger:2020qyi},
in the three orthogonal Time Delay Interferometry channels (A,E,T), assuming stationarity and Gaussianity for all data components, which 
include: (i) the noise of the LISA instrument, simulated starting from a noise model 
with two free parameters $(A,P)$ where $A$ represents the test mass acceleration noise and $P$ represents the optical metrology noise
(the instrument noise at fiducial values for $(A,P)$ is shown by the gray dashed line in \cref{fig:degeneracy});
(ii) two astrophysical foregrounds, the one from in-spiralling binaries in the Milky Way (mainly white dwarf binaries), and the one from extra-galactic, in-spiralling stellar mass black hole binaries: both are simulated starting from  
simple template-based predictions; (iii) finally, 
the SGWB signal, simulated starting from three 
benchmark spectral shapes describing the signal due to, respectively, bubble collision, sound waves and MHD turbulence, and depending on the above listed thermodynamic parameters. 
The reconstruction is performed using Markov Chain Monte Carlo sampling of the likelihood by the \texttt{SGWBinner} code, using \texttt{Cobaya}~\cite{Torrado:2020dgo}, which implements the nested sampler
\texttt{Polychord}~\cite{Handley:2015vkr, Handley:2015fda} and analyses the samples with \texttt{GetDist}~\cite{Lewis:2019xzd}.
The sampled parameters are the noise parameters $(A,P)$, 
with fiducial values $A=3,P=15$;
the amplitudes of the galactic and extra-galactic foregrounds $(\Omega_{\rm Gal},\Omega_{\rm Ext})$, with fiducial values 
$\log_{10} (h^2 \Omega_{\rm Gal})=-7.84$ and 
$\log_{10} (h^2 \Omega_{\rm Ext}) = -12.38 $; and the SGWB parameters that vary depending on the case at hand. 
Information on the priors can be found in \cite{Caprini:2024hue}. 

In \cref{fig:reco_bubble,fig:reco_sw+turb} we show two examples of reconstruction of the thermodynamic parameters taken from \cite{Caprini:2024hue}. 
\cref{fig:reco_bubble}
shows the case of a strong first-order PT, 
for which the thermodynamic parameters are 
$(K,T_*, \beta/H_*)$.
The input values for the latter have been fixed to 
$K=1$, $T_*=200$ GeV and $\beta/H_*=500$.
While the parameters of the instrument noise and the foreground amplitudes are reconstructed relatively well, 
strong degeneracies appear 
in the signal parameters. 
Nevertheless, a detection is made, in the sense that 
$K=0$ is not compatible at more than 95\%. 
\cref{fig:reco_sw+turb}
shows the case of a SGWB due to both sound waves and MHD turbulence, 
for which the two parameters $v_w$ and $\epsilon$
add to those of the previous case. 
The input values are 
$\log_{10}K=-1.1$, $\log_{10}(\ell_*H_*)=-0.6$, $\log_{10}(T_*/{\rm GeV})=2.7$, $v_w=1$, $\epsilon=1$. 
Large parameter degeneracies are still present, but the reconstruction performs better than in the bubbles collision case. 
This is because the SGWB spectrum is richer in features such as spectral breaks and different slopes (cf.~\cref{fig:degeneracy}), which help to break some of the degeneracy. 
Naturally, this improvement make sense within the idealized, template-based framework of Ref.~\cite{Caprini:2024hue}; in a more realistic data-analysis setting, reconstructing a more complex spectral shape is generally more challenging.

The results of \cite{Caprini:2024hue} indicate that, under the conditions considered, it is in principle possible to infer the thermodynamic parameters from a SGWB detection by LISA, albeit with significant degeneracies. 
This conclusion is based on a template-based reconstruction applied to simulated data that include the SGWB signal, the instrumental noise, and two examples of astrophysical foregrounds. It should be emphasized, however, that the simulated data are generated using the same templates for the noise, foregrounds, and SGWB that are adopted in the reconstruction. This implicitly assumes the absence of theoretical and experimental uncertainties in both the signal and noise modelling.
Furthermore, in a realistic search for a SGWB in LISA data, one must contend with the intrinsic degeneracy between the signal and the instrumental noise, both of which are stochastic. Unlike ground-based interferometer networks, which can cross-correlate data streams from noise-independent detectors, and unlike PTAs, which use the Hellings-Downs angular correlation as a characteristic signature of GWs, LISA has no equivalent, unambiguous discriminator to separate a SGWB from the stationary Gaussian component of its instrumental noise. In practice, one must therefore rely on a noise model. However, it is overly optimistic to assume that a pre-flight validated noise model can be applied straightforwardly in a full data-analysis pipeline.
On real data, one would first run analysis pipelines designed to jointly estimate both the noise and the SGWB signal  \cite{Baghi:2023qnq,Karnesis:2019mph,Santini:2025iuj,Pozzoli:2024hkt}. Given the large variety of proposed SGWB sources, each predicting spectra with distinct features, such procedures should ideally search for a SGWB without assuming any specific template (e.g.~as in \cite{Caprini:2019pxz,Flauger:2020qyi}). Instead, they would attempt to infer the spectral shapes of both the SGWB and the noise directly from the measured data. Once a detection has been established through such a template-blind analysis -
hopefully providing preliminary information about the spectral shape of the signal - one could then proceed to apply the template-based reconstruction approach used in Ref.~\cite{Caprini:2024hue} (see also \cite{Gowling:2021gcy,Boileau:2022ter,Gowling:2022pzb,Hindmarsh:2024ttn,Giese:2021dnw}). 
At that stage, however, the analysis of real data will be further complicated by residual contamination arising from the imperfect subtraction of other GW sources in the data stream.

\subsection{Connection with two particle physics scenarios}
\label{sec:twopp}

In the previous section, we highlighted the  degeneracies that arise when reconstructing the thermodynamic parameters from a SGWB detection at LISA, even under the somewhat idealised working conditions of Ref.~\cite{Caprini:2024hue}. 
Ideally, one would even like to go a step further, and translate the reconstructed thermodynamic parameters into constraints on the parameter space of specific BSM scenarios predicting a first-order PT. 
However, mapping a given set of thermodynamic parameters onto the corresponding BSM model parameters is itself a highly non-linear procedure and comes with its own intrinsic degeneracies (for a through discussion and references, see \cite{Caprini:2024hue}). 
These are further compounded by the degeneracies present in the reconstruction from LISA data, implying that, in general, it will not be possible to unambiguously identify the BSM model responsible for the first-order PT solely from the GW signal.

Nevertheless, it remains possible to constrain a specific BSM model once it is assumed. 
The fundamental parameters of the model can be mapped, in a separate step, onto the corresponding set of thermodynamic and/or geometric parameters. For each such set, one can then predict the SGWB signal. After the SGWB has been measured and the thermodynamic and/or geometric parameters have been reconstructed - together with their posterior distributions - these posteriors can be translated into the corresponding  region in the parameter space of the BSM model. 
The LISA measurement can then be placed within a broader view of the model’s parameter space and compared with the regions accessible to current and future particle-physics experiments. 

This strategy was implemented in Ref.~\cite{Caprini:2024hue} for two illustrative scenarios: the Standard Model extended by a real singlet with a $\mathbb{Z}_2$ symmetry, and the Standard Model extended by a $U(1)_{B-L}$ gauge symmetry (see \cref{sec:PT}).
These models are particularly suited as illustrative scenarios, for two reasons: they can give rise to a first-order PT with high SGWBs in the LISA frequency band, and they are characterised by 
few fundamental parameters. 
The strategy proceeds in three steps:
\begin{enumerate}
    \item The first step consists in predicting the GW signal within the given BSM model: one scans the allowed region in the fundamental parameter space of the model 
    and associates to it the thermodynamic and/or geometric parameters of the signal.
    This procedure non trivial to implement: for detail, see \cite{Caprini:2024hue}.
    \item The second step is to simulate the LISA measurement and infer the posteriors on the geometric parameters. The latter are used in this step, rather than the thermodynamic ones, because they reduce the degeneracy in the reconstruction procedure. 
    For this step, in practice one chooses one or more benchmark points for which the SGWB is sufficiently high in LISA. 
    \item In the third step, the LISA measurement via the reconstruction of the geometric parameters can be translated into constraints on the fundamental parameter space of the BSM model, and possibly compared to pre-existing constraints or future forecasts from complementary particle physics experiments. 
    \end{enumerate}

\begin{figure}[ht!]
 \centering
    \begin{minipage}[h]{\textwidth}
        \centering
        \includegraphics[width=0.7\textwidth]{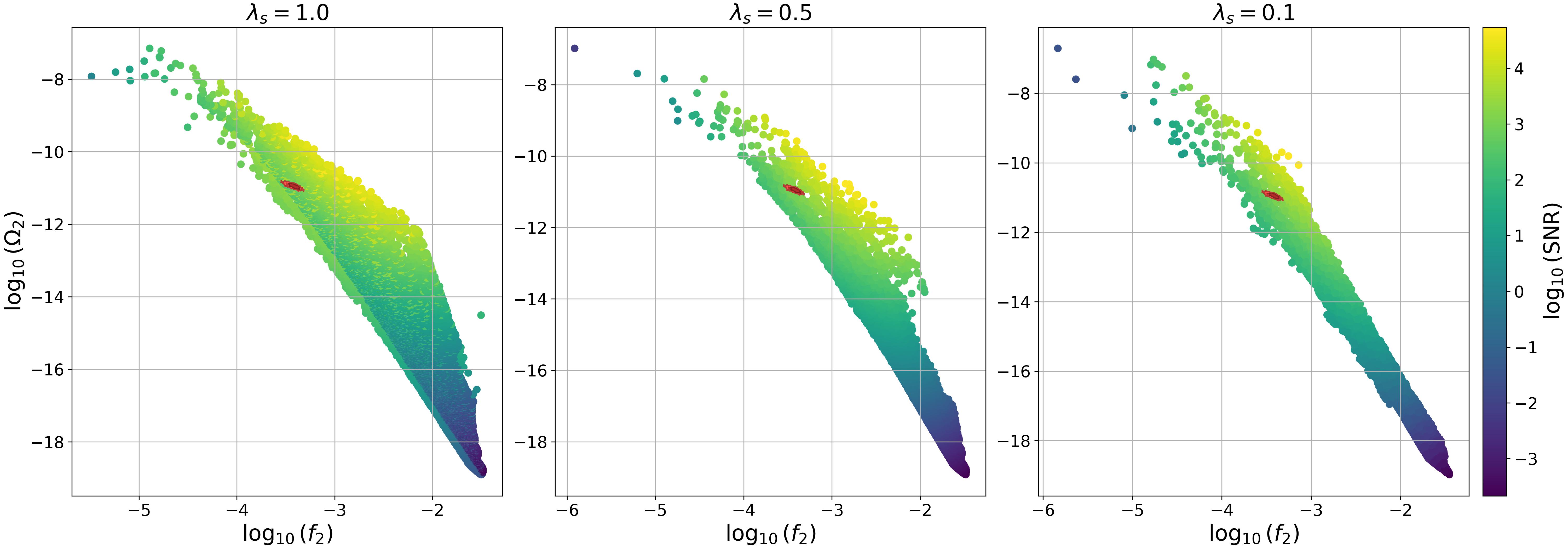}
        \captionof*{figure}{(a) \emph{Step 1:} 
        Geometric parameters $(f_2,\Omega_2)$ of the GW signal from sound waves, obtained by scanning  the singlet parameter space given by the singlet mass and coupling with the Higgs $(m_s,\lambda_{ms})$, for three fixed values of the 
        quartic coupling $\lambda_s$. The colour bar denotes the SNR in LISA of the sound waves SGWB corresponding to each $(f_2,\Omega_2)$ point. The red ellipse on top of each plot comes from the LISA measurement of the chosen benchmark point (see panel (b)). }
           \end{minipage}
    \hfill
    \begin{minipage}[h]{0.46\textwidth}
        \centering
        \vspace*{0.5cm}
        \includegraphics[width=\textwidth]{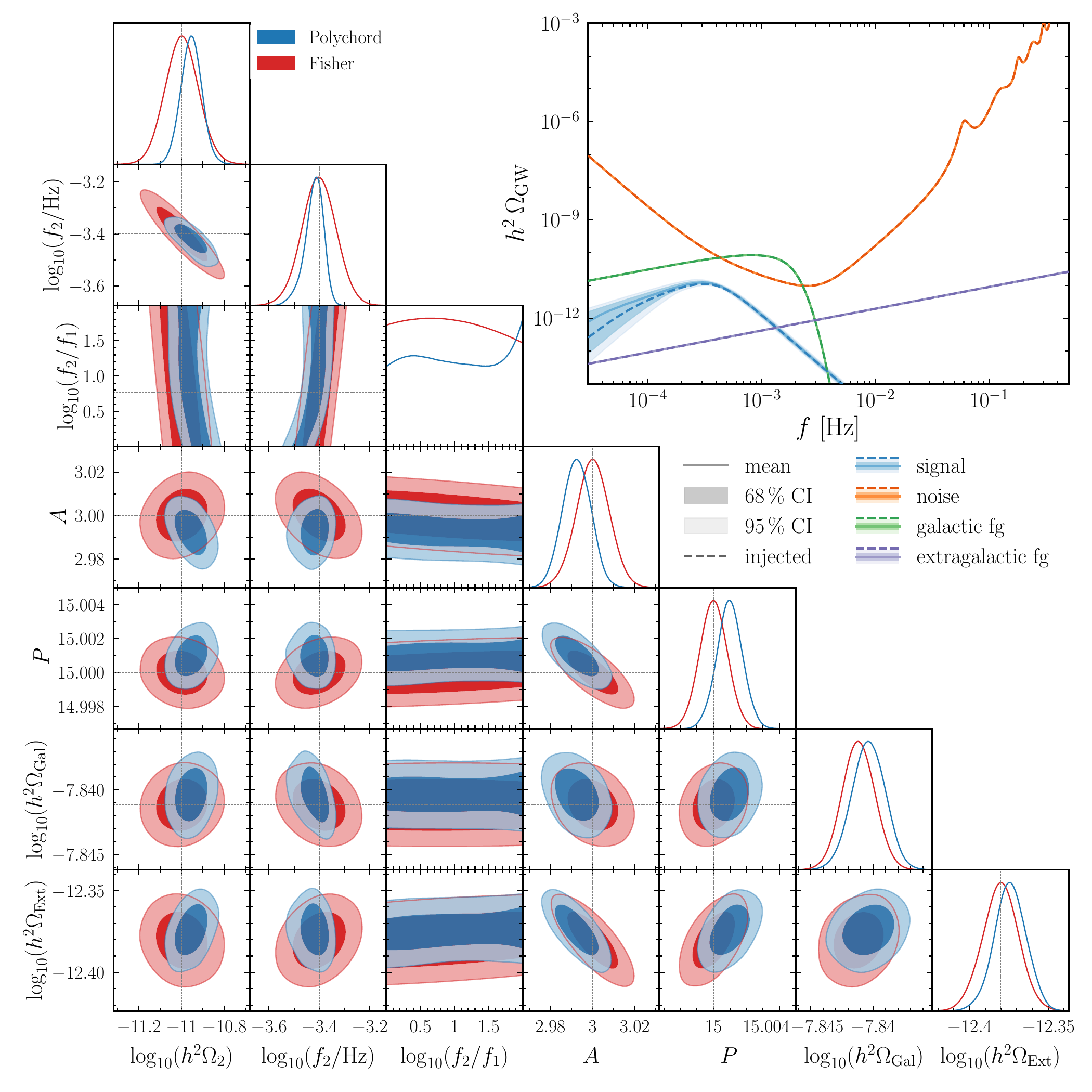}
        \captionof*{figure}{(b) \emph{Step 2:}  Simulated LISA measurement of the benchmark point with geometric parameters $h^2\Omega_2 = 10^{-11}$ and $f_2={0.4}$ mHz (corresponding to the red ellipses in panel (a)) for the SGWB produced by sound waves (concerning $f_2/f_1$, see main text).}
        \label{fig:alpha1}
    \end{minipage}
    \hspace{0.3cm}
    \begin{minipage}[h]{0.5\textwidth}
        \centering
        \includegraphics[width=\textwidth]{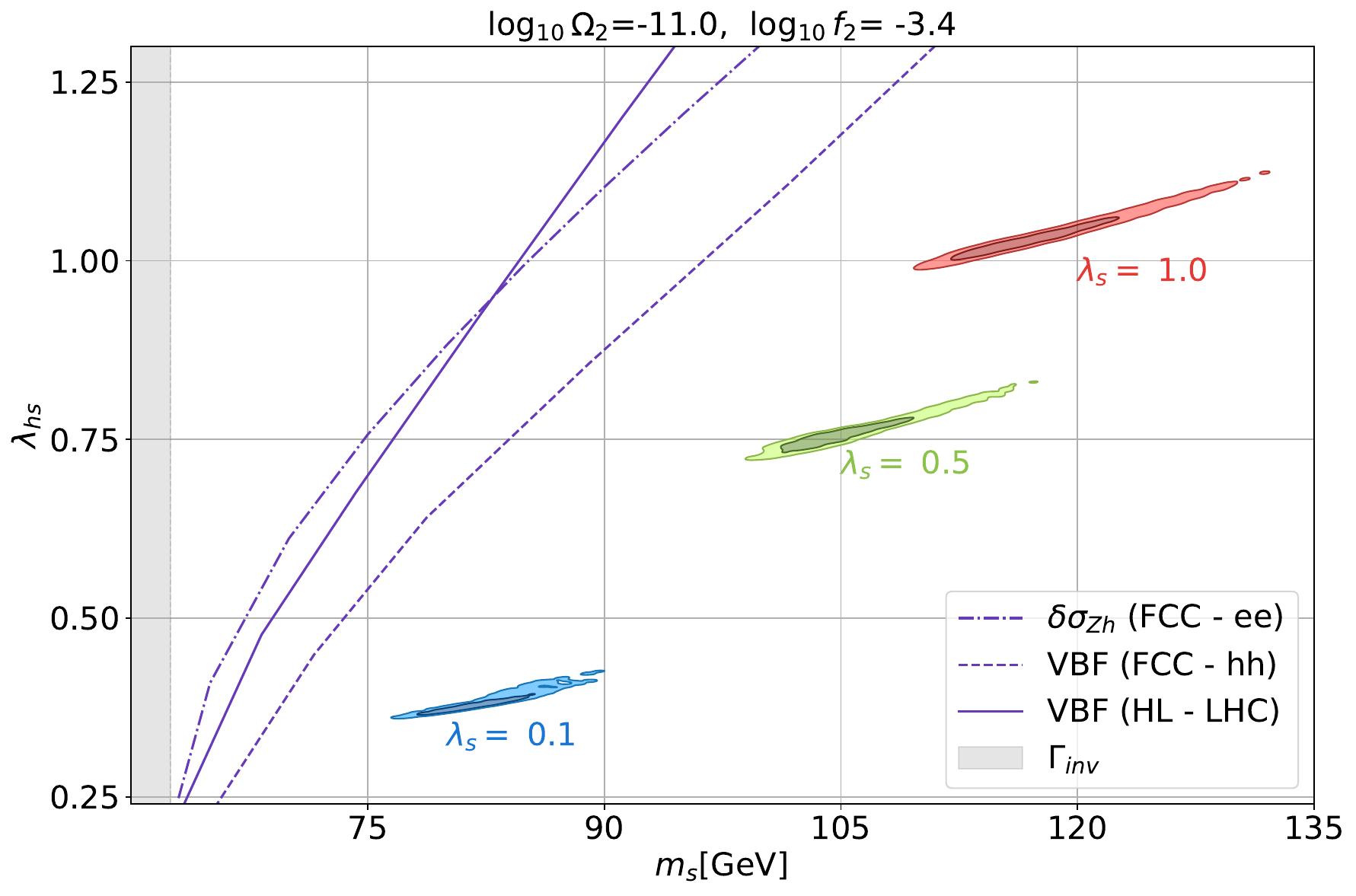}
        \captionof*{figure}{(c) \emph{Step 3:} 
         The posterior contours from the LISA measurement (c.f.~panel (b)) shown as constraints on the fundamental parameter space of the model $(m_s,\lambda_{ms})$, for the three chosen values of the 
        quartic coupling $\lambda_s$ listed in panel (a). 
        The violet curves are experimental upper limits from the FCC and the high luminosity LHC, according to the legend. The gray shaded region is excluded by Higgs invisible decay.}
        \end{minipage}
        \caption{All panels taken from \cite{Caprini:2024hue}: LISA constraints on the SM extended with a $\mathbb{Z}_2$ gauge singlet.}
 \label{fig:scalar}
\end{figure}

The results of this procedure for the $\mathbb{Z}_2$ extension of the SM, obtained in \cite{Caprini:2024hue}, are reported in \cref{fig:scalar}. 
The model is characterised by three parameters: $m_s$ the singlet mass,  $\lambda_{hs}$ the portal coupling between the Higgs and the singlet, and $\lambda_s$ the quartic coupling of the singlet. 
The strength of the PT is determined essentially by   $\lambda_{hs}$ which, however, cannot be too large, to remain within the perturbative regime. 
This in turns means that $m_s$ cannot be too large either: in   \cite{Caprini:2024hue}, the window $65\,{\rm GeV}<m_s<125 \,{\rm GeV}$ is chosen. 
The upper panel of \cref{fig:scalar} shows the geometric parameters $(f_2,\Omega_2)$ of the SGWB generated by sound waves from the first-order PT (double broken power law), obtained by scanning the $(m_s,\lambda_{hs})$ parameter space, for three fixed values of $\lambda_s$.  
To perform step 2, out of the parameter space, one particular benchmark point is chosen, featuring a SGWB high enough to be detected by LISA: in the case at hand, this corresponds to $h^2\Omega_2=10^{-11}$ and $f_2=0.4$ mHz, while the first spectral break is given by the bubble wall velocity, here set to $v_w=1$, and the sound speed, here set to $c_s=1/\sqrt{3}$: $f_1/f_2\simeq 0.4 |v_w-c_s|/{\rm max}(v_w,c_s)$ \cite{Caprini:2024hue}. 
The outcome of the LISA parameter inference on this particular benchmark is shown in the lower left panel of \cref{fig:scalar}. Finally, the lower rigth panel of \cref{fig:scalar} shows the posterior contours obtained from the LISA measurement translated back in the 
$(m_s,\lambda_{hs})$ parameter space, again for three values of $\lambda_s$: LISA can help discriminating the model parameter space, in the region that will not be probed by future colliders. 
This shows the complementary power of LISA in probing this particular scenario. 

It is important to mention that this scenario in its simplest implementation analysed in \cite{Caprini:2024hue} suffers from observational and theoretical challenges, 
linked to the overproduction of dark matter and the presence of domain walls: 
{the latter in particular are formed in the first step of the PT, and despite being annihilated at the second step when the $\mathbb{Z}_2$ symmetry is restored, they would act as seeds for the bubble formation, altering the PT dynamics and the prediction of the GW signal \cite{Blasi:2022woz}.} 

\begin{figure}
    \centering
    \includegraphics[height=0.3\textwidth]{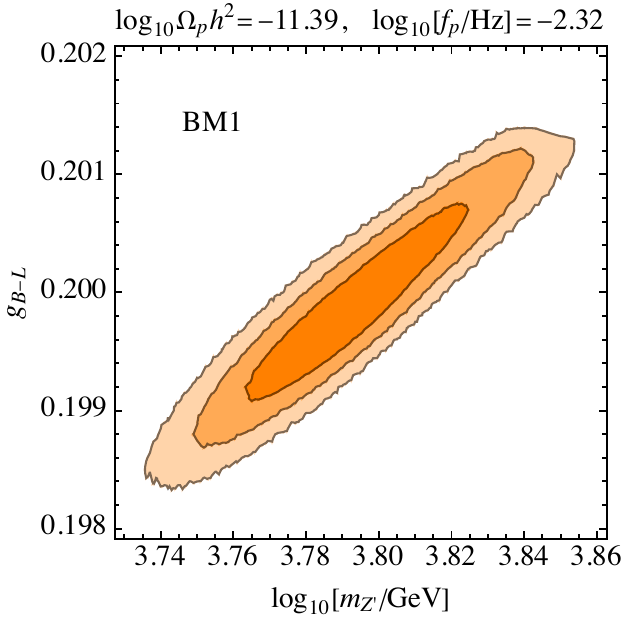} \hspace{2mm}
    \includegraphics[height=0.3\textwidth]{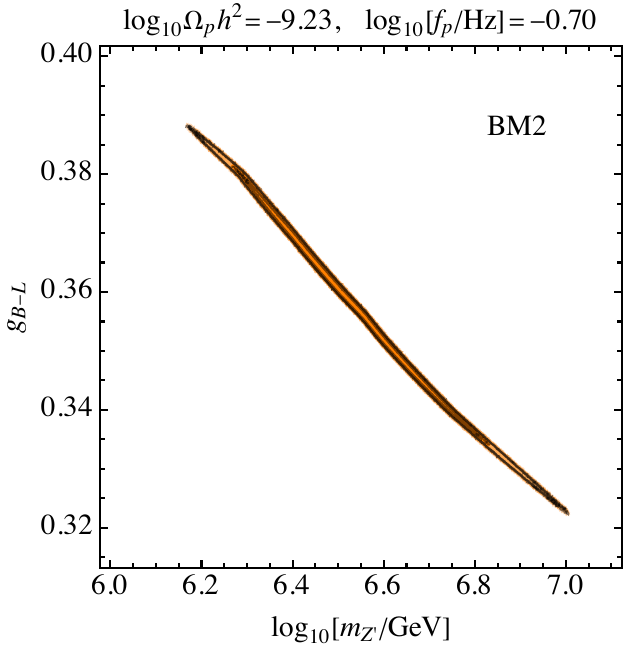}
    \caption{Taken from \cite{Caprini:2024hue}: LISA constraints on the parameters of the SM extended with the $U(1)_{B-L}$ gauge symmetry, i.e.~the gauge coupling $g_{B-L}$ and the boson mass $m_{Z'}$, for two separate benchmark points for the parameters of the SGWB sourced by bubble collisions $(f_b,\Omega_b)$, as given in the plots' upper labels. }
    \label{fig:U1BL}
\end{figure}

The same strategy has been performed in \cite{Caprini:2024hue} also for the case of the SM extended via the $U(1)_{B-L}$ gauge symmetry.
Within this scenario the first-order PT is strong, and therefore the relevant contribution to the SGWB is given by bubbles collision, characterised by a broken power law with the two geometric parameters $(f_b,\Omega_b)$. 
Two different benchmark points have been chosen in \cite{Caprini:2024hue} for step 2: 
$(f_b=5\,{\rm  mHz},h^2\Omega_b = 4\times 10^{-12})$, and $(f_b=0.2 \,{\rm Hz}, h^2\Omega_b = 6\times 10^{-10})$. 
In \cref{fig:U1BL} we reproduce the results of step 3 in the reconstruction procedure: as expected, it appears that 
 different measurements at LISA lead to different constraints on the model parameter space, here provided by the boson mass $m_{Z'}$ and the gauge coupling $g_{B-L}$. 
In particular, the second benchmark point leads to a stronger degeneracy in the reconstruction of the fundamental parameters of the model, despite being characterised by a higher signal amplitude $h^2\Omega_b$ than the first benchmark point. This is because the peak of the signal is in the high-frequency region, where the LISA sensitivity is worst (c.f.~plots and discussion in \cite{Caprini:2024hue}).

\section{Conclusion}
\label{sec:conc}

%
GWs produced in the early Universe behave as a fossil radiation: a stochastic background that can carry valuable information about high-energy physics at energy scales far beyond the reach of both cosmological probes (such as the CMB and BBN) and current particle collider experiments. 
Crucially, the frequency ranges of sensitivity of present and upcoming GW detectors can be associated with the energy scales of various cosmological PTs. 
For instance, ground-based interferometers are sensitive to energies around the Peccei–Quinn transition, LISA to the EW transition and above, and PTAs to the QCD transition.

However, detectability requires the signal amplitude to be sufficiently large. This typically demands that the PT be first-order and/or produce topological defects, which generate the anisotropic stresses sourcing GWs. 
A variety of BSM scenarios exist, in which the otherwise continuous transitions taking place within the SM become first-order, or in which additional symmetry-breakings occurring mainly at higher energy scales generate strong GW signals.
We have provided a concise overview of several representative models of this type.

It would be extremely valuable to probe this diverse phenomenology through GW observations. 
We have explored this prospect in more detail focussing on the specific case of a first-order PT.
We emphasized that making precise predictions for the resulting SGWB spectrum is challenging, owing to the inherently complex and highly non-linear dynamics involved, together with the large freedom available in model building and parameter choices. 
In general, numerical simulations are required for robust predictions and for validating analytical estimates of the signal. 

In addition to the difficulty of forecasting the SGWB spectrum within a given framework, one faces the further challenge of extracting the relevant physical parameters from a potential detection. Following Ref.~\cite{Caprini:2024hue}, we have focused on the case of a SGWB induced by a first-order PT and detectable by LISA. In such case, the SGWB spectrum typically exhibits fewer distinct features than the  thermodynamic parameters that control it, leading to strong degeneracies in parameter reconstruction.
To mitigate this issue, Ref.~\cite{Caprini:2024hue} proposes parametrising the GW signal using geometric parameters characterizing its shape, which allows for a more robust parameter reconstruction from  the observed signal. 
However, the degeneracies re-emerge when one attempts to translate the geometric parameters back into the  thermodynamic ones - an essential step in connecting a measured signal to the physics of the PT.

Ideally, one would like a LISA detection to identify which specific BSM model, among the many capable of producing a first-order PT, is responsible for a given SGWB signal. However, strong degeneracies exist also between the thermodynamic parameters of the PT and those of the underlying BSM model, making such a direct inference extremely challenging. Nevertheless, a detection - or even a non-detection - can still be used to constrain specific models, once chosen beforehand.

To do so, one must first scan the parameter space of the fundamental parameters in the chosen BSM scenario and compute the associated SGWB signal for each parameter point. The expected LISA sensitivity is then applied to simulate a detection (or exclusion), allowing one to infer confidence regions in the geometric and/or thermodynamic parameters of the first-order PT. These confidence regions can subsequently be mapped back onto the corresponding regions in the BSM model parameter space.

We presented the results of Ref.~\cite{Caprini:2024hue}, where this procedure was carried out for two well-motivated BSM scenarios featuring a first-order PT. While the analysis relies on somewhat idealized assumptions and should therefore be viewed mainly as a proof of principle, it nevertheless demonstrates that LISA can provide constraints that are complementary to those expected from future particle colliders.

{In conclusion, the detection of SGWBs originating from BSM processes at high energies in the early Universe holds transformative scientific potential. 
However, substantial progress is still required to fully realise this potential. 
Focusing in particular on first-order PTs, preparedness for a possible detection would ideally rely on a unified and broadly accepted theoretical framework.
Achieving this goal requires progress toward consensus on several key aspects, most notably on the predicted SGWB spectra and their parameter dependencies, as well as on robust methods to connect thermodynamic parameters to the underlying high energy PT models. 
Establishing such a framework would ensure that the link to the fundamental theory can be reconstructed at best in a consistent, transparent, and reusable manner, facilitating community-wide adoption.
This approach offers the most effective strategy for mitigating the intrinsic limitation that GW signals typically exhibit too few distinctive features to unambiguously discriminate among the different underlying scenarios.}

\acknowledgments 
The author acknowledges informative discussions with Germano Nardini.

\appendix 
\section{Summary table of the key parameters} 

\begin{table}[h]
\centering

\caption{Summary table of the key parameters.}
\label{tab:example}
\begin{tabular}{l l}
\hline\hline
Parameter & Meaning \\
\hline
$\ell_*~ (R_*)$  & characteristic length scale of the GW sourcing process (i.e.~bubble size in the case \\ & of a first-orde PT) \\
\hline
$T_*$  & temperature of the early Universe
when the GW sourcing process was operating\\
\hline
${\Pi}/{\rho_{\rm tot}^*} $ & anisotropic
stresses available to source the GWs, normalised to the energy density \\  & in the Universe at the source time \\
\hline
$v_w~(\xi_w)$  & bubble wall velocity \\
\hline
$\beta$  & PT rate parameter \\
\hline
$t_*$  & reference time corresponding to $T_*$ \\
\hline
$K=\rho_s/\rho_{\rm tot}^*$  & GW source energy fraction \\
\hline
$\rho_\varphi$ & gradient energy in the scalar field \\
\hline
$\rho_v$ & kinetic energy of the bulk fluid motion\\
\hline 
$\kappa_{\varphi,v} =\rho_{\varphi,v} /\Delta V_0$ & efficiency parameters (efficiency with which the potential energy is
transformed \\ & into gradient energy of the bubble walls or kinetic energy of the fluid) \\
\hline 
$\alpha= \Delta V_0/[\pi^2g(T_*) T_*^4/30]$~~ & strength of the PT (fraction of the potential to thermal energy) \\
\hline
$\tilde \Omega_{\rm GW}$ & efficiency with which the GW source energy fraction $K$ is converted to GWs\\
\hline
$S(f)$ & normalised SGWB spectral shape \\
\hline 
$\epsilon$ & fraction of kinetic energy in bulk motion that is converted to (MHD) turbulence \\
\hline 
$f_b,\, \Omega_b $ & geometric parameters of the broken power law template: frequency of, \\ &  and amplitude at, the spectral peak\\
\hline 
$f_1,\,f_2,\,\Omega_2$ & geometric parameters of the double broken power law template: frequencies  of \\ &  the two breaks and amplitude at the second break \\
\hline 
$A,\,P$ & parameters of the LISA noise: amplitudes of the acceleration and test mass noises \\
\hline 
$h^2 \Omega_{\rm Gal}$ & amplitude of the Galactic foreground in LISA \\
\hline 
$h^2 \Omega_{\rm Ext}$ & amplitude of the extra-Galactic foreground in LISA \\
\hline 
$m_s$ & singlet mass in the $\mathbb{Z}_2$ extension of the SM \\
\hline 
$\lambda_{hs}$ & portal coupling between the Higgs and the singlet \\
\hline 
$\lambda_s$ & quartic coupling of the singlet \\
\hline
$m_{Z'}$ & boson mass in the SM extended by $U(1)_{B-L}$ gauge symmetry \\
\hline 
$g_{B-L}$ & boson gauge coupling \\
\hline\hline
\end{tabular}
\end{table}

\bibliography{ref}

@book{Durrer:2020fza,
    author = "Durrer, Ruth",
    title = "{The Cosmic Microwave Background}",
    doi = "10.1017/9781316471524",
    isbn = "978-1-316-47152-4, 978-1-107-13522-2",
    publisher = "Cambridge University Press",
    month = "12",
    year = "2020"
}

@article{LISA:2024hlh,
    author = "Colpi, Monica and others",
    collaboration = "LISA",
    title = "{LISA Definition Study Report}",
    eprint = "2402.07571",
    archivePrefix = "arXiv",
    primaryClass = "astro-ph.CO",
    month = "2",
    year = "2024"
}

@article{Karnesis:2019mph,
    author = "Karnesis, Nikolaos and Lilley, Marc and Petiteau, Antoine",
    title = "{Assessing the detectability of a Stochastic Gravitational Wave Background with LISA, using an excess of power approach}",
    eprint = "1906.09027",
    archivePrefix = "arXiv",
    primaryClass = "astro-ph.IM",
    doi = "10.1088/1361-6382/abb637",
    journal = "Class. Quant. Grav.",
    volume = "37",
    number = "21",
    pages = "215017",
    year = "2020"
}

@article{Santini:2025iuj,
    author = "Santini, Alessandro and Muratore, Martina and Gair, Jonathan and Hartwig, Olaf",
    title = "{Flexible, GPU-accelerated approach for the joint characterization of LISA instrumental noise and stochastic gravitational wave backgrounds}",
    eprint = "2507.06300",
    archivePrefix = "arXiv",
    primaryClass = "gr-qc",
    doi = "10.1103/csx9-9trp",
    journal = "Phys. Rev. D",
    volume = "112",
    number = "8",
    pages = "084050",
    year = "2025"
}

@article{Pozzoli:2024hkt,
    author = "Pozzoli, Federico and Gair, Jonathan and Buscicchio, Riccardo and Speri, Lorenzo",
    title = "{Is the stochastic signal really detectable?}",
    eprint = "2412.10468",
    archivePrefix = "arXiv",
    primaryClass = "astro-ph.IM",
    doi = "10.1103/22h4-tqh9",
    journal = "Phys. Rev. D",
    volume = "112",
    number = "6",
    pages = "064035",
    year = "2025"
}

@article{Caprini:2001nb,
    author = "Caprini, Chiara and Durrer, Ruth",
    title = "{Gravitational wave production: A Strong constraint on primordial magnetic fields}",
    eprint = "astro-ph/0106244",
    archivePrefix = "arXiv",
    doi = "10.1103/PhysRevD.65.023517",
    journal = "Phys. Rev. D",
    volume = "65",
    pages = "023517",
    year = "2001"
}

@article{Caprini:2009pr,
    author = "Caprini, Chiara and Durrer, Ruth and Fenu, Elisa",
    title = "{Can the observed large scale magnetic fields be seeded by helical primordial fields?}",
    eprint = "0906.4976",
    archivePrefix = "arXiv",
    primaryClass = "astro-ph.CO",
    doi = "10.1088/1475-7516/2009/11/001",
    journal = "JCAP",
    volume = "11",
    pages = "001",
    year = "2009"
}

@article{Kahniashvili:2009mf,
    author = "Kahniashvili, Tina and Kisslinger, Leonard and Stevens, Trevor",
    title = "{Gravitational Radiation Generated by Magnetic Fields in Cosmological Phase Transitions}",
    eprint = "0905.0643",
    archivePrefix = "arXiv",
    primaryClass = "astro-ph.CO",
    doi = "10.1103/PhysRevD.81.023004",
    journal = "Phys. Rev. D",
    volume = "81",
    pages = "023004",
    year = "2010"
}

@article{Baym:1995fk,
    author = "Baym, Gordon and Bodeker, Dietrich and McLerran, Larry D.",
    title = "{Magnetic fields produced by phase transition bubbles in the electroweak phase transition}",
    eprint = "hep-ph/9507429",
    archivePrefix = "arXiv",
    reportNumber = "TPI-MINN-95-19-T, NUC-MINN-95-15-T, HEP-MINN-95-1344",
    doi = "10.1103/PhysRevD.53.662",
    journal = "Phys. Rev. D",
    volume = "53",
    pages = "662--667",
    year = "1996"
}

@article{Cheng:1994yr,
    author = "Cheng, Bao-lian and Olinto, Angela V.",
    title = "{Primordial magnetic fields generated in the quark - hadron transition}",
    reportNumber = "FERMILAB-PUB-94-081-A",
    doi = "10.1103/PhysRevD.50.2421",
    journal = "Phys. Rev. D",
    volume = "50",
    pages = "2421--2424",
    year = "1994"
}

@article{Sigl:1996dm,
    author = "Sigl, Guenter and Olinto, Angela V. and Jedamzik, Karsten",
    title = "{Primordial magnetic fields from cosmological first order phase transitions}",
    eprint = "astro-ph/9610201",
    archivePrefix = "arXiv",
    doi = "10.1103/PhysRevD.55.4582",
    journal = "Phys. Rev. D",
    volume = "55",
    pages = "4582--4590",
    year = "1997"
}

@article{Figueroa:2012kw,
    author = "Figueroa, Daniel G. and Hindmarsh, Mark and Urrestilla, Jon",
    title = "{Exact Scale-Invariant Background of Gravitational Waves from Cosmic Defects}",
    eprint = "1212.5458",
    archivePrefix = "arXiv",
    primaryClass = "astro-ph.CO",
    doi = "10.1103/PhysRevLett.110.101302",
    journal = "Phys. Rev. Lett.",
    volume = "110",
    number = "10",
    pages = "101302",
    year = "2013"
}

@article{Durrer:2001cg,
    author = "Durrer, R. and Kunz, M. and Melchiorri, A.",
    title = "{Cosmic structure formation with topological defects}",
    eprint = "astro-ph/0110348",
    archivePrefix = "arXiv",
    doi = "10.1016/S0370-1573(02)00014-5",
    journal = "Phys. Rept.",
    volume = "364",
    pages = "1--81",
    year = "2002"
}

@article{Durrer:1999na,
    author = "Durrer, R.",
    title = "{Topological defects in cosmology}",
    doi = "10.1016/S1387-6473(99)00008-1",
    journal = "New Astron. Rev.",
    volume = "43",
    pages = "111--156",
    year = "1999"
}

@article{Caprini:2024ofd,
    author = "Caprini, Chiara and Pujol{\`a}s, Oriol and Quelquejay-Leclere, Hippolyte and Rompineve, Fabrizio and Steer, Dani{\`e}le A.",
    title = "{Primordial gravitational wave backgrounds from phase transitions with next generation ground based detectors}",
    eprint = "2406.02359",
    archivePrefix = "arXiv",
    primaryClass = "astro-ph.CO",
    reportNumber = "CERN-TH-2024-065",
    doi = "10.1088/1361-6382/ad9a48",
    journal = "Class. Quant. Grav.",
    volume = "42",
    number = "4",
    pages = "045015",
    year = "2025"
}

@article{Jinno:2022fom,
    author = "Jinno, Ryusuke and Shakya, Bibhushan and van de Vis, Jorinde",
    title = "{Gravitational Waves from Feebly Interacting Particles in a First Order Phase Transition}",
    eprint = "2211.06405",
    archivePrefix = "arXiv",
    primaryClass = "gr-qc",
    reportNumber = "DESY-22-172, IFT-UAM/CSIC-22-140, MITP-22-095, RESCEU-22/22",
    month = "11",
    year = "2022"
}

@article{Inomata:2024rkt,
    author = "Inomata, Keisuke and Kamionkowski, Marc and Kasai, Kentaro and Shakya, Bibhushan",
    title = "{Gravitational waves from particles produced from bubble collisions in first-order phase transitions}",
    eprint = "2412.17912",
    archivePrefix = "arXiv",
    primaryClass = "astro-ph.CO",
    doi = "10.1103/k4s5-8zqy",
    journal = "Phys. Rev. D",
    volume = "112",
    number = "8",
    pages = "083523",
    year = "2025"
}

@book{Maggiore2,
    author = "Maggiore, Michele",
    title = "{Gravitational Waves. Vol. 2: Astrophysics and Cosmology}",
    isbn = "978-0-19-857089-9",
    publisher = "Oxford University Press",
    month = "3",
    year = "2018"
}

@article{Maggiore:1999vm,
    author = "Maggiore, Michele",
    title = "{Gravitational wave experiments and early universe cosmology}",
    eprint = "gr-qc/9909001",
    archivePrefix = "arXiv",
    reportNumber = "IFUP-TH-20-99",
    doi = "10.1016/S0370-1573(99)00102-7",
    journal = "Phys. Rept.",
    volume = "331",
    pages = "283--367",
    year = "2000"
}

@article{Sharma:2023mao,
    author = "Sharma, Ramkishor and Dahl, Jani and Brandenburg, Axel and Hindmarsh, Mark",
    title = "{Shallow relic gravitational wave spectrum with acoustic peak}",
    eprint = "2308.12916",
    archivePrefix = "arXiv",
    primaryClass = "gr-qc",
    reportNumber = "NORDITA-2023-051, HIP-2023-13/TH",
    doi = "10.1088/1475-7516/2023/12/042",
    journal = "JCAP",
    volume = "12",
    pages = "042",
    year = "2023"
}

@article{Durrer:2013pga,
    author = "Durrer, Ruth and Neronov, Andrii",
    title = "{Cosmological Magnetic Fields: Their Generation, Evolution and Observation}",
    eprint = "1303.7121",
    archivePrefix = "arXiv",
    primaryClass = "astro-ph.CO",
    doi = "10.1007/s00159-013-0062-7",
    journal = "Astron. Astrophys. Rev.",
    volume = "21",
    pages = "62",
    year = "2013"
}

@article{Lewicki:2022pdb,
    author = "Lewicki, Marek and Vaskonen, Ville",
    title = "{Gravitational waves from bubble collisions and fluid motion in strongly supercooled phase transitions}",
    eprint = "2208.11697",
    archivePrefix = "arXiv",
    primaryClass = "astro-ph.CO",
    doi = "10.1140/epjc/s10052-023-11241-3",
    journal = "Eur. Phys. J. C",
    volume = "83",
    number = "2",
    pages = "109",
    year = "2023"
}

@article{Jinno:2017fby,
    author = "Jinno, Ryusuke and Takimoto, Masahiro",
    title = "{Gravitational waves from bubble dynamics: Beyond the Envelope}",
    eprint = "1707.03111",
    archivePrefix = "arXiv",
    primaryClass = "hep-ph",
    reportNumber = "CTPU-17-26, KEK-TH-1986",
    doi = "10.1088/1475-7516/2019/01/060",
    journal = "JCAP",
    volume = "01",
    pages = "060",
    year = "2019"
}

@article{Ellis:2020nnr,
    author = "Ellis, John and Lewicki, Marek and Vaskonen, Ville",
    title = "{Updated predictions for gravitational waves produced in a strongly supercooled phase transition}",
    eprint = "2007.15586",
    archivePrefix = "arXiv",
    primaryClass = "astro-ph.CO",
    reportNumber = "KCL-PH-TH/2020-40, CERN-TH-2020-129",
    doi = "10.1088/1475-7516/2020/11/020",
    journal = "JCAP",
    volume = "11",
    pages = "020",
    year = "2020"
}

@article{Giese:2020rtr,
    author = "Giese, Felix and Konstandin, Thomas and van de Vis, Jorinde",
    title = "{Model-independent energy budget of cosmological first-order phase transitions\textemdash{}A sound argument to go beyond the bag model}",
    eprint = "2004.06995",
    archivePrefix = "arXiv",
    primaryClass = "astro-ph.CO",
    reportNumber = "DESY-20-064",
    doi = "10.1088/1475-7516/2020/07/057",
    journal = "JCAP",
    volume = "07",
    number = "07",
    pages = "057",
    year = "2020"
}

@article{Kainulainen:2019kyp,
    author = "Kainulainen, Kimmo and Keus, Venus and Niemi, Lauri and Rummukainen, Kari and Tenkanen, Tuomas V. I. and Vaskonen, Ville",
    title = "{On the validity of perturbative studies of the electroweak phase transition in the Two Higgs Doublet model}",
    eprint = "1904.01329",
    archivePrefix = "arXiv",
    primaryClass = "hep-ph",
    doi = "10.1007/JHEP06(2019)075",
    journal = "JHEP",
    volume = "06",
    pages = "075",
    year = "2019"
}

@article{Espinosa:2010hh,
    author = "Espinosa, Jose R. and Konstandin, Thomas and No, Jose M. and Servant, Geraldine",
    title = "{Energy Budget of Cosmological First-order Phase Transitions}",
    eprint = "1004.4187",
    archivePrefix = "arXiv",
    primaryClass = "hep-ph",
    reportNumber = "CERN-PH-TH-2010-027",
    doi = "10.1088/1475-7516/2010/06/028",
    journal = "JCAP",
    volume = "06",
    pages = "028",
    year = "2010"
}

@article{Laine:1998jb,
    author = "Laine, M. and Rummukainen, K.",
    editor = "DeGrand, Thomas A. and DeTar, Carleton E. and Sugar, R. and Toussaint, D.",
    title = "{What's new with the electroweak phase transition?}",
    eprint = "hep-lat/9809045",
    archivePrefix = "arXiv",
    doi = "10.1016/S0920-5632(99)85017-8",
    journal = "Nucl. Phys. B Proc. Suppl.",
    volume = "73",
    pages = "180--185",
    year = "1999"
}

@article{Aoki:2006we,
    author = "Aoki, Y. and Endrodi, G. and Fodor, Z. and Katz, S. D. and Szabo, K. K.",
    title = "{The Order of the quantum chromodynamics transition predicted by the standard model of particle physics}",
    eprint = "hep-lat/0611014",
    archivePrefix = "arXiv",
    doi = "10.1038/nature05120",
    journal = "Nature",
    volume = "443",
    pages = "675--678",
    year = "2006"
}

@article{RoperPol:2023dzg,
    author = "Roper Pol, Alberto and Procacci, Simona and Caprini, Chiara",
    title = "{Characterization of the gravitational wave spectrum from sound waves within the sound shell model}",
    eprint = "2308.12943",
    archivePrefix = "arXiv",
    primaryClass = "gr-qc",
    doi = "10.1103/PhysRevD.109.063531",
    journal = "Phys. Rev. D",
    volume = "109",
    number = "6",
    pages = "063531",
    year = "2024"
}

@article{Randall:2006py,
    author = "Randall, Lisa and Servant, Geraldine",
    title = "{Gravitational waves from warped spacetime}",
    eprint = "hep-ph/0607158",
    archivePrefix = "arXiv",
    reportNumber = "CERN-PH-TH-2006-133",
    doi = "10.1088/1126-6708/2007/05/054",
    journal = "JHEP",
    volume = "05",
    pages = "054",
    year = "2007"
}

@article{Caprini:2006jb,
    author = "Caprini, Chiara and Durrer, Ruth",
    title = "{Gravitational waves from stochastic relativistic sources: Primordial turbulence and magnetic fields}",
    eprint = "astro-ph/0603476",
    archivePrefix = "arXiv",
    doi = "10.1103/PhysRevD.74.063521",
    journal = "Phys. Rev. D",
    volume = "74",
    pages = "063521",
    year = "2006"
}

@article{Linde:1981zj,
    author = "Linde, Andrei D.",
    title = "{Decay of the False Vacuum at Finite Temperature}",
    reportNumber = "LEBEDEV-81-265",
    doi = "10.1016/0550-3213(83)90072-X",
    journal = "Nucl. Phys. B",
    volume = "216",
    pages = "421",
    year = "1983",
    note = "[Erratum: Nucl.Phys.B 223, 544 (1983)]"
}

@article{Kosowsky:1991ua,
    author = "Kosowsky, Arthur and Turner, Michael S. and Watkins, Richard",
    title = "{Gravitational radiation from colliding vacuum bubbles}",
    reportNumber = "FERMILAB-PUB-91-323-A",
    doi = "10.1103/PhysRevD.45.4514",
    journal = "Phys. Rev. D",
    volume = "45",
    pages = "4514--4535",
    year = "1992"
}

@article{Hindmarsh:2020hop,
    author = {Hindmarsh, Mark B. and L\"uben, Marvin and Lumma, Johannes and Pauly, Martin},
    title = "{Phase transitions in the early universe}",
    eprint = "2008.09136",
    archivePrefix = "arXiv",
    primaryClass = "astro-ph.CO",
    reportNumber = "MPP-2020-163, HIP-2020-27/TH",
    doi = "10.21468/SciPostPhysLectNotes.24",
    journal = "SciPost Phys. Lect. Notes",
    volume = "24",
    pages = "1",
    year = "2021"
}

@article{Giudice:2024tcp,
    author = "Giudice, Gian F. and Lee, Hyun Min and Pomarol, Alex and Shakya, Bibhushan",
    title = "{Nonthermal Heavy Dark Matter from a First-Order Phase Transition}",
    eprint = "2403.03252",
    archivePrefix = "arXiv",
    primaryClass = "hep-ph",
    reportNumber = "CERN-TH-2024-031, DESY-24-031",
    month = "3",
    year = "2024"
}

@article{Cataldi:2024pgt,
    author = "Cataldi, Martina and Shakya, Bibhushan",
    title = "{Leptogenesis via Bubble Collisions}",
    eprint = "2407.16747",
    archivePrefix = "arXiv",
    primaryClass = "hep-ph",
    reportNumber = "DESY-24-110",
    month = "7",
    year = "2024"
}

@article{Konstandin:2014zta,
    author = "Konstandin, Thomas and Nardini, Germano and Rues, Ingo",
    title = "{From Boltzmann equations to steady wall velocities}",
    eprint = "1407.3132",
    archivePrefix = "arXiv",
    primaryClass = "hep-ph",
    reportNumber = "DESY-14-127, NSF-KITP-14-089",
    doi = "10.1088/1475-7516/2014/09/028",
    journal = "JCAP",
    volume = "09",
    pages = "028",
    year = "2014"
}

@article{Saikawa:2017hiv,
    author = "Saikawa, Ken'ichi",
    title = "{A review of gravitational waves from cosmic domain walls}",
    eprint = "1703.02576",
    archivePrefix = "arXiv",
    primaryClass = "hep-ph",
    reportNumber = "DESY-17-036",
    doi = "10.3390/universe3020040",
    journal = "Universe",
    volume = "3",
    number = "2",
    pages = "40",
    year = "2017"
}

@article{Caprini:2007xq,
    author = "Caprini, Chiara and Durrer, Ruth and Servant, Geraldine",
    title = "{Gravitational wave generation from bubble collisions in first-order phase transitions: An analytic approach}",
    eprint = "0711.2593",
    archivePrefix = "arXiv",
    primaryClass = "astro-ph",
    reportNumber = "CERN-PH-TH-2007-206, SACLAY-T07-142",
    doi = "10.1103/PhysRevD.77.124015",
    journal = "Phys. Rev. D",
    volume = "77",
    pages = "124015",
    year = "2008"
}

@article{vonHarling:2017yew,
    author = "von Harling, Benedict and Servant, Geraldine",
    title = "{QCD-induced Electroweak Phase Transition}",
    eprint = "1711.11554",
    archivePrefix = "arXiv",
    primaryClass = "hep-ph",
    reportNumber = "DESY-17-056",
    doi = "10.1007/JHEP01(2018)159",
    journal = "JHEP",
    volume = "01",
    pages = "159",
    year = "2018"
}

@article{LISACosmologyWorkingGroup:2022kbp,
    author = "Bartolo, Nicola and others",
    collaboration = "LISA Cosmology Working Group",
    title = "{Probing anisotropies of the Stochastic Gravitational Wave Background with LISA}",
    eprint = "2201.08782",
    archivePrefix = "arXiv",
    primaryClass = "astro-ph.CO",
    doi = "10.1088/1475-7516/2022/11/009",
    journal = "JCAP",
    volume = "11",
    pages = "009",
    year = "2022"
}

@article{RoperPol:2019wvy,
    author = "Roper Pol, Alberto and Mandal, Sayan and Brandenburg, Axel and Kahniashvili, Tina and Kosowsky, Arthur",
    title = "{Numerical simulations of gravitational waves from early-universe turbulence}",
    eprint = "1903.08585",
    archivePrefix = "arXiv",
    primaryClass = "astro-ph.CO",
    reportNumber = "NORDITA-2019-024",
    doi = "10.1103/PhysRevD.102.083512",
    journal = "Phys. Rev. D",
    volume = "102",
    number = "8",
    pages = "083512",
    year = "2020"
}

@article{Huber:2008hg,
    author = "Huber, Stephan J. and Konstandin, Thomas",
    title = "{Gravitational Wave Production by Collisions: More Bubbles}",
    eprint = "0806.1828",
    archivePrefix = "arXiv",
    primaryClass = "hep-ph",
    doi = "10.1088/1475-7516/2008/09/022",
    journal = "JCAP",
    volume = "09",
    pages = "022",
    year = "2008"
}

@article{Kosowsky:1992vn,
    author = "Kosowsky, Arthur and Turner, Michael S.",
    title = "{Gravitational radiation from colliding vacuum bubbles: envelope approximation to many bubble collisions}",
    eprint = "astro-ph/9211004",
    archivePrefix = "arXiv",
    reportNumber = "FERMILAB-PUB-92-295-A",
    doi = "10.1103/PhysRevD.47.4372",
    journal = "Phys. Rev. D",
    volume = "47",
    pages = "4372--4391",
    year = "1993"
}

@article{Dillon:2017ctw,
    author = "Dillon, Barry M. and El-Menoufi, Basem Kamal and Huber, Stephan J. and Manuel, Jonathan P.",
    title = "{Rapid holographic phase transition with brane-localized curvature}",
    eprint = "1708.02953",
    archivePrefix = "arXiv",
    primaryClass = "hep-th",
    doi = "10.1103/PhysRevD.98.086005",
    journal = "Phys. Rev. D",
    volume = "98",
    number = "8",
    pages = "086005",
    year = "2018"
}

@article{Ghiglieri:2020mhm,
    author = "Ghiglieri, J. and Jackson, G. and Laine, M. and Zhu, Y.",
    title = "{Gravitational wave background from Standard Model physics: Complete leading order}",
    eprint = "2004.11392",
    archivePrefix = "arXiv",
    primaryClass = "hep-ph",
    doi = "10.1007/JHEP07(2020)092",
    journal = "JHEP",
    volume = "07",
    pages = "092",
    year = "2020"
}

@article{Dolgov:2002ra,
    author = "Dolgov, Alexander D. and Grasso, Dario and Nicolis, Alberto",
    title = "{Relic backgrounds of gravitational waves from cosmic turbulence}",
    eprint = "astro-ph/0206461",
    archivePrefix = "arXiv",
    doi = "10.1103/PhysRevD.66.103505",
    journal = "Phys. Rev. D",
    volume = "66",
    pages = "103505",
    year = "2002"
}

@article{Moore:1995ua,
    author = "Moore, Guy D. and Prokopec, Tomislav",
    title = "{Bubble wall velocity in a first order electroweak phase transition}",
    eprint = "hep-ph/9503296",
    archivePrefix = "arXiv",
    reportNumber = "PUPT-1531, LANCASTER-TH-9503, PUP-TH-1531-(1995), LANCASTER-TH-9503-(1995)",
    doi = "10.1103/PhysRevLett.75.777",
    journal = "Phys. Rev. Lett.",
    volume = "75",
    pages = "777--780",
    year = "1995"
}

@article{BarrosoMancha:2020fay,
    author = "Barroso Mancha, Marc and Prokopec, Tomislav and Swiezewska, Bogumila",
    title = "{Field-theoretic derivation of bubble-wall force}",
    eprint = "2005.10875",
    archivePrefix = "arXiv",
    primaryClass = "hep-th",
    doi = "10.1007/JHEP01(2021)070",
    journal = "JHEP",
    volume = "01",
    pages = "070",
    year = "2021"
}

@article{Cutting:2018tjt,
    author = "Cutting, Daniel and Hindmarsh, Mark and Weir, David J.",
    title = "{Gravitational waves from vacuum first-order phase transitions: from the envelope to the lattice}",
    eprint = "1802.05712",
    archivePrefix = "arXiv",
    primaryClass = "astro-ph.CO",
    reportNumber = "HIP-2018-4-TH",
    doi = "10.1103/PhysRevD.97.123513",
    journal = "Phys. Rev. D",
    volume = "97",
    number = "12",
    pages = "123513",
    year = "2018"
}

@article{Farmer:2003pa,
    author = "Farmer, Alison J. and Phinney, E. Sterl",
    title = "{The gravitational wave background from cosmological compact binaries}",
    eprint = "astro-ph/0304393",
    archivePrefix = "arXiv",
    doi = "10.1111/j.1365-2966.2003.07176.x",
    journal = "Mon. Not. Roy. Astron. Soc.",
    volume = "346",
    pages = "1197",
    year = "2003"
}

@article{Lamberts:2019nyk,
    author = "Lamberts, Astrid and Blunt, Sarah and Littenberg, Tyson B. and Garrison-Kimmel, Shea and Kupfer, Thomas and Sanderson, Robyn E.",
    title = "{Predicting the LISA white dwarf binary population in the Milky Way with cosmological simulations}",
    eprint = "1907.00014",
    archivePrefix = "arXiv",
    primaryClass = "astro-ph.HE",
    doi = "10.1093/mnras/stz2834",
    journal = "Mon. Not. Roy. Astron. Soc.",
    volume = "490",
    number = "4",
    pages = "5888--5903",
    year = "2019"
}

@article{Korol:2021pun,
    author = "Korol, Valeriya and Hallakoun, Na'ama and Toonen, Silvia and Karnesis, Nikolaos",
    title = "{Observationally driven Galactic double white dwarf population for LISA}",
    eprint = "2109.10972",
    archivePrefix = "arXiv",
    primaryClass = "astro-ph.HE",
    doi = "10.1093/mnras/stac415",
    journal = "Mon. Not. Roy. Astron. Soc.",
    volume = "511",
    number = "4",
    pages = "5936--5947",
    year = "2022"
}

@article{Karnesis:2021tsh,
    author = "Karnesis, Nikolaos and Babak, Stanislav and Pieroni, Mauro and Cornish, Neil and Littenberg, Tyson",
    title = "{Characterization of the stochastic signal originating from compact binary populations as measured by LISA}",
    eprint = "2103.14598",
    archivePrefix = "arXiv",
    primaryClass = "astro-ph.IM",
    doi = "10.1103/PhysRevD.104.043019",
    journal = "Phys. Rev. D",
    volume = "104",
    number = "4",
    pages = "043019",
    year = "2021"
}

@article{Babak:2023lro,
    author = "Babak, Stanislav and Caprini, Chiara and Figueroa, Daniel G. and Karnesis, Nikolaos and Marcoccia, Paolo and Nardini, Germano and Pieroni, Mauro and Ricciardone, Angelo and Sesana, Alberto and Torrado, Jes{\'u}s",
    title = "{Stochastic gravitational wave background from stellar origin binary black holes in LISA}",
    eprint = "2304.06368",
    archivePrefix = "arXiv",
    primaryClass = "astro-ph.CO",
    doi = "10.1088/1475-7516/2023/08/034",
    journal = "JCAP",
    volume = "08",
    pages = "034",
    year = "2023"
}

@article{Lehoucq:2023zlt,
    author = "Lehoucq, Leonard and Dvorkin, Irina and Srinivasan, Rahul and Pellouin, Clement and Lamberts, Astrid",
    title = "{Astrophysical uncertainties in the gravitational-wave background from stellar-mass compact binary mergers}",
    eprint = "2306.09861",
    archivePrefix = "arXiv",
    primaryClass = "astro-ph.HE",
    doi = "10.1093/mnras/stad2917",
    journal = "Mon. Not. Roy. Astron. Soc.",
    volume = "526",
    number = "3",
    pages = "4378--4387",
    year = "2023"
}

@article{Piarulli:2024yhj,
    author = "Piarulli, Manuel and Buscicchio, Riccardo and Pozzoli, Federico and Burke, Ollie and Bonetti, Matteo and Sesana, Alberto",
    title = "{Test for LISA foreground Gaussianity and stationarity: Extreme mass-ratio inspirals}",
    eprint = "2410.08862",
    archivePrefix = "arXiv",
    primaryClass = "astro-ph.HE",
    doi = "10.1103/nfn4-pgr5",
    journal = "Phys. Rev. D",
    volume = "111",
    number = "10",
    pages = "103047",
    year = "2025"
}

@article{Ellis:2023oxs,
    author = {Ellis, John and Fairbairn, Malcolm and Franciolini, Gabriele and H{\"u}tsi, Gert and Iovino, Antonio and Lewicki, Marek and Raidal, Martti and Urrutia, Juan and Vaskonen, Ville and Veerm{\"a}e, Hardi},
    title = "{What is the source of the PTA GW signal?}",
    eprint = "2308.08546",
    archivePrefix = "arXiv",
    primaryClass = "astro-ph.CO",
    reportNumber = "KCL-PH-TH/2023-43, CERN-TH-2023-153, AION-REPORT/2023-08",
    doi = "10.1103/PhysRevD.109.023522",
    journal = "Phys. Rev. D",
    volume = "109",
    number = "2",
    pages = "023522",
    year = "2024"
}

@article{Child:2012qg,
    author = "Child, Hillary L. and Giblin, Jr., John T.",
    title = "{Gravitational Radiation from First-Order Phase Transitions}",
    eprint = "1207.6408",
    archivePrefix = "arXiv",
    primaryClass = "astro-ph.CO",
    doi = "10.1088/1475-7516/2012/10/001",
    journal = "JCAP",
    volume = "10",
    pages = "001",
    year = "2012"
}

@article{Jinno:2019bxw,
    author = "Jinno, Ryusuke and Konstandin, Thomas and Takimoto, Masahiro",
    title = "{Relativistic bubble collisions\textemdash{}a closer look}",
    eprint = "1906.02588",
    archivePrefix = "arXiv",
    primaryClass = "hep-ph",
    reportNumber = "DESY-19-102, DESY 19-102",
    doi = "10.1088/1475-7516/2019/09/035",
    journal = "JCAP",
    volume = "09",
    pages = "035",
    year = "2019"
}

@article{Konstandin:2017sat,
    author = "Konstandin, Thomas",
    title = "{Gravitational radiation from a bulk flow model}",
    eprint = "1712.06869",
    archivePrefix = "arXiv",
    primaryClass = "astro-ph.CO",
    reportNumber = "DESY-17-227",
    doi = "10.1088/1475-7516/2018/03/047",
    journal = "JCAP",
    volume = "03",
    pages = "047",
    year = "2018"
}

@article{Cutting:2020nla,
    author = "Cutting, Daniel and Escartin, Elba Granados and Hindmarsh, Mark and Weir, David J.",
    title = "{Gravitational waves from vacuum first order phase transitions II: from thin to thick walls}",
    eprint = "2005.13537",
    archivePrefix = "arXiv",
    primaryClass = "astro-ph.CO",
    reportNumber = "HIP-2020-13/TH",
    doi = "10.1103/PhysRevD.103.023531",
    journal = "Phys. Rev. D",
    volume = "103",
    number = "2",
    pages = "023531",
    year = "2021"
}

@article{Bodeker:2009qy,
    author = "Bodeker, Dietrich and Moore, Guy D.",
    title = "{Can electroweak bubble walls run away?}",
    eprint = "0903.4099",
    archivePrefix = "arXiv",
    primaryClass = "hep-ph",
    doi = "10.1088/1475-7516/2009/05/009",
    journal = "JCAP",
    volume = "05",
    pages = "009",
    year = "2009"
}

@article{Jinno:2022mie,
    author = "Jinno, Ryusuke and Konstandin, Thomas and Rubira, Henrique and Stomberg, Isak",
    title = "{Higgsless simulations of cosmological phase transitions and gravitational waves}",
    eprint = "2209.04369",
    archivePrefix = "arXiv",
    primaryClass = "astro-ph.CO",
    reportNumber = "DESY 22-148, IFT-UAM/CSIC-22-100, TUM-HEP-1416/22",
    doi = "10.1088/1475-7516/2023/02/011",
    journal = "JCAP",
    volume = "02",
    pages = "011",
    year = "2023"
}

@article{Cai:2023guc,
    author = "Cai, Rong-Gen and Wang, Shao-Jiang and Yuwen, Zi-Yan",
    title = "{Hydrodynamic sound shell model}",
    eprint = "2305.00074",
    archivePrefix = "arXiv",
    primaryClass = "gr-qc",
    doi = "10.1103/PhysRevD.108.L021502",
    journal = "Phys. Rev. D",
    volume = "108",
    number = "2",
    pages = "L021502",
    year = "2023"
}

@article{Gould:2019qek,
    author = "Gould, Oliver and Kozaczuk, Jonathan and Niemi, Lauri and Ramsey-Musolf, Michael J. and Tenkanen, Tuomas V. I. and Weir, David J.",
    title = "{Nonperturbative analysis of the gravitational waves from a first-order electroweak phase transition}",
    eprint = "1903.11604",
    archivePrefix = "arXiv",
    primaryClass = "hep-ph",
    reportNumber = "ACFI T19-04, HIP-2019-5/TH",
    doi = "10.1103/PhysRevD.100.115024",
    journal = "Phys. Rev. D",
    volume = "100",
    number = "11",
    pages = "115024",
    year = "2019"
}

@article{Athron:2023xlk,
    author = "Athron, Peter and Bal\'azs, Csaba and Fowlie, Andrew and Morris, Lachlan and Wu, Lei",
    title = "{Cosmological phase transitions: From perturbative particle physics to gravitational waves}",
    eprint = "2305.02357",
    archivePrefix = "arXiv",
    primaryClass = "hep-ph",
    doi = "10.1016/j.ppnp.2023.104094",
    journal = "Prog. Part. Nucl. Phys.",
    volume = "135",
    pages = "104094",
    year = "2024"
}

@article{Megias:2018sxv,
    author = "Meg\'\i{}as, Eugenio and Nardini, Germano and Quir\'os, Mariano",
    title = "{Cosmological Phase Transitions in Warped Space: Gravitational Waves and Collider Signatures}",
    eprint = "1806.04877",
    archivePrefix = "arXiv",
    primaryClass = "hep-ph",
    reportNumber = "UAB-FT-776",
    doi = "10.1007/JHEP09(2018)095",
    journal = "JHEP",
    volume = "09",
    pages = "095",
    year = "2018"
}

@article{Nardini:2007me,
    author = "Nardini, Germano and Quiros, Mariano and Wulzer, Andrea",
    title = "{A Confining Strong First-Order Electroweak Phase Transition}",
    eprint = "0706.3388",
    archivePrefix = "arXiv",
    primaryClass = "hep-ph",
    reportNumber = "UAB-FT-632",
    doi = "10.1088/1126-6708/2007/09/077",
    journal = "JHEP",
    volume = "09",
    pages = "077",
    year = "2007"
}

@article{Kamionkowski:1993fg,
    author = "Kamionkowski, Marc and Kosowsky, Arthur and Turner, Michael S.",
    title = "{Gravitational radiation from first order phase transitions}",
    eprint = "astro-ph/9310044",
    archivePrefix = "arXiv",
    reportNumber = "IASSNS-HEP-93-44, FERMILAB-PUB-93-235-A",
    doi = "10.1103/PhysRevD.49.2837",
    journal = "Phys. Rev. D",
    volume = "49",
    pages = "2837--2851",
    year = "1994"
}

@article{Kajantie:1995kf,
    author = "Kajantie, K. and Laine, M. and Rummukainen, K. and Shaposhnikov, Mikhail E.",
    title = "{The Electroweak phase transition: A Nonperturbative analysis}",
    eprint = "hep-lat/9510020",
    archivePrefix = "arXiv",
    reportNumber = "CERN-TH-95-263, HD-THEP-95-44, HU-TFT-95-57, IUHET-318",
    doi = "10.1016/0550-3213(96)00052-1",
    journal = "Nucl. Phys. B",
    volume = "466",
    pages = "189--258",
    year = "1996"
}

@article{Callan:1977pt,
    author = "Callan, Jr., Curtis G. and Coleman, Sidney R.",
    title = "{The Fate of the False Vacuum. 2. First Quantum Corrections}",
    reportNumber = "HUTP-77-A032",
    doi = "10.1103/PhysRevD.16.1762",
    journal = "Phys. Rev. D",
    volume = "16",
    pages = "1762--1768",
    year = "1977"
}

@article{Kahniashvili:2008pe,
    author = "Kahniashvili, Tina and Campanelli, Leonardo and Gogoberidze, Grigol and Maravin, Yurii and Ratra, Bharat",
    title = "{Gravitational Radiation from Primordial Helical Inverse Cascade MHD Turbulence}",
    eprint = "0809.1899",
    archivePrefix = "arXiv",
    primaryClass = "astro-ph",
    doi = "10.1103/PhysRevD.78.123006",
    journal = "Phys. Rev. D",
    volume = "78",
    pages = "123006",
    year = "2008",
    note = "[Erratum: Phys.Rev.D 79, 109901 (2009)]"
}

@article{Hindmarsh:2015qta,
    author = "Hindmarsh, Mark and Huber, Stephan J. and Rummukainen, Kari and Weir, David J.",
    title = "{Numerical simulations of acoustically generated gravitational waves at a first order phase transition}",
    eprint = "1504.03291",
    archivePrefix = "arXiv",
    primaryClass = "astro-ph.CO",
    reportNumber = "HIP-2015-13-TH",
    doi = "10.1103/PhysRevD.92.123009",
    journal = "Phys. Rev. D",
    volume = "92",
    number = "12",
    pages = "123009",
    year = "2015"
}

@article{Zhou:2022nmt,
    author = "Zhou, Bei and Reali, Luca and Berti, Emanuele and {\c{C}}al{\i}{\c{s}}kan, Mesut and Creque-Sarbinowski, Cyril and Kamionkowski, Marc and Sathyaprakash, B. S.",
    title = "{Subtracting compact binary foregrounds to search for subdominant gravitational-wave backgrounds in next-generation ground-based observatories}",
    eprint = "2209.01310",
    archivePrefix = "arXiv",
    primaryClass = "gr-qc",
    doi = "10.1103/PhysRevD.108.064040",
    journal = "Phys. Rev. D",
    volume = "108",
    number = "6",
    pages = "064040",
    year = "2023"
}

@article{Caprini:2024hue,
    author = "Caprini, Chiara and Jinno, Ryusuke and Lewicki, Marek and Madge, Eric and Merchand, Marco and Nardini, Germano and Pieroni, Mauro and Roper Pol, Alberto and Vaskonen, Ville",
    collaboration = "LISA Cosmology Working Group",
    title = "{Gravitational waves from first-order phase transitions in LISA: reconstruction pipeline and physics interpretation}",
    eprint = "2403.03723",
    archivePrefix = "arXiv",
    primaryClass = "astro-ph.CO",
    reportNumber = "LISA-COSWG-24-01, CERN-TH-2024-029",
    month = "3",
    year = "2024"
}

@article{Auclair:2022jod,
    author = "Auclair, Pierre and Caprini, Chiara and Cutting, Daniel and Hindmarsh, Mark and Rummukainen, Kari and Steer, Dani\`ele A. and Weir, David J.",
    title = "{Generation of gravitational waves from freely decaying turbulence}",
    eprint = "2205.02588",
    archivePrefix = "arXiv",
    primaryClass = "astro-ph.CO",
    reportNumber = "HIP-2021-35/TH",
    doi = "10.1088/1475-7516/2022/09/029",
    journal = "JCAP",
    volume = "09",
    pages = "029",
    year = "2022"
}

@article{Bunk:2017fic,
    author = "Bunk, Don and Hubisz, Jay and Jain, Bithika",
    title = "{A Perturbative RS I Cosmological Phase Transition}",
    eprint = "1705.00001",
    archivePrefix = "arXiv",
    primaryClass = "hep-ph",
    doi = "10.1140/epjc/s10052-018-5529-2",
    journal = "Eur. Phys. J. C",
    volume = "78",
    number = "1",
    pages = "78",
    year = "2018"
}

@article{Niemi:2018asa,
    author = "Niemi, Lauri and Patel, Hiren H. and Ramsey-Musolf, Michael J. and Tenkanen, Tuomas V. I. and Weir, David J.",
    title = "{Electroweak phase transition in the real triplet extension of the SM: Dimensional reduction}",
    eprint = "1802.10500",
    archivePrefix = "arXiv",
    primaryClass = "hep-ph",
    reportNumber = "HIP-2018-7-TH, ACFI-T18-04, HIP-2018-7/TH",
    doi = "10.1103/PhysRevD.100.035002",
    journal = "Phys. Rev. D",
    volume = "100",
    number = "3",
    pages = "035002",
    year = "2019"
}

@article{Banerjee:2024qiu,
    author = "Banerjee, Upalaparna and Chakraborty, Sabyasachi and Prakash, Suraj and Rahaman, Shakeel Ur",
    title = "{The feasibility of ultra-relativistic bubbles in SMEFT}",
    eprint = "2402.02914",
    archivePrefix = "arXiv",
    primaryClass = "hep-ph",
    month = "2",
    year = "2024"
}

@article{Aoki:1996cu,
    author = "Aoki, Yasumichi",
    title = "{Four-dimensional simulation of the hot electroweak phase transition with the SU(2) gauge Higgs model}",
    eprint = "hep-lat/9612023",
    archivePrefix = "arXiv",
    reportNumber = "UTCCP-P-20",
    doi = "10.1103/PhysRevD.56.3860",
    journal = "Phys. Rev. D",
    volume = "56",
    pages = "3860--3865",
    year = "1997"
}

@article{Coleman:1977py,
    author = "Coleman, Sidney R.",
    title = "{The Fate of the False Vacuum. 1. Semiclassical Theory}",
    reportNumber = "HUTP-77-A004",
    doi = "10.1103/PhysRevD.16.1248",
    journal = "Phys. Rev. D",
    volume = "15",
    pages = "2929--2936",
    year = "1977",
    note = "[Erratum: Phys.Rev.D 16, 1248 (1977)]"
}

@article{Hindmarsh:2016lnk,
    author = "Hindmarsh, Mark",
    title = "{Sound shell model for acoustic gravitational wave production at a first-order phase transition in the early Universe}",
    eprint = "1608.04735",
    archivePrefix = "arXiv",
    primaryClass = "astro-ph.CO",
    doi = "10.1103/PhysRevLett.120.071301",
    journal = "Phys. Rev. Lett.",
    volume = "120",
    number = "7",
    pages = "071301",
    year = "2018"
}

@article{Caprini:2018mtu,
    author = "Caprini, Chiara and Figueroa, Daniel G.",
    title = "{Cosmological Backgrounds of Gravitational Waves}",
    eprint = "1801.04268",
    archivePrefix = "arXiv",
    primaryClass = "astro-ph.CO",
    doi = "10.1088/1361-6382/aac608",
    journal = "Class. Quant. Grav.",
    volume = "35",
    number = "16",
    pages = "163001",
    year = "2018"
}

@article{Hindmarsh:2013xza,
    author = "Hindmarsh, Mark and Huber, Stephan J. and Rummukainen, Kari and Weir, David J.",
    title = "{Gravitational waves from the sound of a first order phase transition}",
    eprint = "1304.2433",
    archivePrefix = "arXiv",
    primaryClass = "hep-ph",
    reportNumber = "HIP-2013-07-TH",
    doi = "10.1103/PhysRevLett.112.041301",
    journal = "Phys. Rev. Lett.",
    volume = "112",
    pages = "041301",
    year = "2014"
}

@article{Kosowsky:2001xp,
    author = "Kosowsky, Arthur and Mack, Andrew and Kahniashvili, Tinatin",
    title = "{Gravitational radiation from cosmological turbulence}",
    eprint = "astro-ph/0111483",
    archivePrefix = "arXiv",
    reportNumber = "RAP-334",
    doi = "10.1103/PhysRevD.66.024030",
    journal = "Phys. Rev. D",
    volume = "66",
    pages = "024030",
    year = "2002"
}

@article{Brandenburg:2017neh,
    author = "Brandenburg, Axel and Kahniashvili, Tina and Mandal, Sayan and Roper Pol, Alberto and Tevzadze, Alexander G. and Vachaspati, Tanmay",
    title = "{Evolution of hydromagnetic turbulence from the electroweak phase transition}",
    eprint = "1711.03804",
    archivePrefix = "arXiv",
    primaryClass = "astro-ph.CO",
    reportNumber = "NORDITA-2017-116",
    doi = "10.1103/PhysRevD.96.123528",
    journal = "Phys. Rev. D",
    volume = "96",
    number = "12",
    pages = "123528",
    year = "2017"
}

@article{Schicho:2021gca,
    author = {Schicho, Philipp M. and Tenkanen, Tuomas V. I. and \"Osterman, Juuso},
    title = "{Robust approach to thermal resummation: Standard Model meets a singlet}",
    eprint = "2102.11145",
    archivePrefix = "arXiv",
    primaryClass = "hep-ph",
    doi = "10.1007/JHEP06(2021)130",
    journal = "JHEP",
    volume = "06",
    pages = "130",
    year = "2021"
}

@article{Hogan:1986qda,
    author = "Hogan, C. J.",
    title = "{Gravitational radiation from cosmological phase transitions}",
    journal = "Mon. Not. Roy. Astron. Soc.",
    volume = "218",
    pages = "629--636",
    year = "1986"
}

@article{Gurtler:1997hr,
    author = "Gurtler, M. and Ilgenfritz, Ernst-Michael and Schiller, A.",
    title = "{Where the electroweak phase transition ends}",
    eprint = "hep-lat/9704013",
    archivePrefix = "arXiv",
    reportNumber = "UL-NTZ-10-97, HUB-EP-97-24, DESY-97-086",
    doi = "10.1103/PhysRevD.56.3888",
    journal = "Phys. Rev. D",
    volume = "56",
    pages = "3888--3895",
    year = "1997"
}

@article{Linde:1980tt,
    author = "Linde, Andrei D.",
    title = "{Fate of the False Vacuum at Finite Temperature: Theory and Applications}",
    reportNumber = "LEBEDEV-80-92",
    doi = "10.1016/0370-2693(81)90281-1",
    journal = "Phys. Lett. B",
    volume = "100",
    pages = "37--40",
    year = "1981"
}

@article{Konstandin:2010cd,
    author = "Konstandin, Thomas and Nardini, Germano and Quiros, Mariano",
    title = "{Gravitational Backreaction Effects on the Holographic Phase Transition}",
    eprint = "1007.1468",
    archivePrefix = "arXiv",
    primaryClass = "hep-ph",
    reportNumber = "CERN-PH-TH-2010-151, UAB-FT-683, ULB-TH-10-19",
    doi = "10.1103/PhysRevD.82.083513",
    journal = "Phys. Rev. D",
    volume = "82",
    pages = "083513",
    year = "2010"
}

@article{Moore:1995si,
    author = "Moore, Guy D. and Prokopec, Tomislav",
    title = "{How fast can the wall move? A Study of the electroweak phase transition dynamics}",
    eprint = "hep-ph/9506475",
    archivePrefix = "arXiv",
    reportNumber = "PUPT-1544, PUP-TH-1544, LANCS-TH-9517",
    doi = "10.1103/PhysRevD.52.7182",
    journal = "Phys. Rev. D",
    volume = "52",
    pages = "7182--7204",
    year = "1995"
}

@article{Cutting:2019zws,
    author = "Cutting, Daniel and Hindmarsh, Mark and Weir, David J.",
    title = "{Vorticity, kinetic energy, and suppressed gravitational wave production in strong first order phase transitions}",
    eprint = "1906.00480",
    archivePrefix = "arXiv",
    primaryClass = "hep-ph",
    reportNumber = "HIP-2019-15/TH",
    doi = "10.1103/PhysRevLett.125.021302",
    journal = "Phys. Rev. Lett.",
    volume = "125",
    number = "2",
    pages = "021302",
    year = "2020"
}

@article{VonHarling:2019rgb,
    author = "Von Harling, Benedict and Pomarol, Alex and Pujol\`as, Oriol and Rompineve, Fabrizio",
    title = "{Peccei-Quinn Phase Transition at LIGO}",
    eprint = "1912.07587",
    archivePrefix = "arXiv",
    primaryClass = "hep-ph",
    doi = "10.1007/JHEP04(2020)195",
    journal = "JHEP",
    volume = "04",
    pages = "195",
    year = "2020"
}

@article{Jinno:2020eqg,
    author = "Jinno, Ryusuke and Konstandin, Thomas and Rubira, Henrique",
    title = "{A hybrid simulation of gravitational wave production in first-order phase transitions}",
    eprint = "2010.00971",
    archivePrefix = "arXiv",
    primaryClass = "astro-ph.CO",
    reportNumber = "DESY-20-170, DESY 20-170",
    doi = "10.1088/1475-7516/2021/04/014",
    journal = "JCAP",
    volume = "04",
    pages = "014",
    year = "2021"
}

@article{RoperPol:2018sap,
    author = "Roper Pol, Alberto and Brandenburg, Axel and Kahniashvili, Tina and Kosowsky, Arthur and Mandal, Sayan",
    title = "{The timestep constraint in solving the gravitational wave equations sourced by hydromagnetic turbulence}",
    eprint = "1807.05479",
    archivePrefix = "arXiv",
    primaryClass = "physics.flu-dyn",
    reportNumber = "NORDITA-2018-054",
    doi = "10.1080/03091929.2019.1653460",
    journal = "Geophys. Astrophys. Fluid Dynamics",
    volume = "114",
    number = "1-2",
    pages = "130--161",
    year = "2020"
}

@article{Badger:2022nwo,
    author = "Badger, Charles and others",
    title = "{Probing early Universe supercooled phase transitions with gravitational wave data}",
    eprint = "2209.14707",
    archivePrefix = "arXiv",
    primaryClass = "hep-ph",
    reportNumber = "KCL-PH-TH-2022-43",
    doi = "10.1103/PhysRevD.107.023511",
    journal = "Phys. Rev. D",
    volume = "107",
    number = "2",
    pages = "023511",
    year = "2023"
}

@article{RoperPol:2021xnd,
    author = "Roper Pol, Alberto and Mandal, Sayan and Brandenburg, Axel and Kahniashvili, Tina",
    title = "{Polarization of gravitational waves from helical MHD turbulent sources}",
    eprint = "2107.05356",
    archivePrefix = "arXiv",
    primaryClass = "gr-qc",
    reportNumber = "NORDITA-2021-062",
    doi = "10.1088/1475-7516/2022/04/019",
    journal = "JCAP",
    volume = "04",
    number = "04",
    pages = "019",
    year = "2022"
}

@article{Brandenburg:2021bvg,
    author = "Brandenburg, Axel and Gogoberidze, Grigol and Kahniashvili, Tina and Mandal, Sayan and Roper Pol, Alberto and Shenoy, Nakul",
    title = "{The scalar, vector, and tensor modes in gravitational wave turbulence simulations}",
    eprint = "2103.01140",
    archivePrefix = "arXiv",
    primaryClass = "gr-qc",
    reportNumber = "NORDITA-2021-019",
    doi = "10.1088/1361-6382/ac011c",
    journal = "Class. Quant. Grav.",
    volume = "38",
    number = "14",
    pages = "145002",
    year = "2021"
}

@article{Caldwell:2022qsj,
    author = "Caldwell, Robert and others",
    title = "{Detection of early-universe gravitational-wave signatures and fundamental physics}",
    eprint = "2203.07972",
    archivePrefix = "arXiv",
    primaryClass = "gr-qc",
    doi = "10.1007/s10714-022-03027-x",
    journal = "Gen. Rel. Grav.",
    volume = "54",
    number = "12",
    pages = "156",
    year = "2022"
}

@article{Roshan:2024qnv,
    author = "Roshan, Rishav and White, Graham",
    title = "{Using gravitational waves to see the first second of the Universe}",
    eprint = "2401.04388",
    archivePrefix = "arXiv",
    primaryClass = "hep-ph",
    month = "1",
    year = "2024"
}

@article{Caprini:2019egz,
    author = "Caprini, Chiara and others",
    title = "{Detecting gravitational waves from cosmological phase transitions with LISA: an update}",
    eprint = "1910.13125",
    archivePrefix = "arXiv",
    primaryClass = "astro-ph.CO",
    reportNumber = "DESY-19-159, IPPP/19/27, HIP-2019-14/TH, MITP/19-066, IFT-UAM/CSIC-19-139",
    doi = "10.1088/1475-7516/2020/03/024",
    journal = "JCAP",
    volume = "03",
    pages = "024",
    year = "2020"
}

@article{Gowling:2021gcy,
    author = "Gowling, Chloe and Hindmarsh, Mark",
    title = "{Observational prospects for phase transitions at LISA: Fisher matrix analysis}",
    eprint = "2106.05984",
    archivePrefix = "arXiv",
    primaryClass = "astro-ph.CO",
    doi = "10.1088/1475-7516/2021/10/039",
    journal = "JCAP",
    volume = "10",
    pages = "039",
    year = "2021"
}

@article{Boileau:2022ter,
    author = "Boileau, Guillaume and Christensen, Nelson and Gowling, Chloe and Hindmarsh, Mark and Meyer, Renate",
    title = "{Prospects for LISA to detect a gravitational-wave background from first order phase transitions}",
    eprint = "2209.13277",
    archivePrefix = "arXiv",
    primaryClass = "gr-qc",
    doi = "10.1088/1475-7516/2023/02/056",
    journal = "JCAP",
    volume = "02",
    pages = "056",
    year = "2023"
}

@article{Gowling:2022pzb,
    author = "Gowling, Chloe and Hindmarsh, Mark and Hooper, Deanna C. and Torrado, Jes\'us",
    title = "{Reconstructing physical parameters from template gravitational wave spectra at LISA: first order phase transitions}",
    eprint = "2209.13551",
    archivePrefix = "arXiv",
    primaryClass = "astro-ph.CO",
    doi = "10.1088/1475-7516/2023/04/061",
    journal = "JCAP",
    volume = "04",
    pages = "061",
    year = "2023"
}

@article{Hindmarsh:2024ttn,
    author = "Hindmarsh, Mark and Hooper, Deanna C. and Minkkinen, Tiina and Weir, David J.",
    title = "{Recovering a phase transition signal in simulated LISA data with a modulated galactic foreground}",
    eprint = "2406.04894",
    archivePrefix = "arXiv",
    primaryClass = "astro-ph.CO",
    reportNumber = "HIP-2024-8/TH",
    doi = "10.1088/1475-7516/2025/04/052",
    journal = "JCAP",
    volume = "04",
    pages = "052",
    year = "2025"
}

@article{Giese:2021dnw,
    author = "Giese, Felix and Konstandin, Thomas and van de Vis, Jorinde",
    title = "{Finding sound shells in LISA mock data using likelihood sampling}",
    eprint = "2107.06275",
    archivePrefix = "arXiv",
    primaryClass = "astro-ph.CO",
    reportNumber = "DESY-21-109, DESY 2021-02966",
    doi = "10.1088/1475-7516/2021/11/002",
    journal = "JCAP",
    volume = "11",
    pages = "002",
    year = "2021"
}

@article{Correia:2025qif,
    author = "Correia, Jos{\'e} and Hindmarsh, Mark and Rummukainen, Kari and Weir, David J.",
    title = "{Gravitational waves from strong first order phase transitions}",
    eprint = "2505.17824",
    archivePrefix = "arXiv",
    primaryClass = "astro-ph.CO",
    month = "5",
    year = "2025"
}

@article{Giombi:2025tkv,
    author = "Giombi, Lorenzo and Dahl, Jani and Hindmarsh, Mark",
    title = "{Acoustic gravitational waves beyond leading order in bubble over Hubble radius}",
    eprint = "2504.08037",
    archivePrefix = "arXiv",
    primaryClass = "gr-qc",
    reportNumber = "HIP-2025-13/TH",
    month = "4",
    year = "2025"
}

@article{Dahl:2021wyk,
    author = "Dahl, Jani and Hindmarsh, Mark and Rummukainen, Kari and Weir, David J.",
    title = "{Decay of acoustic turbulence in two dimensions and implications for cosmological gravitational waves}",
    eprint = "2112.12013",
    archivePrefix = "arXiv",
    primaryClass = "gr-qc",
    reportNumber = "HIP-2021-29/TH",
    doi = "10.1103/PhysRevD.106.063511",
    journal = "Phys. Rev. D",
    volume = "106",
    number = "6",
    pages = "063511",
    year = "2022"
}

@article{Caprini:2009yp,
    author = "Caprini, Chiara and Durrer, Ruth and Servant, Geraldine",
    title = "{The stochastic gravitational wave background from turbulence and magnetic fields generated by a first-order phase transition}",
    eprint = "0909.0622",
    archivePrefix = "arXiv",
    primaryClass = "astro-ph.CO",
    doi = "10.1088/1475-7516/2009/12/024",
    journal = "JCAP",
    volume = "12",
    pages = "024",
    year = "2009"
}

@article{Hindmarsh:2017gnf,
    author = "Hindmarsh, Mark and Huber, Stephan J. and Rummukainen, Kari and Weir, David J.",
    title = "{Shape of the acoustic gravitational wave power spectrum from a first order phase transition}",
    eprint = "1704.05871",
    archivePrefix = "arXiv",
    primaryClass = "astro-ph.CO",
    reportNumber = "HIP-2017-02-TH, HIP-2017-02/TH",
    doi = "10.1103/PhysRevD.96.103520",
    journal = "Phys. Rev. D",
    volume = "96",
    number = "10",
    pages = "103520",
    year = "2017",
    note = "[Erratum: Phys.Rev.D 101, 089902 (2020)]"
}

@article{Lewicki:2024xan,
    author = "Lewicki, Marek and Merchand, Marco and Sagunski, Laura and Schicho, Philipp and Schmitt, Daniel",
    title = "{Impact of theoretical uncertainties on model parameter reconstruction from GW signals sourced by cosmological phase transitions}",
    eprint = "2403.03769",
    archivePrefix = "arXiv",
    primaryClass = "hep-ph",
    month = "3",
    year = "2024"
}

@article{Konstandin:2011dr,
    author = "Konstandin, Thomas and Servant, Geraldine",
    title = "{Cosmological Consequences of Nearly Conformal Dynamics at the TeV scale}",
    eprint = "1104.4791",
    archivePrefix = "arXiv",
    primaryClass = "hep-ph",
    doi = "10.1088/1475-7516/2011/12/009",
    journal = "JCAP",
    volume = "12",
    pages = "009",
    year = "2011"
}

@article{Caprini:2015zlo,
    author = "Caprini, Chiara and others",
    title = "{Science with the space-based interferometer eLISA. II: Gravitational waves from cosmological phase transitions}",
    eprint = "1512.06239",
    archivePrefix = "arXiv",
    primaryClass = "astro-ph.CO",
    reportNumber = "DESY-15-246",
    doi = "10.1088/1475-7516/2016/04/001",
    journal = "JCAP",
    volume = "04",
    pages = "001",
    year = "2016"
}

@article{Kajantie:1996mn,
    author = "Kajantie, K. and Laine, M. and Rummukainen, K. and Shaposhnikov, Mikhail E.",
    title = "{Is there a~ hot electroweak phase transition at $m_H \gtrsim m_W$?}",
    eprint = "hep-ph/9605288",
    archivePrefix = "arXiv",
    reportNumber = "CERN-TH-96-126, HD-THEP-96-15, IUHET-333",
    doi = "10.1103/PhysRevLett.77.2887",
    journal = "Phys. Rev. Lett.",
    volume = "77",
    pages = "2887--2890",
    year = "1996"
}

@article{Niksa:2018ofa,
    author = {Niksa, Peter and Schlederer, Martin and Sigl, G\"unter},
    title = "{Gravitational Waves produced by Compressible MHD Turbulence from Cosmological Phase Transitions}",
    eprint = "1803.02271",
    archivePrefix = "arXiv",
    primaryClass = "astro-ph.CO",
    doi = "10.1088/1361-6382/aac89c",
    journal = "Class. Quant. Grav.",
    volume = "35",
    number = "14",
    pages = "144001",
    year = "2018"
}

@article{Croon:2020cgk,
    author = "Croon, Djuna and Gould, Oliver and Schicho, Philipp and Tenkanen, Tuomas V. I. and White, Graham",
    title = "{Theoretical uncertainties for cosmological first-order phase transitions}",
    eprint = "2009.10080",
    archivePrefix = "arXiv",
    primaryClass = "hep-ph",
    reportNumber = "HIP-2020-26/TH",
    doi = "10.1007/JHEP04(2021)055",
    journal = "JHEP",
    volume = "04",
    pages = "055",
    year = "2021"
}

@article{Hindmarsh:2019phv,
    author = "Hindmarsh, Mark and Hijazi, Mulham",
    title = "{Gravitational waves from first order cosmological phase transitions in the Sound Shell Model}",
    eprint = "1909.10040",
    archivePrefix = "arXiv",
    primaryClass = "astro-ph.CO",
    reportNumber = "NORDITA-2019-083, HIP-2019-29/TH",
    doi = "10.1088/1475-7516/2019/12/062",
    journal = "JCAP",
    volume = "12",
    pages = "062",
    year = "2019"
}

@article{Kozaczuk:2015owa,
    author = "Kozaczuk, Jonathan",
    title = "{Bubble Expansion and the Viability of Singlet-Driven Electroweak Baryogenesis}",
    eprint = "1506.04741",
    archivePrefix = "arXiv",
    primaryClass = "hep-ph",
    doi = "10.1007/JHEP10(2015)135",
    journal = "JHEP",
    volume = "10",
    pages = "135",
    year = "2015"
}

@article{Kierkla:2023von,
    author = "Kierkla, Maciej and Swiezewska, Bogumila and Tenkanen, Tuomas V. I. and van de Vis, Jorinde",
    title = "{Gravitational waves from supercooled phase transitions: dimensional transmutation meets dimensional reduction}",
    eprint = "2312.12413",
    archivePrefix = "arXiv",
    primaryClass = "hep-ph",
    doi = "10.1007/JHEP02(2024)234",
    journal = "JHEP",
    volume = "02",
    pages = "234",
    year = "2024"
}

@article{Athron:2022mmm,
    author = "Athron, Peter and Bal\'azs, Csaba and Morris, Lachlan",
    title = "{Supercool subtleties of cosmological phase transitions}",
    eprint = "2212.07559",
    archivePrefix = "arXiv",
    primaryClass = "hep-ph",
    doi = "10.1088/1475-7516/2023/03/006",
    journal = "JCAP",
    volume = "03",
    pages = "006",
    year = "2023"
}

@article{Gould:2023ovu,
    author = "Gould, Oliver and Tenkanen, Tuomas V. I.",
    title = "{Perturbative effective field theory expansions for cosmological phase transitions}",
    eprint = "2309.01672",
    archivePrefix = "arXiv",
    primaryClass = "hep-ph",
    reportNumber = "NORDITA 2023-037",
    doi = "10.1007/JHEP01(2024)048",
    journal = "JHEP",
    volume = "01",
    pages = "048",
    year = "2024"
}

@article{Hoeche:2020rsg,
    author = {H\"oche, Stefan and Kozaczuk, Jonathan and Long, Andrew J. and Turner, Jessica and Wang, Yikun},
    title = "{Towards an all-orders calculation of the electroweak bubble wall velocity}",
    eprint = "2007.10343",
    archivePrefix = "arXiv",
    primaryClass = "hep-ph",
    reportNumber = "FERMILAB-PUB-20-274-T",
    doi = "10.1088/1475-7516/2021/03/009",
    journal = "JCAP",
    volume = "03",
    pages = "009",
    year = "2021"
}

@article{Jinno:2016vai,
    author = "Jinno, Ryusuke and Takimoto, Masahiro",
    title = "{Gravitational waves from bubble collisions: An analytic derivation}",
    eprint = "1605.01403",
    archivePrefix = "arXiv",
    primaryClass = "astro-ph.CO",
    reportNumber = "KEK-TH-1900",
    doi = "10.1103/PhysRevD.95.024009",
    journal = "Phys. Rev. D",
    volume = "95",
    number = "2",
    pages = "024009",
    year = "2017"
}

@article{Chatterjee:2025wbz,
    author = "Chatterjee, Arpan and Frasca, Marco and Ghoshal, Anish and Groote, Stefan",
    title = "{Finite temperature QCD crossover at non-zero chemical potential: A Dyson{\textendash}Schwinger approach}",
    eprint = "2502.02070",
    archivePrefix = "arXiv",
    primaryClass = "hep-ph",
    doi = "10.1016/j.nuclphysb.2025.116972",
    journal = "Nucl. Phys. B",
    volume = "1017",
    pages = "116972",
    year = "2025"
}

@article{Middeldorf-Wygas:2020glx,
    author = {Middeldorf-Wygas, Mandy M. and Oldengott, Isabel M. and B{\"o}deker, Dietrich and Schwarz, Dominik J.},
    title = "{Cosmic QCD transition for large lepton flavor asymmetries}",
    eprint = "2009.00036",
    archivePrefix = "arXiv",
    primaryClass = "hep-ph",
    doi = "10.1103/PhysRevD.105.123533",
    journal = "Phys. Rev. D",
    volume = "105",
    number = "12",
    pages = "123533",
    year = "2022"
}

@article{Zheng:2024tib,
    author = "Zheng, Hui-wen and Gao, Fei and Bian, Ligong and Qin, Si-xue and Liu, Yu-xin",
    title = "{Quantitative analysis of the gravitational wave spectrum sourced from a first-order chiral phase transition of QCD}",
    eprint = "2407.03795",
    archivePrefix = "arXiv",
    primaryClass = "hep-ph",
    doi = "10.1103/PhysRevD.111.L021303",
    journal = "Phys. Rev. D",
    volume = "111",
    number = "2",
    pages = "L021303",
    year = "2025"
}

@article{Binetruy:2012ze,
      author         = "Binetruy, Pierre and Bohe, Alejandro and Caprini, Chiara
                        and Dufaux, Jean-Francois",
      title          = "{Cosmological Backgrounds of Gravitational Waves and
                        eLISA/NGO: Phase Transitions, Cosmic Strings and Other
                        Sources}",
      journal        = "JCAP",
      volume         = "1206",
      year           = "2012",
      pages          = "027",
      doi            = "10.1088/1475-7516/2012/06/027",
      eprint         = "1201.0983",
      archivePrefix  = "arXiv",
      primaryClass   = "gr-qc",
      SLACcitation   = "%%CITATION = ARXIV:1201.0983;%%"
}

@article{ZambujalFerreira:2021cte,
    author = "Zambujal Ferreira, Ricardo and Notari, Alessio and Pujol\`as, Oriol and Rompineve, Fabrizio",
    title = "{High Quality QCD Axion at Gravitational Wave Observatories}",
    eprint = "2107.07542",
    archivePrefix = "arXiv",
    primaryClass = "hep-ph",
    doi = "10.1103/PhysRevLett.128.141101",
    journal = "Phys. Rev. Lett.",
    volume = "128",
    number = "14",
    pages = "141101",
    year = "2022"
}

@article{Peccei:1977hh,
    author = "Peccei, R. D. and Quinn, Helen R.",
    title = "{CP Conservation in the Presence of Instantons}",
    reportNumber = "ITP-568-STANFORD",
    doi = "10.1103/PhysRevLett.38.1440",
    journal = "Phys. Rev. Lett.",
    volume = "38",
    pages = "1440--1443",
    year = "1977"
}

@article{Peccei:1977ur,
    author = "Peccei, R. D. and Quinn, Helen R.",
    title = "{Constraints Imposed by CP Conservation in the Presence of Instantons}",
    reportNumber = "ITP-572-STANFORD",
    doi = "10.1103/PhysRevD.16.1791",
    journal = "Phys. Rev. D",
    volume = "16",
    pages = "1791--1797",
    year = "1977"
}

@article{Dine:1981rt,
    author = "Dine, Michael and Fischler, Willy and Srednicki, Mark",
    title = "{A Simple Solution to the Strong CP Problem with a Harmless Axion}",
    reportNumber = "Print-81-0320 (IAS,PRINCETON)",
    doi = "10.1016/0370-2693(81)90590-6",
    journal = "Phys. Lett. B",
    volume = "104",
    pages = "199--202",
    year = "1981"
}

@article{Kim:1979if,
    author = "Kim, Jihn E.",
    title = "{Weak Interaction Singlet and Strong CP Invariance}",
    reportNumber = "UPR-0120T",
    doi = "10.1103/PhysRevLett.43.103",
    journal = "Phys. Rev. Lett.",
    volume = "43",
    pages = "103",
    year = "1979"
}

@article{Shifman:1979if,
    author = "Shifman, Mikhail A. and Vainshtein, A. I. and Zakharov, Valentin I.",
    title = "{Can Confinement Ensure Natural CP Invariance of Strong Interactions?}",
    reportNumber = "ITEP-64-1979",
    doi = "10.1016/0550-3213(80)90209-6",
    journal = "Nucl. Phys. B",
    volume = "166",
    pages = "493--506",
    year = "1980"
}

@article{Zhitnitsky:1980tq,
    author = "Zhitnitsky, A. R.",
    title = "{On Possible Suppression of the Axion Hadron Interactions. (In Russian)}",
    journal = "Sov. J. Nucl. Phys.",
    volume = "31",
    pages = "260",
    year = "1980"
}

@article{Raffelt:1990yz,
    author = "Raffelt, Georg G.",
    title = "{Astrophysical methods to constrain axions and other novel particle phenomena}",
    reportNumber = "MPI-PAE-PTH-29-90",
    doi = "10.1016/0370-1573(90)90054-6",
    journal = "Phys. Rept.",
    volume = "198",
    pages = "1--113",
    year = "1990"
}

@article{Chang:2018rso,
    author = "Chang, Jae Hyeok and Essig, Rouven and McDermott, Samuel D.",
    title = "{Supernova 1987A Constraints on Sub-GeV Dark Sectors, Millicharged Particles, the QCD Axion, and an Axion-like Particle}",
    eprint = "1803.00993",
    archivePrefix = "arXiv",
    primaryClass = "hep-ph",
    reportNumber = "YITP-SB-18-01, FERMILAB-PUB-17-432-T",
    doi = "10.1007/JHEP09(2018)051",
    journal = "JHEP",
    volume = "09",
    pages = "051",
    year = "2018"
}

@article{Carenza:2020cis,
    author = "Carenza, Pierluca and Fore, Bryce and Giannotti, Maurizio and Mirizzi, Alessandro and Reddy, Sanjay",
    title = "{Enhanced Supernova Axion Emission and its Implications}",
    eprint = "2010.02943",
    archivePrefix = "arXiv",
    primaryClass = "hep-ph",
    reportNumber = "INT-PUB-20-039",
    doi = "10.1103/PhysRevLett.126.071102",
    journal = "Phys. Rev. Lett.",
    volume = "126",
    number = "7",
    pages = "071102",
    year = "2021"
}

@article{Notari:2022ffe,
    author = "Notari, Alessio and Rompineve, Fabrizio and Villadoro, Giovanni",
    title = "{Improved Hot Dark Matter Bound on the QCD Axion}",
    eprint = "2211.03799",
    archivePrefix = "arXiv",
    primaryClass = "hep-ph",
    reportNumber = "CERN-TH-2022-165",
    doi = "10.1103/PhysRevLett.131.011004",
    journal = "Phys. Rev. Lett.",
    volume = "131",
    number = "1",
    pages = "011004",
    year = "2023"
}

@article{Creminelli:2001th,
    author = "Creminelli, Paolo and Nicolis, Alberto and Rattazzi, Riccardo",
    title = "{Holography and the electroweak phase transition}",
    eprint = "hep-th/0107141",
    archivePrefix = "arXiv",
    reportNumber = "CERN-TH-2001-189",
    doi = "10.1088/1126-6708/2002/03/051",
    journal = "JHEP",
    volume = "03",
    pages = "051",
    year = "2002"
}

@article{Gorghetto:2020qws,
    author = "Gorghetto, Marco and Hardy, Edward and Villadoro, Giovanni",
    title = "{More axions from strings}",
    eprint = "2007.04990",
    archivePrefix = "arXiv",
    primaryClass = "hep-ph",
    doi = "10.21468/SciPostPhys.10.2.050",
    journal = "SciPost Phys.",
    volume = "10",
    number = "2",
    pages = "050",
    year = "2021"
}

@article{MillerBertolami:2014rka,
    author = "Miller Bertolami, Marcelo M. and Melendez, Brenda E. and Althaus, Leandro G. and Isern, Jordi",
    title = "{Revisiting the axion bounds from the Galactic white dwarf luminosity function}",
    eprint = "1406.7712",
    archivePrefix = "arXiv",
    primaryClass = "hep-ph",
    doi = "10.1088/1475-7516/2014/10/069",
    journal = "JCAP",
    volume = "10",
    pages = "069",
    year = "2014"
}

@article{Buschmann:2021sdq,
    author = "Buschmann, Malte and Foster, Joshua W. and Hook, Anson and Peterson, Adam and Willcox, Don E. and Zhang, Weiqun and Safdi, Benjamin R.",
    title = "{Dark matter from axion strings with adaptive mesh refinement}",
    eprint = "2108.05368",
    archivePrefix = "arXiv",
    primaryClass = "hep-ph",
    doi = "10.1038/s41467-022-28669-y",
    journal = "Nature Commun.",
    volume = "13",
    number = "1",
    pages = "1049",
    year = "2022"
}

@article{Saikawa:2024bta,
    author = "Saikawa, Ken'ichi and Redondo, Javier and Vaquero, Alejandro and Kaltschmidt, Mathieu",
    title = "{Spectrum of global string networks and the axion dark matter mass}",
    eprint = "2401.17253",
    archivePrefix = "arXiv",
    primaryClass = "hep-ph",
    reportNumber = "KANAZAWA-24-02, MPP-2024-18",
    month = "1",
    year = "2024"
}

@article{Ayala:2014pea,
    author = "Ayala, Adrian and Dom\'\i{}nguez, Inma and Giannotti, Maurizio and Mirizzi, Alessandro and Straniero, Oscar",
    title = "{Revisiting the bound on axion-photon coupling from Globular Clusters}",
    eprint = "1406.6053",
    archivePrefix = "arXiv",
    primaryClass = "astro-ph.SR",
    doi = "10.1103/PhysRevLett.113.191302",
    journal = "Phys. Rev. Lett.",
    volume = "113",
    number = "19",
    pages = "191302",
    year = "2014"
}

@article{Dolan:2022kul,
    author = "Dolan, Matthew J. and Hiskens, Frederick J. and Volkas, Raymond R.",
    title = "{Advancing globular cluster constraints on the axion-photon coupling}",
    eprint = "2207.03102",
    archivePrefix = "arXiv",
    primaryClass = "hep-ph",
    doi = "10.1088/1475-7516/2022/10/096",
    journal = "JCAP",
    volume = "10",
    pages = "096",
    year = "2022"
}

@article{Bar:2019ifz,
    author = "Bar, Nitsan and Blum, Kfir and D'Amico, Guido",
    title = "{Is there a supernova bound on axions?}",
    eprint = "1907.05020",
    archivePrefix = "arXiv",
    primaryClass = "hep-ph",
    reportNumber = "CERN-TH-2019-109",
    doi = "10.1103/PhysRevD.101.123025",
    journal = "Phys. Rev. D",
    volume = "101",
    number = "12",
    pages = "123025",
    year = "2020"
}

@article{Ferreira:2022zzo,
    author = "Ferreira, Ricardo Z. and Notari, Alessio and Pujolas, Oriol and Rompineve, Fabrizio",
    title = "{Gravitational waves from domain walls in Pulsar Timing Array datasets}",
    eprint = "2204.04228",
    archivePrefix = "arXiv",
    primaryClass = "astro-ph.CO",
    reportNumber = "CERN-TH-2022-214",
    doi = "10.1088/1475-7516/2023/02/001",
    journal = "JCAP",
    volume = "02",
    pages = "001",
    year = "2023"
}

@article{DelleRose:2019pgi,
    author = "Delle Rose, Luigi and Panico, Giuliano and Redi, Michele and Tesi, Andrea",
    title = "{Gravitational Waves from Supercool Axions}",
    eprint = "1912.06139",
    archivePrefix = "arXiv",
    primaryClass = "hep-ph",
    doi = "10.1007/JHEP04(2020)025",
    journal = "JHEP",
    volume = "04",
    pages = "025",
    year = "2020"
}

@article{Kahniashvili:2008er,
      author         = "Kahniashvili, Tina and Gogoberidze, Grigol and Ratra,
                        Bharat",
      title          = "{Gravitational Radiation from Primordial Helical MHD
                        Turbulence}",
      journal        = "Phys. Rev. Lett.",
      volume         = "100",
      year           = "2008",
      pages          = "231301",
      doi            = "10.1103/PhysRevLett.100.231301",
      eprint         = "0802.3524",
      archivePrefix  = "arXiv",
      primaryClass   = "astro-ph",
      reportNumber   = "KSUPT_08-1",
      SLACcitation   = "%%CITATION = ARXIV:0802.3524;%%"
}

@article{Huber:2015znp,
      author         = "Huber, Stephan J. and Konstandin, Thomas and Nardini,
                        Germano and Rues, Ingo",
      title          = "{Detectable Gravitational Waves from Very Strong Phase
                        Transitions in the General NMSSM}",
      journal        = "JCAP",
      volume         = "1603",
      year           = "2016",
      number         = "03",
      pages          = "036",
      doi            = "10.1088/1475-7516/2016/03/036",
      eprint         = "1512.06357",
      archivePrefix  = "arXiv",
      primaryClass   = "hep-ph",
      reportNumber   = "DESY-15-247",
      SLACcitation   = "%%CITATION = ARXIV:1512.06357;%%"
}

@article{Schwaller:2015tja,
      author         = "Schwaller, Pedro",
      title          = "{Gravitational Waves from a Dark Phase Transition}",
      journal        = "Phys. Rev. Lett.",
      volume         = "115",
      year           = "2015",
      number         = "18",
      pages          = "181101",
      doi            = "10.1103/PhysRevLett.115.181101",
      eprint         = "1504.07263",
      archivePrefix  = "arXiv",
      primaryClass   = "hep-ph",
      reportNumber   = "CERN-PH-TH-2015-093",
      SLACcitation   = "%%CITATION = ARXIV:1504.07263;%%"
}

@article{Kibble:1976sj,
    author = "Kibble, T. W. B.",
    title = "{Topology of Cosmic Domains and Strings}",
    reportNumber = "ICTP/75/5",
    doi = "10.1088/0305-4470/9/8/029",
    journal = "J. Phys. A",
    volume = "9",
    pages = "1387--1398",
    year = "1976"
}

@book{Vachaspati:2006zz,
    author = "Vachaspati, Tanmay",
    isbn = "978-0-521-14191-8, 978-0-521-83605-0, 978-0-511-24290-8",
    month = "4",
    publisher = "Cambridge University Press",
    title = "{Kinks and domain walls: An introduction to classical and quantum solitons}",
    year = "2010"
}

@article{Kitajima:2023cek,
    author = "Kitajima, Naoya and Lee, Junseok and Murai, Kai and Takahashi, Fuminobu and Yin, Wen",
    title = "{Gravitational Waves from Domain Wall Collapse, and Application to Nanohertz Signals with QCD-coupled Axions}",
    eprint = "2306.17146",
    archivePrefix = "arXiv",
    primaryClass = "hep-ph",
    reportNumber = "TU-1198",
    month = "6",
    year = "2023"
}

@article{Ferreira:2024eru,
    author = "Ferreira, Ricardo Z. and Notari, Alessio and Pujol\`as, Oriol and Rompineve, Fabrizio",
    title = "{Collapsing Domain Wall Networks: Impact on Pulsar Timing Arrays and Primordial Black Holes}",
    eprint = "2401.14331",
    archivePrefix = "arXiv",
    primaryClass = "astro-ph.CO",
    reportNumber = "CERN-TH-2024-020",
    month = "1",
    year = "2024"
}

@article{Schwarz:2009ii,
    author = "Schwarz, Dominik J. and Stuke, Maik",
    title = "{Lepton asymmetry and the cosmic QCD transition}",
    eprint = "0906.3434",
    archivePrefix = "arXiv",
    primaryClass = "hep-ph",
    reportNumber = "BI-TP-2009-14",
    doi = "10.1088/1475-7516/2009/11/025",
    journal = "JCAP",
    volume = "11",
    pages = "025",
    year = "2009",
    note = "[Erratum: JCAP 10, E01 (2010)]"
}

@article{Wygas:2018otj,
    author = {Wygas, Mandy M. and Oldengott, Isabel M. and B\"odeker, Dietrich and Schwarz, Dominik J.},
    title = "{Cosmic QCD Epoch at Nonvanishing Lepton Asymmetry}",
    eprint = "1807.10815",
    archivePrefix = "arXiv",
    primaryClass = "hep-ph",
    doi = "10.1103/PhysRevLett.121.201302",
    journal = "Phys. Rev. Lett.",
    volume = "121",
    number = "20",
    pages = "201302",
    year = "2018"
}

@article{Kitano:2021fdl,
    author = "Kitano, Ryuichiro and Yin, Wen",
    title = "{Strong CP problem and axion dark matter with small instantons}",
    eprint = "2103.08598",
    archivePrefix = "arXiv",
    primaryClass = "hep-ph",
    reportNumber = "KEK-TH-2310",
    doi = "10.1007/JHEP07(2021)078",
    journal = "JHEP",
    volume = "07",
    pages = "078",
    year = "2021"
}

@article{Dimopoulos:1979pp,
    author = "Dimopoulos, Savas",
    title = "{A Solution of the Strong {CP} Problem in Models With Scalars}",
    reportNumber = "CU-TP-154",
    doi = "10.1016/0370-2693(79)91233-4",
    journal = "Phys. Lett. B",
    volume = "84",
    pages = "435--439",
    year = "1979"
}

@article{Treiman:1978ge,
    author = "Treiman, S. B. and Wilczek, Frank",
    title = "{Axion Emission in Decay of Excited Nuclear States}",
    reportNumber = "Print-78-0295 (IAS,PRINCETON)",
    doi = "10.1016/0370-2693(78)90684-6",
    journal = "Phys. Lett. B",
    volume = "74",
    pages = "381--383",
    year = "1978"
}

@article{Tye:1981zy,
    author = "Tye, S. H. H.",
    title = "{A Superstrong Force With a Heavy Axion}",
    reportNumber = "CLNS 81/489",
    doi = "10.1103/PhysRevLett.47.1035",
    journal = "Phys. Rev. Lett.",
    volume = "47",
    pages = "1035",
    year = "1981"
}

@article{Gorghetto:2018myk,
    author = "Gorghetto, Marco and Hardy, Edward and Villadoro, Giovanni",
    title = "{Axions from Strings: the Attractive Solution}",
    eprint = "1806.04677",
    archivePrefix = "arXiv",
    primaryClass = "hep-ph",
    doi = "10.1007/JHEP07(2018)151",
    journal = "JHEP",
    volume = "07",
    pages = "151",
    year = "2018"
}

@article{Holdom:1982ex,
    author = "Holdom, Bob and Peskin, Michael E.",
    title = "{Raising the Axion Mass}",
    reportNumber = "ITP-719-STANFORD",
    doi = "10.1016/0550-3213(82)90228-0",
    journal = "Nucl. Phys. B",
    volume = "208",
    pages = "397--412",
    year = "1982"
}

@article{Holdom:1985vx,
    author = "Holdom, Bob",
    title = "{Strong QCD at High-energies and a Heavy Axion}",
    reportNumber = "Print-85-0092 (TORONTO)",
    doi = "10.1016/0370-2693(85)90371-5",
    journal = "Phys. Lett. B",
    volume = "154",
    pages = "316",
    year = "1985",
    note = "[Erratum: Phys.Lett.B 156, 452 (1985)]"
}

@article{Flynn:1987rs,
    author = "Flynn, Jonathan M. and Randall, Lisa",
    title = "{A Computation of the Small Instanton Contribution to the Axion Potential}",
    reportNumber = "HUTP-87/A015",
    doi = "10.1016/0550-3213(87)90089-7",
    journal = "Nucl. Phys. B",
    volume = "293",
    pages = "731--739",
    year = "1987"
}

@article{Choi:1998ep,
    author = "Choi, Kiwoon and Kim, Hyung Do",
    title = "{Small instanton contribution to the axion potential in supersymmetric models}",
    eprint = "hep-ph/9809286",
    archivePrefix = "arXiv",
    reportNumber = "KAIST-TH-98-18",
    doi = "10.1103/PhysRevD.59.072001",
    journal = "Phys. Rev. D",
    volume = "59",
    pages = "072001",
    year = "1999"
}

@article{Rubakov:1997vp,
    author = "Rubakov, V. A.",
    title = "{Grand unification and heavy axion}",
    eprint = "hep-ph/9703409",
    archivePrefix = "arXiv",
    reportNumber = "INR-97-231",
    doi = "10.1134/1.567390",
    journal = "JETP Lett.",
    volume = "65",
    pages = "621--624",
    year = "1997"
}

@article{Agrawal:2017ksf,
    author = "Agrawal, Prateek and Howe, Kiel",
    title = "{Factoring the Strong CP Problem}",
    eprint = "1710.04213",
    archivePrefix = "arXiv",
    primaryClass = "hep-ph",
    reportNumber = "FERMILAB-PUB-17-500-T",
    doi = "10.1007/JHEP12(2018)029",
    journal = "JHEP",
    volume = "12",
    pages = "029",
    year = "2018"
}

@article{NANOGrav:2023gor,
    author = "Agazie, Gabriella and others",
    collaboration = "NANOGrav",
    title = "{The NANOGrav 15 yr Data Set: Evidence for a Gravitational-wave Background}",
    eprint = "2306.16213",
    archivePrefix = "arXiv",
    primaryClass = "astro-ph.HE",
    doi = "10.3847/2041-8213/acdac6",
    journal = "Astrophys. J. Lett.",
    volume = "951",
    number = "1",
    pages = "L8",
    year = "2023"
}

@article{Reardon:2023gzh,
    author = "Reardon, Daniel J. and others",
    title = "{Search for an Isotropic Gravitational-wave Background with the Parkes Pulsar Timing Array}",
    eprint = "2306.16215",
    archivePrefix = "arXiv",
    primaryClass = "astro-ph.HE",
    doi = "10.3847/2041-8213/acdd02",
    journal = "Astrophys. J. Lett.",
    volume = "951",
    number = "1",
    pages = "L6",
    year = "2023"
}

@article{Xu:2023wog,
    author = "Xu, Heng and others",
    title = "{Searching for the Nano-Hertz Stochastic Gravitational Wave Background with the Chinese Pulsar Timing Array Data Release I}",
    eprint = "2306.16216",
    archivePrefix = "arXiv",
    primaryClass = "astro-ph.HE",
    doi = "10.1088/1674-4527/acdfa5",
    journal = "Res. Astron. Astrophys.",
    volume = "23",
    number = "7",
    pages = "075024",
    year = "2023"
}

@article{EPTA:2023fyk,
    author = "Antoniadis, J. and others",
    collaboration = "EPTA, InPTA:",
    title = "{The second data release from the European Pulsar Timing Array - III. Search for gravitational wave signals}",
    eprint = "2306.16214",
    archivePrefix = "arXiv",
    primaryClass = "astro-ph.HE",
    doi = "10.1051/0004-6361/202346844",
    journal = "Astron. Astrophys.",
    volume = "678",
    pages = "A50",
    year = "2023"
}

@article{NANOGrav:2023hfp,
    author = "Agazie, Gabriella and others",
    collaboration = "NANOGrav",
    title = "{The NANOGrav 15 yr Data Set: Constraints on Supermassive Black Hole Binaries from the Gravitational-wave Background}",
    eprint = "2306.16220",
    archivePrefix = "arXiv",
    primaryClass = "astro-ph.HE",
    doi = "10.3847/2041-8213/ace18b",
    journal = "Astrophys. J. Lett.",
    volume = "952",
    number = "2",
    pages = "L37",
    year = "2023"
}

@article{NANOGrav:2023hvm,
    author = "Afzal, Adeela and others",
    collaboration = "NANOGrav",
    title = "{The NANOGrav 15 yr Data Set: Search for Signals from New Physics}",
    eprint = "2306.16219",
    archivePrefix = "arXiv",
    primaryClass = "astro-ph.HE",
    reportNumber = "FERMILAB-PUB-23-589-T",
    doi = "10.3847/2041-8213/acdc91",
    journal = "Astrophys. J. Lett.",
    volume = "951",
    number = "1",
    pages = "L11",
    year = "2023"
}

@article{Antoniadis:2022pcn,
    author = "Antoniadis, J. and others",
    title = "{The International Pulsar Timing Array second data release: Search for an isotropic gravitational wave background}",
    eprint = "2201.03980",
    archivePrefix = "arXiv",
    primaryClass = "astro-ph.HE",
    doi = "10.1093/mnras/stab3418",
    journal = "Mon. Not. Roy. Astron. Soc.",
    volume = "510",
    number = "4",
    pages = "4873--4887",
    year = "2022"
}

@article{EPTA:2023xxk,
    author = "Antoniadis, J. and others",
    collaboration = "EPTA",
    title = "{The second data release from the European Pulsar Timing Array: V. Implications for massive black holes, dark matter and the early Universe}",
    eprint = "2306.16227",
    archivePrefix = "arXiv",
    primaryClass = "astro-ph.CO",
    month = "6",
    year = "2023"
}

@article{Gherghetta:2020ofz,
    author = "Gherghetta, Tony and Nguyen, Minh D.",
    title = "{A Composite Higgs with a Heavy Composite Axion}",
    eprint = "2007.10875",
    archivePrefix = "arXiv",
    primaryClass = "hep-ph",
    reportNumber = "UMN--TH--3923/20",
    doi = "10.1007/JHEP12(2020)094",
    journal = "JHEP",
    volume = "12",
    pages = "094",
    year = "2020"
}

@article{Hook:2019qoh,
    author = "Hook, Anson and Kumar, Soubhik and Liu, Zhen and Sundrum, Raman",
    title = "{High Quality QCD Axion and the LHC}",
    eprint = "1911.12364",
    archivePrefix = "arXiv",
    primaryClass = "hep-ph",
    reportNumber = "UMD-PP-019-07",
    doi = "10.1103/PhysRevLett.124.221801",
    journal = "Phys. Rev. Lett.",
    volume = "124",
    number = "22",
    pages = "221801",
    year = "2020"
}

@article{Dimopoulos:2016lvn,
    author = "Dimopoulos, Savas and Hook, Anson and Huang, Junwu and Marques-Tavares, Gustavo",
    title = "{A collider observable QCD axion}",
    eprint = "1606.03097",
    archivePrefix = "arXiv",
    primaryClass = "hep-ph",
    doi = "10.1007/JHEP11(2016)052",
    journal = "JHEP",
    volume = "11",
    pages = "052",
    year = "2016"
}

@article{Damour:2000wa,
    author = "Damour, Thibault and Vilenkin, Alexander",
    title = "{Gravitational wave bursts from cosmic strings}",
    eprint = "gr-qc/0004075",
    archivePrefix = "arXiv",
    reportNumber = "IHES-P-00-32",
    doi = "10.1103/PhysRevLett.85.3761",
    journal = "Phys. Rev. Lett.",
    volume = "85",
    pages = "3761--3764",
    year = "2000"
}

@inproceedings{Battye:1997jk,
    author = "Battye, R. A. and Shellard, E. P. S.",
    title = "{Recent perspectives on axion cosmology}",
    booktitle = "{1st International Heidelberg Conference on Dark Matter in Astro and Particle Physics}",
    eprint = "astro-ph/9706014",
    archivePrefix = "arXiv",
    reportNumber = "IMPERIAL-TP-95-96-31A",
    pages = "554--579",
    month = "6",
    year = "1997"
}

@article{PhysRevD.32.3172,
  title = {Goldstone bosons in string models of galaxy formation},
  author = {Davis, Richard Lynn},
  journal = {Phys. Rev. D},
  volume = {32},
  issue = {12},
  pages = {3172--3177},
  numpages = {0},
  year = {1985},
  month = {Dec},
  publisher = {American Physical Society},
  doi = {10.1103/PhysRevD.32.3172},
  url = {https://link.aps.org/doi/10.1103/PhysRevD.32.3172}
}

@ARTICLE{1987PhLB..195..361H,
       author = {{Harari}, Diego and {Sikivie}, P.},
        title = "{On the evolution of global strings in the early universe}",
      journal = {Physics Letters B},
         year = 1987,
        month = sep,
       volume = {195},
       number = {3},
        pages = {361-365},
          doi = {10.1016/0370-2693(87)90032-3},
       adsurl = {https://ui.adsabs.harvard.edu/abs/1987PhLB..195..361H}
}

@article{DAVIS1988219,
title = {Antisymmetric tensors and spontaneous symmetry breaking},
journal = {Physics Letters B},
volume = {214},
number = {2},
pages = {219-222},
year = {1988},
issn = {0370-2693},
doi = {https://doi.org/10.1016/0370-2693(88)91472-4},
url = {https://www.sciencedirect.com/science/article/pii/0370269388914724},
author = {R.L. Davis and E.P.S. Shellard}
}

@article{PhysRevD.9.2273,
  title = {Classical direct interstring action},
  author = {Kalb, Michael and Ramond, P.},
  journal = {Phys. Rev. D},
  volume = {9},
  issue = {8},
  pages = {2273--2284},
  numpages = {0},
  year = {1974},
  month = {Apr},
  publisher = {American Physical Society},
  doi = {10.1103/PhysRevD.9.2273},
  url = {https://link.aps.org/doi/10.1103/PhysRevD.9.2273}
}

@article{Aggarwal:2020olq,
    author = "Aggarwal, Nancy and others",
    title = "{Challenges and opportunities of gravitational-wave searches at MHz to GHz frequencies}",
    eprint = "2011.12414",
    archivePrefix = "arXiv",
    primaryClass = "gr-qc",
    reportNumber = "CERN-TH-2020-185, HIP-2020-28/TH, DESY 20-195, CERN-TH-2020-185, HIP-2020-28/TH, DESY 20-195",
    doi = "10.1007/s41114-021-00032-5",
    journal = "Living Rev. Rel.",
    volume = "24",
    number = "1",
    pages = "4",
    year = "2021"
}

@misc{Virgoweb,
    url = "https://www.virgo-gw.eu/"
}

@misc{LIGOweb,
    url = "https://www.ligo.caltech.edu/"
}

@misc{KAGRAweb,
  url = "https://gwcenter.icrr.u-tokyo.ac.jp/en/"
}

@misc{CEweb,
  url = "https://cosmicexplorer.org/"
}

@misc{ETweb,
  url = "https://www.et-gw.eu/"
}

@misc{LISAweb,
  url = "https://www.esa.int/Science_Exploration/Space_Science/LISA"
}

@misc{Cosmogw,
  url = "https://github.com/cosmoGW/cosmoGW"
}

@article{Torrado:2020dgo,
    author = "Torrado, Jesus and Lewis, Antony",
    title = "{Cobaya: Code for Bayesian Analysis of hierarchical physical models}",
    eprint = "2005.05290",
    archivePrefix = "arXiv",
    primaryClass = "astro-ph.IM",
    reportNumber = "TTK-20-15",
    doi = "10.1088/1475-7516/2021/05/057",
    journal = "JCAP",
    volume = "05",
    pages = "057",
    year = "2021"
}

@article{Ekstedt:2024fyq,
    author = "Ekstedt, Andreas and Gould, Oliver and Hirvonen, Joonas and Laurent, Benoit and Niemi, Lauri and Schicho, Philipp and van de Vis, Jorinde",
    title = "{How fast does the WallGo? A package for computing wall velocities in first-order phase transitions}",
    eprint = "2411.04970",
    archivePrefix = "arXiv",
    primaryClass = "hep-ph",
    reportNumber = "CERN-TH-2024-174, DESY-24-162, HIP-2024-21/TH",
    doi = "10.1007/JHEP04(2025)101",
    journal = "JHEP",
    volume = "04",
    pages = "101",
    year = "2025"
}

@article{Handley:2015vkr,
    author = "Handley, W. J. and Hobson, M. P. and Lasenby, A. N.",
    title = "{polychord: next-generation nested sampling}",
    eprint = "1506.00171",
    archivePrefix = "arXiv",
    primaryClass = "astro-ph.IM",
    doi = "10.1093/mnras/stv1911",
    journal = "Mon. Not. Roy. Astron. Soc.",
    volume = "453",
    number = "4",
    pages = "4385--4399",
    year = "2015"
}

@article{Handley:2015fda,
    author = "Handley, W. J. and Hobson, M. P. and Lasenby, A. N.",
    title = "{PolyChord: nested sampling for cosmology}",
    eprint = "1502.01856",
    archivePrefix = "arXiv",
    primaryClass = "astro-ph.CO",
    doi = "10.1093/mnrasl/slv047",
    journal = "Mon. Not. Roy. Astron. Soc.",
    volume = "450",
    number = "1",
    pages = "L61--L65",
    year = "2015"
}

@article{Lewis:2019xzd,
    author = "Lewis, Antony",
    title = "{GetDist: a Python package for analysing Monte Carlo samples}",
    eprint = "1910.13970",
    archivePrefix = "arXiv",
    primaryClass = "astro-ph.IM",
    doi = "10.1088/1475-7516/2025/08/025",
    journal = "JCAP",
    volume = "08",
    pages = "025",
    year = "2025"
}

@article{Maggiore:2024cwf,
    author = "Maggiore, Michele and Iacovelli, Francesco and Belgacem, Enis and Mancarella, Michele and Muttoni, Niccol{\`o}",
    title = "{Comparison of global networks of third-generation gravitational-wave detectors}",
    eprint = "2411.05754",
    archivePrefix = "arXiv",
    primaryClass = "gr-qc",
    doi = "10.1088/1361-6382/ae110b",
    journal = "Class. Quant. Grav.",
    volume = "42",
    number = "21",
    pages = "215004",
    year = "2025"
}

@article{Binetruy:2009vt,
    author = "Binetruy, P. and Bohe, A. and Hertog, T. and Steer, Daniele A.",
    title = "{Gravitational Wave Bursts from Cosmic Superstrings with Y-junctions}",
    eprint = "0907.4522",
    archivePrefix = "arXiv",
    primaryClass = "hep-th",
    doi = "10.1103/PhysRevD.80.123510",
    journal = "Phys. Rev. D",
    volume = "80",
    pages = "123510",
    year = "2009"
}

@article{Damour:2001bk,
    author = "Damour, Thibault and Vilenkin, Alexander",
    title = "{Gravitational wave bursts from cusps and kinks on cosmic strings}",
    eprint = "gr-qc/0104026",
    archivePrefix = "arXiv",
    reportNumber = "IHES-P-01-15",
    doi = "10.1103/PhysRevD.64.064008",
    journal = "Phys. Rev. D",
    volume = "64",
    pages = "064008",
    year = "2001"
}

@article{Gherghetta:2016fhp,
    author = "Gherghetta, Tony and Nagata, Natsumi and Shifman, Mikhail",
    title = "{A Visible QCD Axion from an Enlarged Color Group}",
    eprint = "1604.01127",
    archivePrefix = "arXiv",
    primaryClass = "hep-ph",
    reportNumber = "UMN-TH-3522-16, FTPI-MINN-16-11",
    doi = "10.1103/PhysRevD.93.115010",
    journal = "Phys. Rev. D",
    volume = "93",
    number = "11",
    pages = "115010",
    year = "2016"
}

@article{Berezhiani:2000gh,
    author = "Berezhiani, Zurab and Gianfagna, Leonida and Giannotti, Maurizio",
    title = "{Strong CP problem and mirror world: The Weinberg-Wilczek axion revisited}",
    eprint = "hep-ph/0009290",
    archivePrefix = "arXiv",
    reportNumber = "DFAQ-TH-2000-04",
    doi = "10.1016/S0370-2693(00)01392-7",
    journal = "Phys. Lett. B",
    volume = "500",
    pages = "286--296",
    year = "2001"
}

@book{Vilenkin:2000jqa,
    author = "Vilenkin, A. and Shellard, E. P. S.",
    title = "{Cosmic Strings and Other Topological Defects}",
    isbn = "978-0-521-65476-0",
    publisher = "Cambridge University Press",
    month = "7",
    year = "2000"
}

@article{Hindmarsh:1994re,
  ids = {hindmarsh_cosmic_1995},
  title = {Cosmic Strings},
  author = {Hindmarsh, M. B. and Kibble, T. W. B.},
  year = {1995},
  journal = {Rept. Prog. Phys.},
  volume = {58},
  number = {SUSX-TP-94-74, IMPERIAL-TP-94-95-5, NI-94025},
  eprint = {hep-ph/9411342},
  eprinttype = {arxiv},
  pages = {477--562},
  doi = {10.1088/0034-4885/58/5/001},
  archiveprefix = {arXiv}
}

@article{Steer:2010jk,
    author = "Steer, Daniele A. and Vachaspati, Tanmay",
    title = "{Light from Cosmic Strings}",
    eprint = "1012.1998",
    archivePrefix = "arXiv",
    primaryClass = "hep-th",
    doi = "10.1103/PhysRevD.83.043528",
    journal = "Phys. Rev. D",
    volume = "83",
    pages = "043528",
    year = "2011"
}

@article{Vachaspati:2015cma,
    author = "Vachaspati, Tanmay and Pogosian, Levon and Steer, Daniele",
    title = "{Cosmic Strings}",
    eprint = "1506.04039",
    archivePrefix = "arXiv",
    primaryClass = "astro-ph.CO",
    doi = "10.4249/scholarpedia.31682",
    journal = "Scholarpedia",
    volume = "10",
    number = "2",
    pages = "31682",
    year = "2015"
}

@article{Ade:2013xla,
      author         = "Ade, P. A. R. and others",
      title          = "{Planck 2013 results. XXV. Searches for cosmic strings
                        and other topological defects}",
      collaboration  = "Planck",
      journal        = "Astron. Astrophys.",
      volume         = "571",
      year           = "2014",
      pages          = "A25",
      doi            = "10.1051/0004-6361/201321621",
      eprint         = "1303.5085",
      archivePrefix  = "arXiv",
      primaryClass   = "astro-ph.CO",
      reportNumber   = "CERN-PH-TH-2013-138",
      SLACcitation   = "%%CITATION = ARXIV:1303.5085;%%"
}

@article{Caprini:2010xv,
    author = "Caprini, Chiara and Durrer, Ruth and Siemens, Xavier",
    title = "{Detection of gravitational waves from the QCD phase transition with pulsar timing arrays}",
    eprint = "1007.1218",
    archivePrefix = "arXiv",
    primaryClass = "astro-ph.CO",
    doi = "10.1103/PhysRevD.82.063511",
    journal = "Phys. Rev. D",
    volume = "82",
    pages = "063511",
    year = "2010"
}

@article{Neronov:2020qrl,
    author = "Neronov, Andrii and Roper Pol, Alberto and Caprini, Chiara and Semikoz, Dmitri",
    title = "{NANOGrav signal from magnetohydrodynamic turbulence at the QCD phase transition in the early Universe}",
    eprint = "2009.14174",
    archivePrefix = "arXiv",
    primaryClass = "astro-ph.CO",
    doi = "10.1103/PhysRevD.103.L041302",
    journal = "Phys. Rev. D",
    volume = "103",
    number = "4",
    pages = "041302",
    year = "2021"
}

@article{RoperPol:2022iel,
    author = "Roper Pol, Alberto and Caprini, Chiara and Neronov, Andrii and Semikoz, Dmitri",
    title = "{Gravitational wave signal from primordial magnetic fields in the Pulsar Timing Array frequency band}",
    eprint = "2201.05630",
    archivePrefix = "arXiv",
    primaryClass = "astro-ph.CO",
    doi = "10.1103/PhysRevD.105.123502",
    journal = "Phys. Rev. D",
    volume = "105",
    number = "12",
    pages = "123502",
    year = "2022"
}

@article{Cline:2025bwe,
    author = "Cline, James M. and Laurent, Benoit",
    title = "{Bubble wall velocity for first-order QCD phase transition}",
    eprint = "2502.12321",
    archivePrefix = "arXiv",
    primaryClass = "hep-ph",
    doi = "10.1103/PhysRevD.111.083522",
    journal = "Phys. Rev. D",
    volume = "111",
    number = "8",
    pages = "083522",
    year = "2025"
}

@article{Planck:2018vyg,
    author = "Aghanim, N. and others",
    collaboration = "Planck",
    title = "{Planck 2018 results. VI. Cosmological parameters}",
    eprint = "1807.06209",
    archivePrefix = "arXiv",
    primaryClass = "astro-ph.CO",
    doi = "10.1051/0004-6361/201833910",
    journal = "Astron. Astrophys.",
    volume = "641",
    pages = "A6",
    year = "2020",
    note = "[Erratum: Astron.Astrophys. 652, C4 (2021)]"
}

@article{Ringeval:2010ca,
      author         = "Ringeval, Christophe",
      title          = "{Cosmic strings and their induced non-Gaussianities in
                        the cosmic microwave background}",
      journal        = "Adv. Astron.",
      volume         = "2010",
      year           = "2010",
      pages          = "380507",
      doi            = "10.1155/2010/380507",
      eprint         = "1005.4842",
      archivePrefix  = "arXiv",
      primaryClass   = "astro-ph.CO",
      SLACcitation   = "%%CITATION = ARXIV:1005.4842;%%"
}

@article{Vachaspati:2009kq,
      author         = "Vachaspati, Tanmay",
      title          = "{Cosmic Rays from Cosmic Strings with Condensates}",
      journal        = "Phys. Rev.",
      volume         = "D81",
      year           = "2010",
      pages          = "043531",
      doi            = "10.1103/PhysRevD.81.043531",
      eprint         = "0911.2655",
      archivePrefix  = "arXiv",
      primaryClass   = "astro-ph.CO",
      SLACcitation   = "%%CITATION = ARXIV:0911.2655;%%"
}

@article{Long:2014mxa,
      author         = "Long, Andrew J. and Hyde, Jeffrey M. and Vachaspati,
                        Tanmay",
      title          = "{Cosmic Strings in Hidden Sectors: 1. Radiation of
                        Standard Model Particles}",
      journal        = "JCAP",
      volume         = "1409",
      year           = "2014",
      number         = "09",
      pages          = "030",
      doi            = "10.1088/1475-7516/2014/09/030",
      eprint         = "1405.7679",
      archivePrefix  = "arXiv",
      primaryClass   = "hep-ph",
      SLACcitation   = "%%CITATION = ARXIV:1405.7679;%%"
}

@article{MacGibbon:1989kk,
    author = "MacGibbon, Jane H. and Brandenberger, Robert H.",
    title = "{High-energy neutrino flux from ordinary cosmic strings}",
    reportNumber = "BROWN-HET-697",
    doi = "10.1016/0550-3213(90)90020-E",
    journal = "Nucl. Phys. B",
    volume = "331",
    pages = "153--172",
    year = "1990"
}

@article{Brandenberger:1986vj,
      author         = "Brandenberger, Robert H.",
      title          = "{On the Decay of Cosmic String Loops}",
      journal        = "Nucl. Phys.",
      volume         = "B293",
      year           = "1987",
      pages          = "812-828",
      doi            = "10.1016/0550-3213(87)90092-7",
      reportNumber   = "PRINT-86-1356 (CAMBRIDGE)",
      SLACcitation   = "%%CITATION = NUPHA,B293,812;%%"
}

@article{Auclair:2019wcv,
    author = "Auclair, Pierre and others",
    title = "{Probing the gravitational wave background from cosmic strings with LISA}",
    eprint = "1909.00819",
    archivePrefix = "arXiv",
    primaryClass = "astro-ph.CO",
    doi = "10.1088/1475-7516/2020/04/034",
    journal = "JCAP",
    volume = "04",
    pages = "034",
    year = "2020"
}

@article{Figueroa:2023zhu,
    author = "Figueroa, Daniel G. and Pieroni, Mauro and Ricciardone, Angelo and Simakachorn, Peera",
    title = "{Cosmological Background Interpretation of Pulsar Timing Array Data}",
    eprint = "2307.02399",
    archivePrefix = "arXiv",
    primaryClass = "astro-ph.CO",
    reportNumber = "CERN-TH-2023-132",
    month = "7",
    year = "2023"
}

@article{Romano:2016dpx,
    author = "Romano, Joseph D. and Cornish, Neil J.",
    title = "{Detection methods for stochastic gravitational-wave backgrounds: a unified treatment}",
    eprint = "1608.06889",
    archivePrefix = "arXiv",
    primaryClass = "gr-qc",
    doi = "10.1007/s41114-017-0004-1",
    journal = "Living Rev. Rel.",
    volume = "20",
    number = "1",
    pages = "2",
    year = "2017"
}

@article{Janssen:2014dka,
    author = "Janssen, Gemma and others",
    editor = "Bourke, Tyler L. and others",
    title = "{Gravitational wave astronomy with the SKA}",
    eprint = "1501.00127",
    archivePrefix = "arXiv",
    primaryClass = "astro-ph.IM",
    doi = "10.22323/1.215.0037",
    journal = "PoS",
    volume = "AASKA14",
    pages = "037",
    year = "2015"
}

@article{Zhou:2022otw,
    author = "Zhou, Bei and Reali, Luca and Berti, Emanuele and \c{C}al\i{}\c{s}kan, Mesut and Creque-Sarbinowski, Cyril and Kamionkowski, Marc and Sathyaprakash, B. S.",
    title = "{Compact Binary Foreground Subtraction in Next-Generation Ground-Based Observatories}",
    eprint = "2209.01221",
    archivePrefix = "arXiv",
    primaryClass = "gr-qc",
    month = "9",
    year = "2022"
}

@article{Branchesi_2023,
doi = {10.1088/1475-7516/2023/07/068},
url = {https://dx.doi.org/10.1088/1475-7516/2023/07/068},
year = {2023},
month = {jul},
publisher = {IOP Publishing},
volume = {2023},
number = {07},
pages = {068},
author = {Marica Branchesi and others},
title = {Science with the Einstein Telescope: a comparison of different designs},
journal = {Journal of Cosmology and Astroparticle Physics}
}

\end{document}